\documentclass[11pt]{article}

\usepackage[usenames,dvipsnames,svgnames,table]{xcolor} 
\usepackage[obeyspaces,hyphens,spaces]{url}
\usepackage{jcapmod}
\usepackage[english]{babel}
\usepackage{amsmath, amssymb, amsbsy, amstext, amsthm,mathtools}
\usepackage{graphicx}
\usepackage{amsfonts}
\usepackage{fouridx}
\usepackage{cases}
\usepackage{bm}
\usepackage{bbm}
\usepackage{cleveref}
\usepackage{slashed}
\usepackage{mdframed}
\usepackage{tensor}

\usepackage{setspace}

\usepackage{braket}

\usepackage{subcaption}

\usepackage{tikz}
\usetikzlibrary{calc,shadings,patterns,tikzmark,fadings}
\usetikzlibrary{decorations.pathreplacing,decorations.markings}

\definecolor{Blue}{rgb}{0.25, 0.41, 0.88}
\definecolor{Red}{rgb}{0.92,0.,0.}
\definecolor{darkorange}{rgb}{1.0,0.549,0.}
\definecolor{cobalt}{RGB}{44, 98, 120}
\definecolor{Mathematica1}{rgb}{0.368417, 0.506779, 0.709798}
\definecolor{Mathematica2}{rgb}{0.880722, 0.611041, 0.142051}
\definecolor{Mathematica3}{rgb}{0.560181, 0.691569, 0.194885}
\definecolor{Mathematica4}{rgb}{0.922526, 0.385626, 0.209179}
\definecolor{Mathematica5}{rgb}{0.528488, 0.470624, 0.701351}
\definecolor{Mathematica6}{rgb}{0.772079, 0.431554, 0.102387}
\definecolor{Mathematica7}{rgb}{0.363898, 0.618501, 0.782349}
\definecolor{Mathematica8}{rgb}{1, 0.75, 0}
\definecolor{Mathematica9}{rgb}{0.647624, 0.37816, 0.614037}
\definecolor{plotBlue}{RGB}{94, 130, 181}
\definecolor{plotRed}{RGB}{233, 85, 54}
\definecolor{plotGreen}{RGB}{142, 176, 50}
\definecolor{plotPurple}{RGB}{135, 120, 178}
\definecolor{cornellRed}{HTML}{B31B1B}
\definecolor{cornellBlue}{HTML}{0068AC}
\definecolor{cornellGreen}{HTML}{6EB43F}
\definecolor{regal}{RGB}{90,0,120}          

\newcolumntype{C}[1]{>{\centering\let\newline\\\arraybackslash\hspace{0pt}}m{#1}}

\def\e{{\lab{e}}}

\makeatletter
\newlength{\apb@width}
\newcommand{\autoparbox}[2][c]{\settowidth{\apb@width}{#2}\parbox[#1]{\apb@width}{#2}}

\makeatother

\makeatletter
\newsavebox\myboxA
\newsavebox\myboxB
\newlength\mylenA

\newcommand*\xoverline[2][0.75]{
    \sbox{\myboxA}{$\m@th#2$}%
    \setbox\myboxB\null
    \ht\myboxB=\ht\myboxA%
    \dp\myboxB=\dp\myboxA%
    \wd\myboxB=#1\wd\myboxA
    \sbox\myboxB{$\m@th\overline{\copy\myboxB}$}
    \setlength\mylenA{\the\wd\myboxA}
    \addtolength\mylenA{-\the\wd\myboxB}%
    \ifdim\wd\myboxB<\wd\myboxA%
       \rlap{\hskip 0.5\mylenA\usebox\myboxB}{\usebox\myboxA}%
    \else
        \hskip -0.5\mylenA\rlap{\usebox\myboxA}{\hskip 0.5\mylenA\usebox\myboxB}%
    \fi}
\makeatother

\numberwithin{equation}{section}

\newcommand{\cev}[1]{\reflectbox{\ensuremath{\vec{\reflectbox{\ensuremath{#1}}}}}}

\newcommand{\ud}{\mathrm{d}}
\newcommand{\lab}[1]{{\mathrm{#1}}}
\newcommand{\slab}[1]{{\textsc{#1}}}
\newcommand{\mb}[1]{{\mathbf{#1}}}
\newcommand{\minus}{{\scalebox {0.75}[1.0]{$-$}}}
\newcommand{\sminus}{{\scalebox {0.6}[0.75]{$-$}}}
\newcommand{\ssminus}{{\scalebox {0.5}[0.65]{$-$}}}

\newcommand{\nord}[1]{{:\mathrel{#1}:}}
\newcommand{\floq}[1]{{\scalebox{0.75}{$(#1)$}}}
\newcommand{\floqq}[1]{{\scalebox{0.65}{$(#1)$}}}

\theoremstyle{definition}

\DeclareRobustCommand{\SkipTocEntry}[4]{}

\newcommand{\es}{\hspace{0.5pt}}
\newcommand{\ess}{\hspace{1pt}}

\definecolor{blue2}{cmyk}{1, 0.1, 0.1, 0.1}

\definecolor{pyBlue}{RGB}{31, 119, 180}
\definecolor{pyRed}{RGB}{214, 39, 40}
\definecolor{pyGreen}{RGB}{44, 160, 44}
\definecolor{pyBlue2}{RGB}{0, 111, 237}
\definecolor{pyRed2}{RGB}{224, 52, 36}

\newcommand{\subp}{{\scriptscriptstyle +}}
\newcommand{\subm}{{\scriptscriptstyle -}}
\newcommand{\condp}[2]{p(\es #1 \ess | \ess #2 \es )}
\newcommand{\cond}[3]{#1(\es #2 \ess | \ess #3 \es )}
\newcommand{\dvg}[3]{\mathcal{D}_{\slab{#1}}[\ess #2 \, |\, #3 \ess ]}

\begin{document}

\pagenumbering{roman}
\begin{titlepage}
\baselineskip=15.5pt \thispagestyle{empty}

\bigskip\

\vspace{1cm}
\begin{center}
{\fontsize{20}{24}\selectfont  {\bfseries Infinite Distances and Factorization}}
\end{center}
\vspace{0.1cm}
\begin{center}
{John Stout} 
\end{center}

\begin{center}
\vskip8pt

\textit{Department of Physics, Harvard University, Cambridge, MA 02138, USA}

\end{center}

\vspace{1.4cm}
\hrule \vspace{0.3cm}
\noindent {\bf Abstract}\\[0.1cm]
	The information metric provides a unique notion of distance along any continuous family of physical theories, placing two theories further apart the more distinguishable they are. As such, the information metric is typically singular at quantum critical points, reflecting the fact they make qualitatively different scale-invariant predictions when compared to other theories in the family. However, such singularities are always at finite distance. The goal of this paper is to study infinite distance metric singularities and understand what type of physics generates them. We argue that infinite distance limits in the information metric correspond to theories in which expectation values factorize, and that unitarity restricts the asymptotic form of the metric to have a universal logarithmic singularity whose coefficient only depends on how the factorization limit is approached. 
    We illustrate this relationship in a set of wide-ranging examples. Furthermore, we argue that it provides a particularly simple bottom-up motivation for the seemingly universal behavior of quantum gravity in asymptotic regions of moduli space described by the Swampland Distance Conjecture: gravity universally couples to stress-energy and thus abhors factorization limits. The towers of light fields that appear in these limits in so many examples serve to decouple gravity and consistently allow for factorization. We make this more precise via holography and suggest a way to consistently realize infinite distance limits without a tower of light fields.
\vskip12pt
\hrule
\vskip10pt

\end{titlepage}

\thispagestyle{empty}
\setcounter{page}{2}
\begin{spacing}{1}
\tableofcontents
\end{spacing}

\clearpage
\pagenumbering{arabic}
\setcounter{page}{1}

\newpage

\section{Introduction} \label{sec:intro}

    A major goal of theoretical physics is to characterize the space of theories consistent with a given set of principles and understand the range of physical behaviors they permit. It has proved particularly fruitful to focus on the extremes of this space---the much smaller set of exceptional or qualitatively unique theories---as a way of understanding the entirety of it. Indeed, much like the columns propping up the canopy of an enormous tent, our modern understanding and definition of the space of relativistic quantum field theories is braced by quantum critical points and the conformal field theories that live at them.

    It has also proved useful to encode the relations among different theories geometrically~\cite{Douglas:2010ic}. This is most familiar in continuous families of supersymmetric or conformal field theories, in which case the symmetry group identifies a natural Riemannian metric called the moduli space~\cite{Strominger:1990pd,Seiberg:1994bz,Seiberg:1994rs,Intriligator:1995au,Dijkgraaf:1997ip,DHoker:1999yni,Strassler:2003qg,Freedman:2012zz} or Zamolodchikov metric~\cite{Zamolodchikov:1986gt,Seiberg:1988pf,Kutasov:1988xb,dijkgraaf1989geometrical, Cvetic:1989ii,Candelas:1989qn,Candelas:1989ug,Ranganathan:1993vj,Cecotti:1991me,Leigh:1995ep,Seiberg:1999xz,deBoer:2008ss,Papadodimas:2009eu,Green:2010da,Gerchkovitz:2014gta,Gomis:2015yaa,Behan:2017mwi}, respectively. However, these are special cases of a more general structure: the information metric~\cite{rao1992information,provost1980,Wootters:1981ki,amari2016information,cover2006elements,bengtsson2017geometry,Caticha:2008eso,nielsen2020elementary} provides a general notion of distance on any continuous family of theories, which happily reduces to the moduli space or Zamolodchikov metric when restricted to the appropriate class of theories. This information metric quantifies how different two nearby theories are based on the predictions that they make, and places them further apart the easier it is to distinguish between them. Accordingly, this information metric is \emph{singular} at quantum critical points~\cite{Venuti:2007qcs,Carollo:2019ygj}, reflecting the fact that these theories display scale-invariant fluctuations and are thus qualitatively different than those gapped theories around them. However, these critical points are always associated with finite distance metric singularities~\cite{Stout:2021ubb}. 

    This paper is about infinite distance metric singularities. Given the importance of their finite distance counterparts, which are so crucial to our understanding of the space of quantum field theories, one naturally wonders whether theories at infinite distance enjoy a similar role. Infinite distance limits have recently become a subject of intense theoretical interest, as they arise ubiquitously in the asymptotic limits of continuous families of quantum gravitational theories. Consistent quantum gravitational theories are found to behave universally in these limits~\cite{Ooguri:2006in,Brennan:2017rbf,Palti:2019pca,vanBeest:2021lhn,Grana:2021zvf,Grimm:2018ohb,Lee:2019wij,Baume:2020dqd,Perlmutter:2020buo}, where an infinite tower of fields become light and local effective field theory necessarily breaks down. At the same time, infinite distance points arise in the simplest quantum mechanical models~\cite{Stout:2021ubb}. What, then, does the simple harmonic oscillator have in common with F-theory compactified along a degenerating elliptically fibered Calabi-Yau threefold?  Why does gravity demand this seemingly universal behavior?  Is there an organizational principle---like the conformal invariance of critical points---that can help us understand these infinite distance points?

    The goal of this paper is to argue that infinite distance limits in the information metric correspond to theories in which expectation values factorize, probability collapses onto a discrete set of events, and statistical ambiguity vanishes.  Moreover, we argue that unitarity restricts the asymptotic form of the metric in this limit to have a universal logarithmic singularity. 
    We illustrate this relationship in several examples, and argue that it provides a particularly simple bottom-up motivation for why quantum gravity behaves the way it does near infinite distance points: gravity universally couples to stress-energy and thus abhors factorization limits. The towers of light fields that appear in these limits in so many examples serve to decouple gravity and consistently allow for factorization. We make this more precise via holography.

    \paragraph{Outline}
        
    We begin in Section~\ref{sec:review} by reviewing the various aspects of information theory and geometry relevant to our main story. In \S\ref{sec:classicalIG}, we introduce the classical information metric, which defines a notion of distance along arbitrary continuous families of classical probability distributions, and discuss its operational interpretation in terms of the distinguishability of nearby theories. We introduce its quantum mechanical analog in \S\ref{sec:quantumIG}, which defines a similar notion of distance between quantum states. We use this in \S\ref{sec:ftInfoGeo} to define a metric on continuous families of quantum theories via their vacuum states. As we discuss there, the quantum information metric is extensive in the number of degrees of freedom in the theory and must be regulated. We thus define the intensive quantum information metric, which is proportional to the metric on moduli space when it exists, and describe how it can be singular at a quantum critical point. However, such singularities are always at finite distance.

    In Section~\ref{sec:infFact}, we consider infinite distance singularities in the information metric. In \S\ref{sec:classicalInfFact}, we study how a one-parameter family of classical probability distributions $\condp{x}{\epsilon}$ must behave in the limit $\epsilon \to 0$ in order for $\epsilon = 0$ to be at infinite distance in the information metric. We argue that an infinite distance singularity signals that the family $\condp{x}{\epsilon}$ degenerates and approaches a factorization limit. In these limits, the statistical ambiguity in samples drawn from the distribution disappears and connected correlation functions (or cumulants) vanish. Under fairly relaxed assumptions, unitarity forces the metric to have a universal logarithmic metric singularity, $\ud s^2 \sim \mathcal{C}\,  \ud \epsilon^2/\epsilon^2$, where the coefficient $\mathcal{C}$ depends on precisely how the family $\condp{x}{\epsilon}$ approaches this factorization limit. The reason for this universality is simple: probability distributions are normalized, so any increase in their height must come at the expense of their width. Degenerating families thus obey a scaling symmetry as $\epsilon \to 0$, resulting in the universal logarithmic singularity. We extend this to quantum states and the quantum information metric in \S\ref{sec:quantumInfFact}.

    In Section~\ref{sec:examples}, we illustrate the relationship between factorization and infinite distance in a variety of examples. In Section~\ref{sec:largeN}, we discuss how to define the information metric on a family of theories whose number $\propto N$ of fundamental degrees of freedom changes, illustrating this for probability distributions, quantum mechanical theories, and quantum field theories in \S\ref{sec:largeNClassical}, \S\ref{sec:quantLargeN}, and \S\ref{sec:ftLargeN}, respectively.  We expect that large-$N$ factorization implies that all large-$N$ limits, as long as they exist, are at infinite distance in this information metric and we show this explicitly for $\lab{O}(N)$-invariant vector models.

    In Section~\ref{sec:swamp}, we discuss the implications this relationship has for quantum gravity. As mentioned above, consistent quantum gravitational theories seem to universally feature a tower of fields that become exponentially light with respect to the Planck mass as we approach an infinite distance point. The Swampland Distance Conjecture proposes that this behavior holds for all consistent quantum gravitational theories. We argue that this behavior is a consequence of gravity's universal interaction with stress-energy. It is extremely difficult to consistently realize a factorization limit in quantum gravity because gravity couples \emph{everything} and fluctuations of spacetime always introduce inherent ambiguities or noise to any observation. Gravity must then decouple to consistently realize a factorization limit, and this is the functional role of the tower of light fields. By requiring that the Planck mass diverges and that the information metric has the universal logarithmic singularity discussed above as we approach an infinite distance limit, we derive a form of the Swampland Distance Conjecture from purely bottom-up reasoning.

    We make this picture more precise using holography, where gravity's universal obstruction to factorization can be seen from the fact that a conformal field theory's stress-energy tensor, which is dual to gravity in the bulk, provides a universal contribution to $n$-point functions whose form is entirely determined by conformal symmetry. This contribution can only be removed (barring a familiar loophole which we discuss) by sending the central charge of the theory, and thus the bulk Planck mass, to infinity. This points to a potential class of counterexamples to the Swampland Distance Conjecture that consistently realizes a factorization limit without the infinite tower of exponentially light fields.

    Finally, in Section~\ref{sec:conclusions} we conclude with some open questions and future directions.

    \paragraph{Conventions}

    We work in $(d+1)$-dimensional spacetime, where $d$ is the spatial dimension. We will generally work in Euclidean signature and use $\tau$ to denote Euclidean time, while bold-faced Roman letters like $\mb{x}$ denote spatial points. We use Greek indices at the end of the alphabet, $\mu, \nu, \ldots = 0, \ldots\!\es\es\es, d$, as spacetime indices. 
    We use $M$ to denote the number of samples taken from a distribution or theory, $N$ to denote the total number of degrees of freedom or, in Section~\ref{sec:largeN}, the rank of a symmetry group, and $n$ to denote the dimension of parameter space. We use $\varphi^a$, with indices from Roman letters at the beginning of the alphabet ${a = 1, \ldots\!\es\es\es, n}$ to denote coordinates on parameter or moduli space. When there is no chance of ambiguity, we suppress this index---for example, we will denote a continuous family of probability distributions as $\condp{x}{\varphi^a} \equiv \condp{x}{\varphi}$, though we will quickly restrict to families in a single non-negative parameter $\varphi = \epsilon \geq 0$.

\section{The Classical and Quantum Information Metrics} \label{sec:review}
	
	In this section, we review some relevant aspects of classical and quantum information theory. In \S\ref{sec:classicalIG}, we introduce the classical information metric and discuss its precise operational interpretation. In \S\ref{sec:quantumIG}, we introduce the quantum information metric and discuss its relationship to its classical counterpart. Finally, in \S\ref{sec:ftInfoGeo}, we discuss how to use the quantum information metric to define a notion of distance between different quantum field theories. We describe this metric's relationship to the more familiar metrics defined for supersymmetric or conformal field theories and then discuss its behavior near quantum critical points.

    \subsection{Classical Information Geometry} \label{sec:classicalIG}

      	Any family of probability distributions $\condp{x}{\varphi}$ for a continuous random variable $x \in \mathbb{R}^N$ defines a \emph{statistical manifold}, parameterized by the continuous coordinates $\varphi \in \mathbb{R}^n$, in which each point~$\varphi$ represents a single probability distribution $\condp{x}{\varphi}$. Can we define a natural notion of distance between two points on this manifold, say $\condp{x}{\varphi}$ and $\condp{x}{\varphi'}$, which properly treats these points as \emph{probability distributions}? A distance that places two distributions further apart if they make very different predictions, and close together if they are nearly the same? This is the focus of classical information geometry~\cite{amari2016information,nielsen2020elementary,cover2006elements,Caticha:2008eso,bengtsson2017geometry}, whose basics we review here.

      	\newpage
        One of the most useful measures of separation between any two probability distributions $\condp{x}{\varphi}$ and $\condp{x}{\varphi'}$ is the relative entropy or Kullback-Leibler (KL) divergence, defined as
        \begin{equation}
            \dvg{kl}{\varphi}{\varphi'} = \int\!\ud^N x\, \condp{x}{\varphi}\ess \log \frac{\condp{x}{\varphi}}{\condp{x}{\varphi'}}\,. \label{eq:klDivergence}
        \end{equation}
        This is not a distance between the two distributions since it is asymmetric in its arguments, but it provides a particularly natural way (which is, under reasonable assumptions~\cite{Hobson:1969ant}, unique up to an overall constant) of characterizing how different two probability distributions are based on the predictions they make. Importantly, this divergence is both positive definite, $\dvg{kl}{\varphi}{\varphi'} \geq 0$, and vanishes if and only if the two distributions are the identical, $\condp{x}{\varphi} = \condp{x}{\varphi'}$.

        The KL divergence has a useful operational interpretation~\cite{cover2006elements} in terms of a classification problem. Suppose we are given a set of $M$ random samples $\{x_i\}$ which are all drawn independently and identically from either the ``reference'' distribution $\condp{x}{\varphi_0}$ or the ``test'' distribution $\condp{x}{\varphi}$. We are tasked with designing a test that determines whether or not these samples were drawn from the reference distribution $\condp{x}{\varphi_0}$. That is, this test determines the validity of the statement ``\emph{The samples $\{x_i\}$ were drawn independently and identically from $\condp{x}{\varphi_0}$.}'' when evaluated on the set $\{x_i\}$. This test should return \texttt{True} if the $\{x_i\}$ were indeed drawn from $\condp{x}{\varphi'}$ and \texttt{False} if they were drawn from $\condp{x}{\varphi}$. Furthermore, an optimal test will return \texttt{True} on a set of samples that were drawn from $\condp{x}{\varphi_0}$ with a probability that approaches $1$ as that set becomes large enough to suppress statistical noise, $M \to \infty$. 

        Such a test cannot be perfect, however, and it will return \texttt{True} on samples actually drawn from the test distribution $\condp{x}{\varphi}$ with probability $F$. This is the probability our test fails to properly distinguish between the two distributions. Sanov's theorem~\cite{cover2006elements} states that the best possible tests---those with the smallest possible $F$---fail with a probability that shrinks to $0$ in the large sample size limit $M \to \infty$ at a rate governed by the KL divergence,
        \begin{equation}
        	F \sim \mathcal{C} \exp\big(- M \dvg{kl}{\varphi}{\varphi_0}\big)\,, \mathrlap{\qquad M \to \infty\,,} \label{eq:sanovs}
        \end{equation}
        where $\mathcal{C}$ is an $M$-independent constant. The KL divergence thus measures the theoretically optimal rate at which we can distinguish between two distributions in the large sample size limit.

        By convention, we say that the two distributions $\condp{x}{\varphi}$ and $\condp{x}{\varphi'}$ are \emph{distinguishable in $M$ measurements} if
        \begin{equation}
        	M \dvg{kl}{\varphi}{\varphi'} \geq 1\,. \label{eq:klDistinguishability}
        \end{equation}
        The KL divergence thus defines a measure of separation between two probability distributions based on how easy it is to distinguish between them via the predictions they make, in terms of the needed number of samples, using the best possible tests. However, it is not a distance on the statistical manifold because there is an inherent asymmetry between the reference distribution $\condp{x}{\varphi_0}$ (i.e. the null hypothesis) and the test distribution $\condp{x}{\varphi}$.\footnote{Another way of interpreting the KL divergence $\dvg{kl}{\varphi}{\varphi'}$ is that it is the average surprise of seeing data that looks like it is drawn from $\condp{x}{\varphi'}$ when you expected it to be drawn from $\condp{x}{\varphi}$. Seeing data that is distributed with a very heavy tail $p(x) \propto (1 + x^2)^{\sminus1}$ is extremely surprising if we expect it to be normally distributed, $p(x) \propto \exp(\minus x^2/2)$. Conversely, seeing normally distributed data is not so surprising if we expect it to be distributed with a heavy tail. Said simply, seeing many events which we expect to be rare is much more surprising than seeing many events which we expect to be common. This is why the KL divergence is asymmetric.}

        Though the KL divergence is not a distance itself, it can be used to define one. The divergence's most important property is that it is convex, and so its Hessian
        \begin{equation}
            g_{ab}(\varphi) = \left.-\frac{\partial^2 \dvg{kl}{\varphi}{\varphi'}}{\partial \varphi^a \es \partial \varphi'^b}\right|_{\varphi'=\varphi}\,, \label{eq:classInfMetHess}
        \end{equation}
        is positive definite. We can interpret this Hessian as a Riemannian metric on the statistical manifold~\cite{rao1992information}, called the Fisher or \emph{classical information metric}, defined explicitly in terms of the distribution $\condp{x}{\varphi}$ by
        \begin{equation}
             g_{ab}(\varphi) = \int\!\ud^N x\, \condp{x}{\varphi} \, \frac{\partial \log \condp{x}{\varphi}}{\partial \varphi^a} \frac{\partial \log \condp{x}{\varphi}}{\partial \varphi^b} =  -\!\ess\int\!\ud^N x\, \condp{x}{\varphi}\ess \frac{\partial^2 \log \condp{x}{\varphi}}{\partial \varphi^a \, \partial \varphi^b}\,. \label{eq:fimDef}
        \end{equation}
        Since $\dvg{kl}{\varphi + \ud \varphi}{\varphi}\approx \frac{1}{2} g_{ab}(\varphi)\,  \ud \varphi^a\, \ud \varphi^b$, this information metric inherits many of the KL divergence's most useful functional and interpretational aspects.

        For instance, two distributions are far apart from one another in the distance defined by (\ref{eq:fimDef}) if they make very different predictions, and so this notion of distance also appropriately treats each point $\varphi$ on the statistical manifold as a probability distribution. Following (\ref{eq:klDistinguishability}),  we say that the distribution $\condp{x}{\varphi}$ can be distinguished from the nearby distribution $\condp{x}{\varphi+\ud \varphi}$ in $M$ measurements if
        \begin{equation}
            \frac{1}{2} \ess g_{ab}(\varphi)\ess \ud \varphi^a \ess \ud \varphi^b \gtrsim \frac{1}{M}\,. \label{eq:metDistinguishability}
        \end{equation}
        This condition defines an ellipsoid in the statistical manifold, inside of which lie distributions that cannot be distinguished from $\condp{x}{\varphi}$ in $M$ measurements. The size of this ellipsoid shrinks as $M \to \infty$, matching our intuition that we should be able to distinguish between different distributions, no matter how similar they are, as long as we take enough data. In fact, it can be shown~\cite{cencov2000statistical,Cambpell:1986aec} that this is the unique Riemannian metric on the statistical manifold, up to an overall constant, that properly treats each point as a probability distribution.

\begin{figure}
                \centering
                \includegraphics{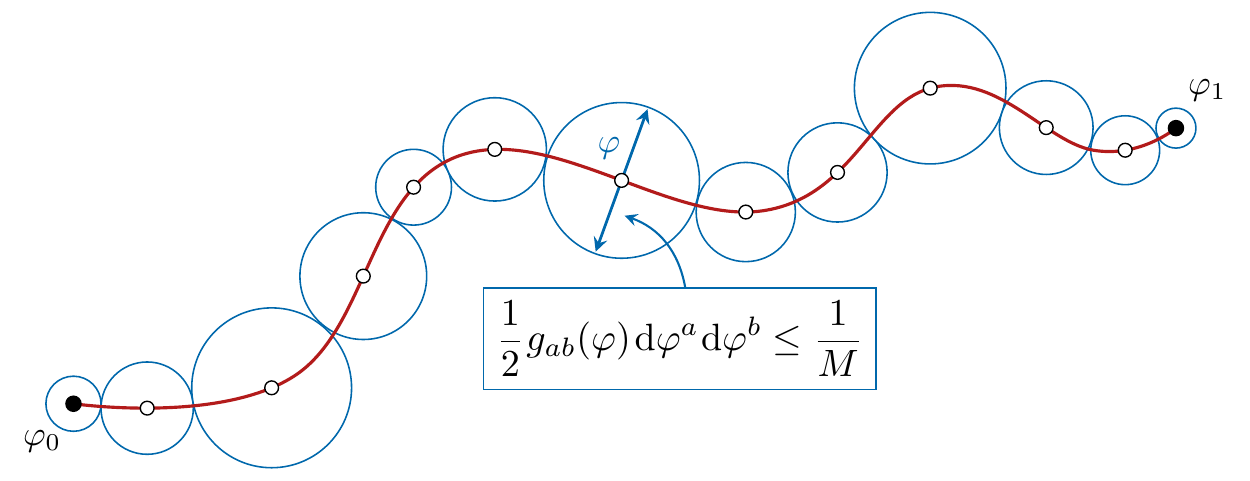}
                \caption{For finitely many samples, we can draw an ``ellipsoid of distinguishability'' around the point $\varphi$, defined by $M \mathcal{D}_\slab{kl}[\varphi + \ud \varphi \es | \es \varphi] \approx \frac{1}{2}M g_{ab}(\varphi)\es \ud \varphi^a \ud \varphi^b = 1$. Here, we have assumed that the metric is isotropic so that these ellipsoids are circles.  The statistical length of the curve the number of such ellipsoids that we can fit along the curve, or equivalently the number of distinguishable distributions, divided by a conventional factor of~$\sqrt{M/2}$. \label{fig:statLength}}
        \end{figure}

        This provides a direct operational interpretation of the \emph{statistical length} along a curve inside the statistical manifold in terms of the number of distributions that can be distinguished along it. As illustrated in Figure~\ref{fig:statLength}, we can segment the curve $\varphi^a(t)$, with $t \in [0, 1]$, connecting the endpoint distributions $\varphi^a_0 = \varphi^a(0)$ and $\varphi^a_1 = \varphi^a(1)$,  with non-overlapping ``ellipsoids of indistinguishability''~(\ref{eq:metDistinguishability}) and count the number $N_\lab{dist}(M)$ that can be fit along it. The statistical length is then directly related to this count in the limit of large sample size,
         \begin{equation}
            d(\varphi_1, \varphi_0) = \int_{0}^{1}\!\ud t\, \sqrt{g_{ab}\big(\varphi(t)\big) \, \dot{\varphi}^a(t)\, \dot{\varphi}^b(t)} = \lim_{M \to \infty} \frac{N_\lab{dist}(M)}{\sqrt{M/2}}\,. \label{eq:statLength}
        \end{equation}
       	We divide by a conventional $\sqrt{M/2}$, which can be thought of as the ``standard'' statistical ambiguity or root-mean-square noise associated with trying to distinguish between normal distributions of different mean but unit variance, $\condp{x}{\mu} = \e^{\sminus (x-\mu)^2/2}/\sqrt{2\pi}$.
        
        \newpage
        Crucially, this metric is independent of which degrees of freedom we choose to express the probability distribution in terms of. We are free to perform an arbitrary change of basis $x \to y = y(x)$ without affecting the metric as long as this coordinate change does not depend on the parameters $\varphi$. This invariance of the classical information metric under redefinitions of the random variables is tightly related to the fact that the metric characterizes how distinguishable members of the family $\condp{x}{\varphi}$ are from one another \emph{using the most discerning measurements possible}. If a $\varphi$-dependent basis transformation is used, we must be careful in defining exactly what is held fixed when we compute variations $\partial/\partial \varphi^a$. Otherwise, the information metric would change simply because it would then be characterizing an entirely different family of distributions. 

        The main focus of this work is on infinite distance limits in the information metric. If the information metric assigns infinite distance between two infinitesimally separated distributions, they must be extremely easy to distinguish between in the sense that only finitely many samples are needed to decide \emph{with certainty} which distribution those samples came from.\footnote{This interpretation relies on the equivalence of the KL divergence and statistical length for nearby distributions. However, two equivalent distributions can be separated by an infinite statistical length if the path taken in parameter space, for example, is either extremely circuitous or passes through an infinite distance singularity. In these cases, the higher-order differences between the statistical length and KL divergence can accumulate and~diverge. } A distribution that is at infinite distance is thus \emph{hyper-distinguishable}. In Section~\ref{sec:infFact}, we will study how the family of distributions $\condp{x}{\varphi}$ must behave to generate an infinite distance point, and argue that the only way this can happen is if the family degenerates and assigns non-zero probability to a discrete set of points. In this limit, expectation values for \emph{some} degrees of freedom must factorize and any statistical ambiguity in those observations vanishes. An analogous argument applies to the quantum information metric, the quantum analog of the classical information metric, whose definition and properties we now review.

        \newpage 
    \subsection{Quantum Information Geometry} \label{sec:quantumIG}

        As its name suggests, the quantum information metric~\cite{Wootters:1981ki,provost1980,bengtsson2017geometry} is the quantum mechanical analog of the classical information metric.
   	 	There are many ways to motivate its definition, but perhaps the most natural is the following. 

   	 	Let us consider a family of quantum states $|\Psi(\varphi)\rangle$ in an $(N+1)$-dimensional Hilbert space. This is the quantum mechanical analog of the family of probability distributions $\condp{x}{\varphi}$ discussed in the previous section. Since the Hilbert space is spanned by a set of complex vectors $\psi_i \in \mathbb{C}^{N+1}$, $i = 1, \ldots, N+1$, modulo an overall complex rescaling $\psi_i \sim \lambda \psi_i$ by $\lambda \in \mathbb{C}$, it can be identified with $N$-dimensional complex projective space~$\mathbb{CP}^{N}$, which is endowed with a natural metric called the \emph{Fubini-Study metric}. The family of states~$|\Psi(\varphi)\rangle$ then defines a submanifold in $\mathbb{CP}^N$, onto which we can pull back the Fubini-Study metric with line element 
        \begin{equation}
            \ud s^2 = g_{ab}(\varphi) \, \ud \varphi^a \, \ud \varphi^b = \frac{\langle \ud \Psi | \ud \Psi \rangle}{\langle \Psi | \Psi \rangle} - \frac{\langle \ud \Psi | \Psi \rangle}{\langle \Psi | \Psi \rangle} \frac{\langle \Psi | \ud \Psi \rangle}{\langle \Psi | \Psi \rangle}\,. \label{eq:qimDef}
        \end{equation} 
        This is the \emph{quantum information metric}.
        Usefully, it can also be computed from a potential,
        \begin{equation}
            g_{ab}(\varphi) = \left.\frac{\partial^2 \mathcal{K}(\varphi, \varphi')}{\partial \varphi^a \, \partial \varphi^b}\right|_{\varphi = \varphi'} \label{eq:qimKahler}
        \end{equation}
        where we have defined the log-fidelity or the K\"{a}hler potential,
        \begin{equation}
            \mathcal{K}(\varphi, \varphi') = \log \, \langle \Psi(\varphi) | \Psi(\varphi') \rangle\,. \label{eq:kahlerPotential}
        \end{equation}
        In general,~$|\Psi(\varphi)\rangle$ need not define a complex submanifold in $\mathbb{CP}^N$, nor do the coordinates $\varphi^a$ need to be complex. In the case that they are real, (\ref{eq:kahlerPotential}) defines a real-valued analog of K\"{a}hler geometry, called a Hessian geometry~\cite{nielsen2020elementary,shima2007geometry,zhang2014divergence}. However, when $|\Psi(\varphi)\rangle$ defines a complex manifold it is legitimately K\"{a}hler, and so we will use this familiar terminology for both cases even if it is not necessarily precise.

        This definition may seem a bit ad hoc but, aside from its geometric motivation, it is also natural from an information-theoretic perspective. By projecting the family of states $|\Psi(\varphi)\rangle$ onto a fixed basis $|x \rangle$ with $x \in \mathbb{R}^N$,
        \begin{equation}
            \langle x |\Psi(\varphi) \rangle = \sqrt{\condp{x}{\varphi}}\, \e^{i \alpha(x\,|\,\varphi)}\,,
        \end{equation}
        we may define a corresponding family of classical probability distributions $\condp{x}{\varphi} \equiv |\langle x | \Psi(\varphi)\rangle|^2$ and phase functions $\exp\!\big(i\alpha(\es x\ess|\ess \varphi)\big)$. The quantum information metric is then expressed as~\cite{Facchi_2010}
        \begin{equation}
            \ud s^2 = g_{ab}(\varphi)\, \ud \varphi^a \,\ud \varphi^b = \frac{1}{4} \ess \mathbb{E}_p\big[(\ud \log p)^2\big] + \mathbb{E}_p\big[(\ud \alpha)^2\big] - \big(\mathbb{E}_p[\ud \alpha]\big)^2\,, \label{eq:qimCim}
        \end{equation}
        where we denote expectation values with respect to $\condp{x}{\varphi}$ as $\mathbb{E}_p\big[f(x)\big] = \int\!\ud x\, \condp{x}{\varphi} f(x)$. For trivial phase functions, $\ud \alpha = 0$, the quantum information metric reduces to the classical one for $\condp{x}{\varphi}$,  up to a constant factor of $1/4$.

        Furthermore, the contribution from a non-trivial phase $\exp\!\big(i\cond{\alpha}{x}{\varphi}\big)$ is necessarily non-negative, $\mathbb{E}_p\big[(\ud \alpha)^2\big] - \big(\mathbb{E}_p[\ud \alpha]\big)^2 \geq 0$. The associated classical information metric for the distribution thus provides a lower bound on the full quantum information metric. Intuitively, this makes sense---access to the phase information contained in the state (or knowledge of the ``optimal'' basis in which $\alpha = 0$ in which to discern between members of $|\Psi(\varphi) \rangle$) provides \emph{additional} information and thus \emph{more} power to distinguish between nearby quantum states than that which is contained solely in the classical distribution $\condp{x}{\varphi}$. As in the classical case, the quantum information metric characterizes how distinguishable a given quantum state is from its neighbors using measurements from the most discerning possible operators or measurements~\cite{Wootters:1981ki,bengtsson2017geometry}.

    \subsection{Field Theoretic Information Geometry} \label{sec:ftInfoGeo}
    	
    	While the quantum information metric is defined on families of quantum states, we can also use it to define a notion of distance between different \emph{quantum field theories} by using the quantum information metric associated to their vacua, which we denote $|\Omega(\varphi)\rangle$.

        There are a variety of ways to compute this information metric. For simple theories, it is often easiest to explicitly construct the ground state and compute the K\"{a}hler potential (\ref{eq:kahlerPotential}). However, for more complicated theories it can be better to use a prescription that is more amenable to perturbation theory. For instance, we may consider a family of $(d+1)$-dimensional theories, defined by the Euclidean action $S_\slab{e}[\varphi, \eta]$, which is deformed by the local operators
        \begin{equation}
            S_\slab{e}[\varphi + \ud \varphi, \eta] = S_\slab{e}[\varphi, \eta] + \ud \varphi^a \left[\Lambda_\slab{uv}^{d+1 - \Delta_a} \int\!\ud^{d+1} x\, \mathcal{O}_a(x)\right]\,, \label{eq:actionDeformation}
        \end{equation}
        as we vary the parameters $\varphi^a$. Here, $\eta$ denotes the ``fundamental'' fields over which we path integrate and the operators $\mathcal{O}_a(x)$ are taken to have scaling dimensions $\Delta_a$. For convenience, we will assume that the operators $\mathcal{O}_a(x)$ are time-reversal symmetric as this simplifies our presentation, though this assumption is not necessary. We introduce a conventional factor of $\Lambda_\slab{uv}^{d+1 - \Delta_a}$ to ensure that the parameters $\varphi^a$ are dimensionless. The quantum information metric $g_{ab}(\varphi)$ can then be extracted from the Euclidean connected two-point function $\langle \mathcal{O}_a(x) \mathcal{O}_b(y)\rangle_\lab{c} \equiv \langle \mathcal{O}_a(x) \mathcal{O}_b(y) \rangle - \langle \mathcal{O}_a(x) \rangle \langle \mathcal{O}_b(y)\rangle $ in the theory described by $S_\slab{e}[\varphi, \eta]$ \cite{MIyaji:2015mia,Alvarez-Jimenez:2017gus,Trivella:2016brw,Bak:2015jxd,Bak:2017rpp,Belin:2018bpg,Stout:2021ubb}
        \begin{equation}
            g_{ab}(\varphi) = \Lambda_{\slab{uv}}^{2 d + 2 - \Delta_a - \Delta_b}\int_{\varepsilon}^{\infty}\!\ud \tau_1 \int_{\sminus \infty}^{\sminus \varepsilon}\!\ud \tau_2 \int_{L^d}\!\ud^d x_1 \, \ud^d x_2\, \langle \mathcal{O}_a(\tau_1, \mb{x}_1) \mathcal{O}_b(\tau_2, \mb{x}_2)\rangle_\lab{c}\,. \label{eq:infMetCorr}
        \end{equation}
        This expression diverges in both the UV and IR, and so we must regulate it by introducing the UV and IR cutoffs $\Lambda_\slab{uv} = \varepsilon^{\sminus 1}$ and $\Lambda_\slab{ir} = L^{\sminus 1}$, respectively.

        Except when exactly at a quantum critical point, the information metric for a family of quantum field theories is \emph{extensive} in the number of degrees of freedom, and so it diverges as we try to remove the UV and IR cutoffs, $g_{ab}(\varphi) \propto (L \Lambda_\slab{uv})^d$. This divergence arises because the square of the vacuum wavefunctional of a quantum field theory, $\big|\Psi[\eta(\mb{x})]\big|^2$, yields a distribution over \emph{field configurations}. Each draw from this distribution provides $M \sim (L \Lambda_\slab{uv})^d$ pieces of information~\cite{Balasubramanian:2014bfa} which we can use to distinguish between nearby theories. From~(\ref{eq:metDistinguishability}), we thus expect that the information metric scales extensively in the number of spatial sites, $g_{ab} \propto (L \Lambda_\slab{uv})^d$, much in the same way the total energy of a system scales extensively in the volume, $E \sim L^d$. We often use intensive quantities like the energy density to characterize a physical system,  and it will be similarly useful to study the \emph{intensive quantum information metric},
        \begin{equation}
            \tilde{g}_{ab}(\varphi) \equiv \lim_{\substack{L \to \infty \\  \Lambda_\slab{uv} \to \infty}} \mathcal{C} (L \Lambda_\slab{uv})^{\sminus d}\,  g_{ab}(\varphi) \label{eq:intensiveMetric}
        \end{equation}
        and its associated line element $\ud \tilde{s}^2 = \tilde{g}_{ab}\,\ud \varphi^a \, \ud \varphi^b$. Here, $\mathcal{C}$ is an overall constant that depends precisely on how we regulate the theory, and we will discuss how to define it in Section~\ref{sec:examples}.

        One of the major benefits of using this information metric to define a notion of distance between arbitrary QFTs is that it reduces to known cases when restricted to those highly-symmetric classes of theories in which a natural metric has been identified. It thus provides a unifying notion of distance between QFTs that incorporates these more specialized cases. For instance, in a family of conformal field theories (CFTs) that are related to one another as in (\ref{eq:actionDeformation}) by the exactly marginal operators $\mathcal{O}_a(x)$, each with scaling dimension $\Delta_a = d+1$, there is a natural metric on this \emph{conformal manifold} called the \emph{Zamolodchikov metric} $G^{\slab{z}}_{\smash{ab}}(\varphi)$~\cite{Zamolodchikov:1986gt,Seiberg:1988pf,Kutasov:1988xb,dijkgraaf1989geometrical, Cvetic:1989ii,Candelas:1989qn,Candelas:1989ug,Ranganathan:1993vj,Cecotti:1991me,Leigh:1995ep,Seiberg:1999xz,deBoer:2008ss,Papadodimas:2009eu,Green:2010da,Gerchkovitz:2014gta,Gomis:2015yaa,Behan:2017mwi}, which is defined by the normalization of the marginal operator's two-point functions
        \begin{equation}
        	\langle \mathcal{O}_a(x) \mathcal{O}_b(y) \rangle = \frac{G^{\slab{z}}_{\smash{ab}}(\varphi)}{|x - y|^{2 d +2}}\,.
        \end{equation}
        By using (\ref{eq:infMetCorr}), the extensive quantum information metric is then~\cite{MIyaji:2015mia}
        \begin{equation}
        	g_{ab}(\varphi) = \int_{\varepsilon}^{\infty}\!\ud \tau_1 \int_{\sminus \infty}^{\sminus \varepsilon}\!\ud \tau_2 \int_{L^d} \!\!\ud^d x_1 \ess \ud^d x_2 \, \frac{G^{\slab{z}}_{\smash{ab}}(\varphi)}{|x_1 - x_2|^{2 d+2}} \sim \mathcal{C} (L \Lambda_\slab{uv})^d  G^{\slab{z}}_{\smash{ab}}(\varphi)
        \end{equation}
        with $\mathcal{C} = {(\pi/4)^{\frac{1}{2}(d+1)}}/{\big[d \ess \Gamma\big(\frac{d+3}{2}\big)}\big]$ as $L, \Lambda_\slab{uv} \to \infty$, and so the intensive quantum information metric (\ref{eq:intensiveMetric}) is directly proportional to the Zamolodchikov metric, $\tilde{g}_{ab}(\varphi) \propto G^\slab{z}_{\smash{ab}}(\varphi)$.\footnote{In fact, in four-dimensional $\mathcal{N}=2$ superconformal field theories where the Zamolodchikov metric can be computed exactly~\cite{Gerchkovitz:2014gta,Gomis:2015yaa}, many of these information-theoretic quantities have natural conformal field theory analogs. For example, the K\"{a}hler potential (\ref{eq:kahlerPotential}) is related to the partition function $\mathcal{Z}_{\lab{S}^4} = \langle \Omega(\varphi) | \Omega(\varphi)\rangle$ of the theory on the four-sphere~$\lab{S}^4$, and the Zamolodchikov metric with respect to holomorphic and antiholomorphic marginal deformations ($\ud \varphi^a$ and $\ud \bar{\varphi}^{\bar{b}}$, respectively)  can be computed from $G_{\smash{a\bar{b}}}^\slab{z}(\varphi) = 192 \ess \partial_{a} \partial_{\smash{\bar{b}}} \log \mathcal{Z}_{\lab{S}^4}$ in the conventions of~\cite{Baggio:2014ioa}. We can interpret this as a specific example of a more general information-theoretic structure.}

        Similarly, in a family of nonlinear sigma models defined by the Euclidean action 
        \begin{equation}
        	S_\slab{e}[\phi] = \frac{1}{2} \ess \Lambda_\slab{uv}^{d-1}\! \int\!\ud^{d+1} x\,\ess  G^\slab{ms}_{\smash{ab}}(\phi) \, \partial_\mu \phi^a \es \partial_\mu \phi^b\,,
        \end{equation}
        the metric on moduli space $G^\slab{ms}_{\smash{ab}}(\varphi)$ provides a natural notion of distance between theories defined by the vacuum expectation values $\langle \phi^a \rangle = \varphi^a$. The intensive quantum information metric associated to this family is again proportional to this natural metric~\cite{Trivella:2016brw,Stout:2021ubb}
        \begin{equation}
        	\tilde{g}_{ab}(\varphi) \propto  G^\slab{ms}_{\smash{ab}}(\varphi)\,. \label{eq:intMetMS}
        \end{equation} 
        In both cases, the intensive quantum information metric has the same singularity structure as these more specialized metrics, and so any divergence or infinite distance point in the Zamolodchikov or moduli space metrics also inherits an information-theoretic interpretation.

        Because of its relationship to the classical information metric, the quantum information metric places theories which make very different predictions very far apart. We thus expect that the information metric can \emph{diverge} as we vary the parameters~$\varphi^a$ through a quantum critical point, since there this family of theories displays a rapid, qualitative change in its behavior and predictions. Indeed, this is what a phase transition means---qualitatively distinct phases make very different predictions, and so we expect that the information metric should place these phases very far apart from one another even if they are ``nearby'' in parameter space. We can confirm that this is the case as follows. If we work near a quantum critical point and assume that the operators $\mathcal{O}_a(x)$ and $\mathcal{O}_b(x)$ in (\ref{eq:actionDeformation}) have the same scaling dimensions $\Delta_a = \Delta_b = \Delta$, the Euclidean correlator that determines the metric (\ref{eq:infMetCorr}) behaves as
        \begin{equation}
            \langle \mathcal{O}_a(x_1) \mathcal{O}_b(x_2) \rangle \sim \frac{\e^{-|x_1 - x_2|/\xi}}{|x_1 - x_2|^{2 \Delta}}\,, \mathrlap{\quad |x_1 - x_2| \to \infty\,,}
        \end{equation}
        at large distances, where $\xi$ is the correlation length in the theory. The metric then scales as~\cite{Venuti:2007qcs,Carollo:2019ygj}
        \begin{equation}
            g_{ab}(\varphi) \propto 
            \begin{dcases}
                (L \Lambda_{\slab{uv}})^d & 2 \Delta > d+2 \\
                (L \Lambda_{\slab{uv}})^{d} \log(\xi \Lambda_{\slab{uv}}) & 2 \Delta = d+2 \\
                (L \Lambda_{\slab{uv}})^{d} (\xi \Lambda_\slab{uv})^{2 + d - 2 \Delta} & 2 \Delta < d + 2
            \end{dcases}\,. \label{eq:metScaling}
        \end{equation}
        That is, as long as the scaling dimension of the corresponding operator is $2 \Delta\leq d+2$, the information metric will diverge as we approach the quantum critical point and the correlation length of the theory diverges, $\xi \to \infty$.

        Since we will be primarily concerned with the interpretation and properties of infinite distance singularities in the intensive quantum information metric, a natural question arises: can the metric singularities generated by a quantum critical point ever be at infinite distance? The answer is no~\cite{Stout:2021ubb}, under very reasonable assumptions. Without loss of generality, we can restrict to a one-dimensional family of theories parameterized by $\epsilon\geq 0$. If we approach a quantum critical point as $\epsilon \to 0$, then the correlation length of the theory behaves as~\cite{Goldenfeld:2018lop,Sachdev:2011qpt,Kardar:2007spf} 
        \begin{equation}
            \xi(\epsilon) \propto |\epsilon|^{\sminus \nu}\,,
        \end{equation}
        where $\nu$ is called the correlation length critical exponent. Assuming that the local operator (\ref{eq:actionDeformation}) associated to $\epsilon$ has scaling dimension $2 \Delta < d +2$ so that the metric diverges as $\xi \to \infty$, from (\ref{eq:metScaling}) the intensive information metric scales as
        \begin{equation}
            \ud \tilde{s}^2 \propto \frac{\ud \epsilon^2}{\epsilon^{\nu(d + 2 - 2 \Delta)}}\,.
        \end{equation}
        The point $\epsilon = 0$ is thus at infinite distance whenever
        \begin{equation}
            \nu(d + 2 - 2 \Delta) \geq 2\,. \label{eq:infDistIneq0}
        \end{equation}
        However, quantum critical points typically obey scaling relations, and the critical exponent~\cite{Venuti:2007qcs} is usually related to the scaling dimension $\Delta$ by $\nu^{\sminus 1} = d + 1 - \Delta$. Using this scaling relation, we can rewrite (\ref{eq:infDistIneq0}) as
        \begin{equation}
            \frac{(d + 2 - 2 \Delta)}{d + (d + 2 - 2 \Delta)} \geq 1\,. \label{eq:infDistInequality}
        \end{equation}
        This inequality can never be satisfied in field theories with non-zero dimension $d > 0 $, as the terms in parentheses must always be non-negative if the metric is to diverge. We thus see that quantum critical points which display typical scaling behavior can never be at infinite distance in the intensive quantum information metric. 

        As we will see in the next section, this is a simple consequence of the fact that infinite distance points always correspond to factorization limits in which statistical or quantum fluctuations vanish. Quantum critical points still have fluctuations and non-trivial correlations, and so the information metric places them at finite distance from nearby theories.

\section{Infinite Distance Limits are Factorization Limits} \label{sec:infFact}

      		The goal of this section is to understand how a family of probability distributions or quantum states must behave in order to generate an infinite distance singularity in the information metric. 

      		Before we proceed to the main analysis, it will be helpful to first understand when infinite statistical length might occur from the perspective of the KL divergence, as the former can be thought of as the infinitesimal version of the latter (\ref{eq:classInfMetHess}). It will be more helpful to work in terms of a symmetrized KL divergence, called the Jeffreys' divergence,
            \begin{equation}
                \begin{aligned}
                    \dvg{j}{\varphi}{\varphi'} 
                    = \!\frac{1}{2}\int\!\ud^N x\!\left[\ess \condp{x}{\varphi} \log \frac{\condp{x}{\varphi}}{\condp{x}{\varphi'}} + \condp{x}{\varphi'} \log \frac{\condp{x}{\varphi'}}{\condp{x}{\varphi}}\right], \label{eq:jeffreysDivergence}
                \end{aligned}
            \end{equation}
            whose Hessian is also the classical information metric,
            \begin{equation}
                g_{ab}(\varphi) = \left.-\frac{\partial^2 \dvg{j}{\varphi}{\varphi'}}{\partial \varphi^a \,\partial \varphi'^b}\right|_{\varphi'=\varphi},
            \end{equation}
            since the KL divergence is symmetric at leading order, $\dvg{kl}{\varphi+ \ud \varphi}{\varphi} \approx \frac{1}{2} g_{ab}(\varphi)\, \ud \varphi^a \, \ud \varphi^b + \cdots$.

            When can the Jeffreys' divergence (\ref{eq:jeffreysDivergence}) become infinite? The fact that $\condp{x}{\varphi}$ is a \emph{probability distribution}---and is thus a non-negative function with unit integral---is severely restrictive, and there are only two qualitatively distinct scenarios we know of that can produce an infinite Jeffreys' divergence. The first is when one of the distributions has a ``heavy'' tail and the other does not, like when $\condp{x}{\varphi} = \pi^{\sminus 1} (1+x^2)^{\sminus 1}$ is the Cauchy distribution and $\condp{x}{\varphi'} = \e^{\sminus x^2/2}/\sqrt{2\pi}$ is Gaussian. In light of Sanov's theorem (\ref{eq:sanovs}), we can view this infinity as a reflection of the fact that higher moments of the Cauchy distribution do not converge, and thus neither do empirical estimations like those found by taking $M$ samples $\{x_i\}$ and computing, for instance, $\frac{1}{M-1} \sum_{i = 1}^{M} x_i^2$.

            The other way (\ref{eq:jeffreysDivergence}) can blow up is if the two distributions do not share the same support. That is, there are a set of outcomes with non-zero measure\footnote{However, the Jeffreys divergence is not infinite if one of the distributions assigns zero probability to a discrete set of points, like if $\condp{x}{\varphi} \propto \sin^2 2 x \, \e^{-x^2/2}$ and $\condp{x}{\varphi'} \propto \e^{-x^2/2}$.} in $\condp{x}{\varphi}$ that $\condp{x}{\varphi'}$ would say \emph{never} occur, or vice versa, and so a single observation of one of these forbidden events allows us to immediately rule out one of the distributions with certainty.

            In this section, we will argue that infinite distance singularities in the information metric are a symptom of the latter---they represent the case in which a family of probability distributions $\condp{x}{\varphi}$ \emph{degenerate} and there is a ``collapse'' of probability onto a measure zero set of outcomes. More plainly, $\condp{x}{\varphi}$ approaches a Dirac delta function (or a superposition of them) in this infinite distance limit and so expectation values factorize, $\langle x^n \rangle = \langle x \rangle^n$. We will first understand this for families of classical probability distributions and then for families of quantum states.
 			
 		 \subsection{Classical Infinite Distance Limits} \label{sec:classicalInfFact}

            The goal of this section is to argue that an infinite distance point in the classical information metric corresponds to a factorization limit, in which the limiting distribution degenerates and probability ``collapses'' onto a discrete set of events. Under reasonable assumptions, the information metric associated to a factorization limit takes a universal ``logarithmic'' form, up to an overall constant which depends on how the family degenerates.

            We will focus on families of classical probability distributions $\condp{x}{\epsilon}$ in a single parameter~$\epsilon$, in the limit as $\epsilon \to 0^+$. We will always take this limit from above, and thus suppress the $+$ superscript. This is equivalent to considering a family of distributions along a curve in a higher-dimensional statistical manifold. For pedagogical purposes, we will first assume that $x \in \mathbb{R}$ is a single continuous random variable, but will later extend our discussion to distributions in arbitrary numbers of continuous variables.  We will also assume that the family $\condp{x}{\epsilon}$ admits an asymptotic expansion in (not necessarily integer) powers of $\epsilon$, though this is mainly for presentational purposes and we will relax this assumption later. 

            Finally, we will assume that all but a finite set of the predictions of the family $\condp{x}{\epsilon}$, in the form of expectation values $\langle f(x) \rangle$ with $f(x)$ an arbitrary but well-behaved function,\footnote{By well-behaved, we mean that we want to exclude extremely rapidly growing functions like $e^{e^{e^x}}$ whose expectation values typically diverge for all values of $\epsilon$. If they do actually converge in the statistical family, this means that the probability distributions must approach zero extremely rapidly as $|x| \to \infty$. It would instead be more natural to work in a different set of variables instead, $x \to x' = \log \log \log x$. This type of restriction to ``simple'' observables will also be necessary when we discuss classical limits in \S\ref{sec:classicalLimits} and large $N$ limits in \S\ref{sec:ftLargeN}. } approach a finite limit as $\epsilon \to 0$. This is our strongest assumption and will allow us to argue for the existence of a universal logarithmic infinite distance singularity $\ud s^2 \propto \ud \epsilon^2/\epsilon^2$, whose coefficient depends on how the family $\condp{x}{\epsilon}$ approaches factorization. We will later relax this assumption as well and discuss how more severe infinite distance singularities can arise, though these too are always associated with factorization limits.

            What behavior yields an infinite distance singularity in the classical information metric? Let us first assume that the limit $\epsilon \to 0$ is ``regular,'' in the sense that the family approaches a fixed probability distribution $p_0(x)$. In this limit, we are free to parameterize this family as\footnote{Since $\condp{x}{\epsilon}$ is a member of $L^2(\mathbb{R})$ for all $\epsilon$, we can choose an $\epsilon$-independent basis of $L^2(\mathbb{R})$ to expand $\condp{x}{\epsilon}$ in such as the Hermite functions $\psi_n(x) = [2^n n! \sqrt{\pi} \sigma]^{\sminus 1/2} H_n(x/\sigma) e^{-x^2/(2 \sigma^2)}$, with $\sigma$ the variance of $p_0(x)$. This is also called an Edgeworth expansion. In this case $\cond{f}{x}{\epsilon} = \sum_{n} c_n(\epsilon) \psi_n(x)$ and the $\epsilon$-expansion can be understood as a continuous sequence of vectors in the Hilbert space $L^2(\mathbb{R})$.  \label{fn:hermite}}
             \begin{equation}
                p(x \, |\, \epsilon) = p_0(x)\left[1 + \epsilon^\alpha \cond{f}{x}{\epsilon}\right]\,, \label{eq:probExp} 
            \end{equation}
            where the fiducial distribution $p_0(x)$ is both normalized, $\int\ud x\, p_0(x) = 1$, and orthogonal to the correction $\cond{f}{x}{\epsilon}$, $\int\!\ud x\, p_0(x) \cond{f}{x}{\epsilon} = 0$. We have also introduced the leading exponent $\alpha > 0$ of the $\epsilon$-expansion so that $\lim_{\epsilon \to 0} \cond{f}{x}{\epsilon} = f_0(x)$ is an $\epsilon$-independent function. We may view (\ref{eq:probExp}) as an $\epsilon$-expansion around each point $x$. For instance, a simple example of such a decomposition is the ``anharmonic'' family of distributions,
            \begin{equation}
                \begin{aligned}
                    \condp{x}{\epsilon} &= \sqrt{\frac{\sigma^2 \epsilon}{3}} \,  \left[K_{\frac{1}{4}}\!\left(\tfrac{3}{4 \sigma^4 \epsilon}\right)\right]^{\sminus 1} \!\exp\!\left(-\frac{x^2}{2 \sigma^2} - \frac{ \epsilon x^4}{4!} -\frac{3}{4 \sigma^4 \epsilon} \right)  \\
                    &\sim \frac{1}{\sqrt{2 \pi \sigma^2}}\e^{\sminus \frac{x^2}{2 \sigma^2}} \left[1 + \frac{\epsilon}{4!}\big(3 \sigma^4 - x^4\big) + O(\epsilon^2)\right]
                \end{aligned}
            \end{equation}
            where $K_\nu(z)$ is the modified Bessel function. In this case, the leading exponent is $\alpha = 1$ and the leading-order correction is $\cond{f}{x}{\epsilon} = \epsilon^{\sminus 1} \left[\condp{x}{\epsilon}/\condp{x}{0} - 1\right] \sim \frac{1}{4!}\big(3 \sigma^4 - x^4\big) + O(\epsilon)$.

            It is easy to see that such families can never have an infinite distance point. We can compute the information metric from (\ref{eq:fimDef}), whose leading behavior is
            \begin{equation}
                \ud s^2  \sim \frac{\alpha^2\ess \ud \epsilon^2}{\epsilon^{2 - 2 \alpha}} \int_{\sminus \infty}^{\infty}\!\ud x\, f_0(x)^2 p_0(x)\,,\mathrlap{\qquad \epsilon \to 0\,.} \label{eq:notInfDist}
            \end{equation}
            Since $\alpha > 0$, the statistical distance to the distribution at $\epsilon = 0$ is finite. Interestingly, if the dependence on $\epsilon$ is non-analytic, $0 < \alpha < 1$, then $\epsilon \to 0$ is a \emph{finite distance singularity}. We can think of this as a simple analog of the fact that quantum critical points that obey scaling can only generate finite distance singularities in the information metric, cf. (\ref{eq:infDistInequality}), even if their properties---for instance, the appearance of massless degrees of freedom---allow for non-analytic dependence on the control parameter $\epsilon$, such that $0 < \alpha < 1$.

            The approximation (\ref{eq:notInfDist}) is valid when the coefficient $\int\!\ud x\, f_0(x)^2 p_0(x)$ converges, so that the leading $\epsilon$-scaling can be trivially extracted. The leading-order correction $f_0(x)$ must be fairly well-behaved since $\condp{x}{\epsilon}$ is a family of probability distributions. Specifically, this means that $f_0(x)$ is not singular because $\condp{x}{\epsilon}$ is necessarily positive for all $x$ and $\int\ud x\, f_0(x) p_0(x) = 0$. However, this coefficient could diverge if $p_0(x)$ is heavy-tailed enough, as it is for the Cauchy distribution. While this would violate our assumption that expectation values $\langle f(x) \rangle$ remain finite as $\epsilon \to 0$, it is natural to ask whether these types of distributions can generate an infinite distance singularity anyway. Though we have not proved that this can never happen, we believe that this is very unlikely. For instance, consider the family of $t$-distributions defined by
            \begin{equation}
                \condp{x}{\epsilon} = \frac{1}{\sqrt{\epsilon} B\big(\frac{\epsilon}{2}, \frac{1}{2}\big)}\left(\frac{\epsilon}{x^2 + \epsilon}\right)^{\frac{1 + \epsilon}{2}}\,,
            \end{equation}
            where $B(\mu, \nu) = \int_0^1\ud t\, t^{\mu - 1} (1 - t)^{\nu - 1}$ is the Euler beta function. These distributions become extremely heavy-tailed as $\epsilon \to 0$ and include the Cauchy distribution when $\epsilon = 1$.  Indeed, as $\epsilon \to 0$ the limiting distribution is non-normalizable and is thus not a well-defined probability distribution. The information metric for this family can be explicitly computed,
            \begin{equation}
                \ud s^2 = \frac{1}{2} \ud \epsilon^2\left((1 + \epsilon)\Big[2 \pi \csc \pi \epsilon + \psi\big(\tfrac{1-\epsilon}{2}\big) - \psi\big(\tfrac{2 - \nu}{2}\big)\Big] + \log \pi \epsilon + 2 \log\!\Big[\Gamma\big(\tfrac{\epsilon}{2}\big)/{\Gamma\big(\tfrac{\epsilon + 1}{2}\big)}\Big]\right)\,,
            \end{equation}
            where $\psi(z) = \Gamma'(z)/\Gamma(z)$ is the digamma function.
            As $\epsilon \to 0$, this behaves as $\ud s^2 \sim \ud \epsilon^2/\epsilon$, and is thus \emph{not} at infinite distance, and we thus expect that such ``heavy-tailed'' troubles are not dramatic enough to generate an infinite distance singularity. This matches with our intuition coming from the Jeffreys' divergence, which is finite between two distributions that are both~heavy-tailed.

\begin{figure}
                \centering
                \includegraphics{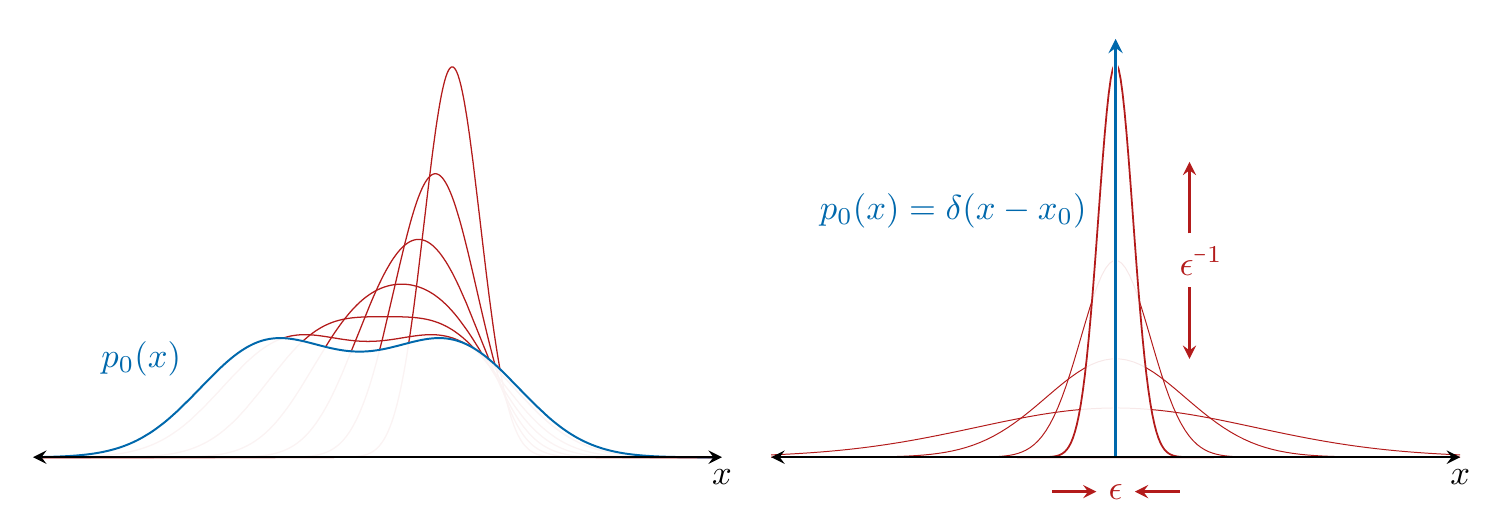}
                \caption{A regular family of probability distributions $\condp{x}{\epsilon}$ [\emph{left, {\color{cornellRed}red}}\,] approaches a fixed fiducial distribution $p_0(x)$ [\emph{left, {\color{cornellBlue}blue}}\ess], and can never generate an infinite distance singularity in the information metric because the family's predictions do not change dramatically as $\epsilon \to 0$. In contrast, the predictions of a degenerating family of distributions [\emph{right, {\color{cornellRed}red}}\,] change so dramatically that a single sample $x \neq x_0$ can immediately rule out the $\epsilon = 0$ distribution $p_0(x) = \delta(x - x_0)$ [\emph{right, {\color{cornellBlue} blue}}\ess], without statistical ambiguity. Unitarity restricts the possible ways this type of limit can be approached, as $\int\!\ud x\, \condp{x}{\epsilon} =1$, which forces the information metric to have a \emph{universal} infinite distance singularity $\ud s^2 \propto \ud \epsilon^2/\epsilon^2$.  \label{fig:comparison}}
            \end{figure}

            How, then, can an infinite distance singularity arise? In (\ref{eq:probExp}), we assumed that the fiducial distribution $p_0(x)$ was independent of $\epsilon$. If $\condp{x}{\epsilon}$ were a general function of $x$ and $\epsilon$, we would be also be able to include \emph{negative} powers of $\epsilon$ in our expansion. For instance, we could write
            \begin{equation}
                \condp{x}{\epsilon} = \frac{1}{\epsilon^{\gamma}}\, p_0(x)\left[1  + \epsilon^{\alpha} \cond{f}{x}{\epsilon}\right]
            \end{equation}
            where $\gamma$ is another positive exponent. Clearly, this is not allowed for probability distributions since the family must be properly normalized for all $\epsilon$. However, it could be that the leading order distribution's width shrinks at the same time as its amplitude diverges so that its integral is unchanged, as illustrated in Figure~\ref{fig:comparison}. Ignoring the higher-order corrections, we could have that
            \begin{equation}
                \condp{x}{\epsilon} \sim \frac{1}{\epsilon^{\gamma}} \eta(x/\epsilon^{\gamma})\,,\mathrlap{\qquad \epsilon \to 0\,,} \label{eq:irregularFamily}
            \end{equation}
            where $\eta(y)$, sometimes called a ``bump function,'' is a non-negative function with unit integral, $\int_{\sminus \infty}^{\infty}\!\ud y\, \eta(y) = 1$. That is, the fact that $\condp{x}{\epsilon}$ is a family of \emph{probability distributions} restricts any distribution whose maximum value diverges as $\epsilon \to 0$ to simultaneously shrink in width. Otherwise, it is not a probability distribution. One might call this a restriction from unitarity, as it follows from the fact that probabilities must sum to one. 

            Since the family $\condp{x}{\epsilon}$ approaches a function with infinite height, zero width, and integral one, we have that
            \begin{equation}
                \lim_{\epsilon \to 0} \condp{x}{\epsilon} = \lim_{\epsilon \to 0} \frac{1}{\epsilon^{\gamma}} \eta(x/\epsilon^\gamma) = \delta(x - x_0)\,,
            \end{equation}
            where $x_0$ is some $\epsilon$-independent constant. Indeed, functions like $\eta(x)$ are often called \emph{nascent delta functions}, as they provide a useful definition of a Dirac delta function in terms of the limit of a smooth function. In this limit, the statistical family \emph{degenerates} to a delta function and all probability collapses to a single point. This $\epsilon \to 0$ limit thus corresponds to a \emph{factorization limit}, in which
            \begin{equation}
                \langle x^n \rangle = x_0^n\,,
            \end{equation}  
            and all connected expectation values except for $\langle x \rangle = x_0$ vanish.

            If we allow for this type of $\epsilon$-dependence in the family $\condp{x}{\epsilon}$,
            \begin{equation}
                \condp{x}{\epsilon} = \frac{1}{\epsilon^{\gamma}}\, \eta(x/\epsilon^\gamma)\left[1  + \epsilon^{\alpha} \cond{f}{x}{\epsilon}\right]\,,
            \end{equation}
            we find that the information metric indeed has an infinite distance singularity,\footnote{A similar expression has appeared in~\cite{Clingman:2015lxa}, see also \cite{Erdmenger:2020vmo}, which attempted to reconstruct the family of distributions $\condp{x}{\epsilon}$ directly from the information metric assuming that the family of distributions enjoys a scaling symmetry, as it does near a degenerate point (\ref{eq:irregularFamily}).}
            \begin{equation}
                \ud s^2 \sim \frac{\gamma^2 \, \ud \epsilon^2}{\epsilon^2} \int_{\sminus \infty}^{\infty}\!\ud y \!\ess \left[1 + y \frac{\ud \log \eta(y)}{\ud y}\right]^2 \!\ess\eta(y)\,,\mathrlap{\qquad \epsilon \to 0\,.} \label{eq:classMetSing}
            \end{equation}
            Importantly, the logarithmic form of the infinite distance singularity, i.e. that $\ud s^2 \propto \ud \epsilon^2/\epsilon^2$, is independent of \emph{how} this factorization limit is approached. This scaling of the metric with $\epsilon$ is thus a \emph{universal} property of these factorization limits.\footnote{This universal structure has been argued for in models with complexified parameter or moduli spaces by using asymptotic Hodge theory, see e.g.~\cite{Grimm:2018ohb,Grimm:2020cda}. We interpret the above as an alternative information-theoretic explanation for why this structure appears broadly that does not rely on K\"{a}hlerity of the parameter space. However, we stress that this asymptotic Hodge-theoretic analysis is much more detailed, and it would be interesting to understand what additional information-theoretic constraints can be derived from that structure.} However, the positive coefficient in front of the metric does depend on both the precise algebraic dependence of $\condp{x}{\epsilon}$ (i.e. the exponent $\gamma$) and the bump function $\eta(y)$ used. For instance, if $\condp{x}{\epsilon} \sim \e^{-x^2/(2 \epsilon^2)}/\sqrt{2 \pi \epsilon^2}$ as $\epsilon\to 0$ is a Gaussian, then $\ud \epsilon^2 = 2 \ess \ud \epsilon^2/\epsilon^2$, while  $\ud s^2 = \ud \epsilon^2/(2 \epsilon^2)$  if $\condp{x}{\epsilon} \sim (\epsilon/\pi)/(\epsilon^2 + x^2)$ as $\epsilon \to 0$ is a Lorentzian. It is thus natural to integrate by parts and rewrite (\ref{eq:classMetSing}) as\footnote{As is clear from (\ref{eq:classMetSing}), the term in parenthesis is always positive.}
            \begin{equation}
                \ud s^2 \sim \frac{2 \gamma^2 \ess  \ud \epsilon^2}{\epsilon^2} \left(1 - \frac{1}{2} \int_{\sminus \infty}^{\infty}\!\ud y \left[1 + y^2\ess \frac{\ud^2 \log \eta(y)}{\ud y^2}\right] \eta(y)\right),\mathrlap{\quad \epsilon \to 0\,,} \label{eq:classMetSing2}
            \end{equation}
            so that the integral vanishes when evaluated on any Gaussian bump function $\eta(y)$.

            It is easy to see that a factorization limit $\langle x^n \rangle = x_0^n + O(\epsilon^{2 \gamma})$ also implies that the information metric diverges. The reconstruction of a probability distribution over $x \in \mathbb{R}$ from its moments is the well-studied and savory Hamburger moment problem~\cite{schmudgen2017moment}. It can be solved by constructing the moment-generating function 
            \begin{equation}
                \lim_{\epsilon \to 0} \int_{\sminus \infty}^{\infty}\!\ud x\, \e^{x t}\ess  \condp{x}{\epsilon} =  \sum_{n = 0}^{\infty} \frac{\langle x^n \rangle t^n}{n!} = \e^{x_0 t}\,,
            \end{equation} 
            and taking the inverse Laplace transform with respect to $t$ to find $\lim_{\epsilon \to 0} \condp{x}{\epsilon} = \delta(x - x_0)$. Since the moment generating function has an infinite radius of convergence in $t$, this solution is unique so we know that the family $\condp{x}{\epsilon}$ must asymptote to a delta function as $\epsilon \to 0$. From the above discussion, this means that this factorization limit is at infinite distance and $\ud s^2 \propto \ud \epsilon^2/\epsilon^2$.

            Note that the point $\epsilon = 0$ is also at infinite distance when $\gamma < 0$, in which case the distribution is broadening and not collapsing, approaching a uniform distribution in $x$ rather than a delta function. This is still a factorization limit,\footnote{Note that expectation values of well-behaved functions like $\langle f(x) \rangle$ could also diverge in this limit. A simple example are the moments of the Gaussian $\condp{x}{\sigma} = \exp(\minus x^2/2 \sigma^2)/\sqrt{2 \pi \sigma^2}$, whose expectation values $\langle x^{2k} \rangle \propto \sigma^{2k}$ diverge in the infinite distance limit $\sigma \to \infty$, with $\ud s^2 = 2 \,\ud \sigma^2/\sigma^2$.} in the sense that the probability distribution associated with derivatives of $x$ collapses to a delta function. This interpretation is stronger for families of quantum states, which we discuss in the next section, where $\gamma > 0$ or $\gamma < 0$ determines whether predictions are factorizing in either the coordinate $x$ or its conjugate momentum, respectively.

            There is a way out in avoiding a ``strict'' factorization limit $\langle x^n \rangle \to \langle x \rangle^n$ at an infinite distance point. In the above discussion, we assumed that probability localized about a particular value $x = x_0$, but it could be that it instead collapses to a discrete set $\{x_i\}$ as $\epsilon \to 0$, yielding expectation values that behave as $\langle x^n \rangle = x_1^n + x_2^n + \cdots$. A familiar example of this is given the family of distributions
            \begin{equation}
                \condp{x}{\beta} = \frac{1}{\mathcal{Z}(\beta)} \e^{-\beta (x^2 - 1)^2}\,, \quad \text{with}\quad  \mathcal{Z}(\beta) = \frac{\pi}{2} \e^{\sminus \frac{\beta}{2}}\left[ I_{\sminus \frac{1}{4}}\big(\tfrac{\beta}{2}\big) + I_{\frac{1}{4}} \big(\tfrac{\beta}{2}\big)\right]\,, \label{eq:doubleWell}
            \end{equation}
            which collapses  around the minima of the potential $V(x) = (x^2 - 1)^2$ at $x = \pm 1$ as $\beta \to \infty$. Here, we have introduced $\beta \propto \epsilon^{\sminus 2}$ because, for this particular type of $\beta$-dependence, the information metric is trivial to compute (\ref{eq:fimDef}),
            \begin{equation}
                \ud s^2 =  \big[\partial_\beta^2 \log \mathcal{Z}(\beta)\big] \, \ud \beta^2\,. \label{eq:infMetPF}
            \end{equation}
            This example is also useful in understanding the universal nature of the logarithmic metric singularity $\ud s^2 \propto \ud \beta^2/\beta^2$.  As $\beta \to \infty$, we can compute the partition function $\mathcal{Z}(\beta)$ using the saddle-point approximation. In our example, the two saddles have identical properties and we can approximate the partition function as
            \begin{equation}
                \mathcal{Z}(\beta) \sim 2 \int_{\sminus \infty}^{\infty} \!\ud z\, \e^{\sminus 4 \beta z^2} = \sqrt{\frac{\pi}{\beta}}\,.
            \end{equation}
            Crucially, the ``fluctuation determinant'' around each saddle contributes a factor of $\beta^{\sminus 1/2}$. This is the only term that contributes to the information metric (\ref{eq:infMetPF}), and is responsible for the logarithmic form of the information metric, $\ud s^2 = \ud \beta^2/(2 \beta^2) = 2 \ess \ud \epsilon^2/\epsilon^2$. 

            This conclusion is unchanged for more complicated potentials. In the general case that the potential has multiple degenerate minima $x_i$, with $i = 1, \ldots, m$, with $V(x) \sim V_* + V^{(p_i)}(x_i) (x - x_i)^{p_i}$ and $p_i \geq 2$, we have~\cite{Bender:2013amm}
            \begin{equation}
                \mathcal{Z}(\beta) \sim e^{-\beta V_*} \sum_{i = 1}^{m} 
                 \frac{2 \Gamma(1/p_i)}{p_i} \left[\frac{p_i!}{\beta V^\floq{p_i}(r_i)}\right]^{1/p_i} ,\mathrlap{\qquad \beta \to \infty\,.} \label{eq:generalSaddle}
            \end{equation}
            As $\beta \to \infty$, the minima with the maximal degree $p_i$, which we will denote $p$, dominate the partition function and so we may approximate it as as
            \begin{equation}
                \mathcal{Z}(\beta) \sim \frac{\mathcal{C} e^{-\beta V_*}}{\beta^{1/p}}\,,
            \end{equation} 
            with $\mathcal{C}$ an unimportant constant that depends on the number and properties of the saddles with maximal $p_i$. The information metric in this general case behaves as
            \begin{equation}
                \ud s^2 = \frac{\ud \beta^2}{p \es\beta^2}. \label{eq:universalCIM}
            \end{equation}
            Again, we see that \emph{its logarithmic form is a universal consequence of probability localizing to a set of points}, while the overall coefficient depends on \emph{how} the family of distributions localizes about these points. Furthermore, it makes intuitive sense that the overall coefficient only depends on the maximal $p_i$, as those are the points about which probability is localizing fastest and distinguishability is most enhanced.

            There is another way of avoiding a strict factorization limit while still generating an infinite distance singularity.\footnote{We thank Matt Reece for suggesting this example.} If the family of distributions consists of a mixture of both a degenerating and non-degenerating distribution, as it does for the pair of Gaussians,
            \begin{equation}
                \condp{x}{\epsilon} = \frac{1}{2} \frac{1}{\sqrt{2 \pi}} \e^{\sminus x^2/2} + \frac{1}{2} \frac{1}{\sqrt{2 \pi \epsilon^2}}\ess  \e^{\sminus {(x - \mu)^2}/({2 \epsilon^2})}\,,\label{eq:pathological}
            \end{equation} 
            then it will also display an infinite distance metric singularity $\ud s^2 \sim \ud \epsilon^2/\epsilon^2$ as $\epsilon \to 0$. Here, only a single contribution to expectation values factorizes as $\epsilon \to 0$, $\langle x^{2k} \rangle \sim \mu^{2k} + \frac{1}{2}(2 k -1)!!$. Much like in (\ref{eq:doubleWell}), it is not completely correct to say that an infinite distance limit always corresponds to a strict factorization limit. Rather, it always corresponds to a degeneration limit in which probability, in part, localizes to a measure zero set of events. Typically, the signal of this degeneration is the strict factorization of expectation values, $\langle x^k \rangle = \langle x \rangle^k$, as will be the case in all of the examples we discuss in Section~\ref{sec:examples}. We expect that distributions like (\ref{eq:pathological}) do not appear in well-behaved physical systems. For example, realizing (\ref{eq:pathological}) from something like a Boltzmann distribution, $\condp{x}{\epsilon} \propto \exp(-\beta V(x))$, would require that the potential $V(x)$ be both unbounded from below and to have structure on multiple arbitrarily different length scales---it would need an arbitrarily deep albeit thin minimum at $x = \mu$. Neither do we expect distributions like (\ref{eq:pathological}) to arise as the ground state distributions of well-behaved quantum mechanical systems, as this would require potentials that are both arbitrarily negative and singular. We thus regard these as pathological and will ignore them. 

            That an infinite distance point corresponds to a collapse of probability matches with our intuition coming from the Jeffreys' divergence discussed at the beginning of this section, and our general intuition that distributions at infinite distance are hyper-distinguishable. Here, the support of the distribution collapses to a discrete, measure zero, set of points.\footnote{In the previous counterexample, this statement is true \emph{at leading order in $\epsilon$}, since in that limit the first Gaussian in (\ref{eq:pathological}) can be ignored. This explains why such pathological examples are still at infinite distance in the information metric.} Accordingly, the Jeffreys' divergence is infinite and so we expect that the statistical length is too.  However, we should note that it is not the case that the statistical length must be infinite when the Jeffreys' divergence is. For instance, consider the family
            \begin{equation}
                \condp{x}{\epsilon} = \frac{1}{2} \Big[\lab{erf}\big([x + \tfrac{1}{2}]/\epsilon\big) - \lab{erf}\big([x - \tfrac{1}{2}]/\epsilon\big) \Big]
            \end{equation}
            which asymptotes to the uniform distribution as $\epsilon \to 0$,
            \begin{equation}
                \lim_{\epsilon \to 0} \condp{x}{\epsilon} = \begin{cases} 1 & |x| < \frac{1}{2}  \\ 0 & |x| > \frac{1}{2} \end{cases}
            \end{equation}
            with $\lim_{\epsilon \to 0} \condp{\pm \frac{1}{2}}{\epsilon} = \frac{1}{2}$. Since the support of the family changes from $x \in (\minus \infty,\infty)$ to the interval $x \in [-\frac{1}{2}, \frac{1}{2}]$ as $\epsilon \to 0$, the Jeffreys' divergence between $\epsilon = 0$ and any other point $\epsilon > 0$ diverges. However, the information metric only displays a finite distance singularity,
            \begin{equation}
                \ud s^2 \sim \frac{\ud \epsilon^2}{\epsilon} \left[\frac{4}{\pi} \int_{\sminus \infty}^{\infty}\!\ud y\, \frac{y^2 e^{\sminus 2 y^2}}{\lab{erfc}\, y}\right] \approx \frac{1.5605}{\epsilon}\,,\mathrlap{\qquad \epsilon \to 0\,.}
            \end{equation}
            We should have expected this, as this family approaches a fixed-amplitude distribution as $\epsilon \to 0$ and thus belongs to the family $(\ref{eq:probExp})$.  Similarly, if we consider a family which is a mixture of a Gaussian and a Lorentzian,
            \begin{equation}
                \condp{x}{\epsilon} = \frac{\epsilon}{\pi}\frac{1}{1 + x^2} + \frac{1-\epsilon}{\sqrt{2 \pi}}\,  \e^{\sminus x^2/2}\,,
            \end{equation} 
            the Jeffreys' divergence between the two endpoints at $\epsilon = 0$ and $\epsilon = 1$ is infinite since the $\epsilon = 1$ distribution is very heavy-tailed. However, one can show by numerical integration that the information metric only has a finite distance singularity as $\epsilon \to 0$, since this family is again of the form (\ref{eq:probExp}). Only factorization limits generate infinite distances.

            Before we discuss infinite distance limits in the quantum information metric, it will be useful to discuss and relax some of our initial assumptions. Very little of this discussion has relied on the assumption that $\condp{x}{\epsilon}$ is a distribution for a single random variable $x$. Indeed, our argument that families (\ref{eq:probExp}) that approach a fixed distribution never produce an infinite distance point did not rely on the dimensionality of the random variable, and so should be applicable even for random fields, $N \to \infty$. However, the explicit form of the information metric in the factorization limit does change. Let us assume that $N$ degrees of freedom factorize as $\epsilon \to 0$, so that
            \begin{equation}
                \condp{\mb{x}, \mb{z}}{\epsilon} \sim \frac{1}{\epsilon^{\gamma N}} \eta(\mb{x}/\epsilon^\gamma) p_0(\mb{z}) \,,\mathrlap{\qquad \epsilon \to 0\,,}
            \end{equation}
            where $p_0(\mb{z})$ represents the probability distribution for an arbitrary number of non-factorizing degrees of freedom. 
            Happily, the infinite distance singularity,
            \begin{equation}
                \ud s^2 \sim \frac{2 N \gamma^2\, \ud \epsilon^2}{\epsilon^2} \left(1 - \frac{1}{2} \int_{\mathbb{R}^N}\!\!\ud^{N} \!\es y\, \eta(\mb{y}) \left[1 + \frac{1}{N}\ess y^i y^j\ess  \frac{\partial^2  \log \eta(\mb{y})}{\partial y^i \ess \partial y^j}\right]\right)\,,\quad \epsilon \to 0\,, \label{eq:cimInfRewrite}
            \end{equation}
            only depends on the factorizing degrees of freedom, as those drive the hyper-distinguishability between members of the family $\condp{\mb{x}, \mb{z}}{\epsilon}$.
            This is the higher-dimensional generalization of (\ref{eq:classMetSing2}), and we have again arranged it so that the integral vanishes when evaluated on any Gaussian bump function. Importantly, the overall coefficient of the metric is proportional to the number of degrees of freedom that factorize. This agrees with the expected extensive scaling of the information metric (\ref{eq:intensiveMetric}), a conclusion we expect to survive for any bump function~$\eta(\mb{y})$.

            Furthermore, we assumed that expectation values of $\condp{x}{\epsilon}$ remain finite as $\epsilon \to 0$. Relaxing this assumption allows for a more general class of infinite distance limits, which can be illustrated in the simple Gaussian family,
            \begin{equation}
                \condp{x}{\sigma, \mu} = \frac{1}{\sqrt{2 \pi \sigma^2}} \exp\!\left(-\frac{(x - \mu)^2}{2 \sigma^2}\right)
            \end{equation}
            whose information metric is that of Euclidean $\lab{AdS}_2$,
            \begin{equation}
                \ud s^2 = \frac{2 \ess \ud \sigma^2 + \ud \mu^2}{\sigma^2}\,.
            \end{equation}
            Obviously, if $\mu = \mu(\epsilon)$ is an arbitrary function of $\epsilon$ and the variance $\sigma$ is fixed, we can make the metric $\ud s^2 = \sigma^{\sminus 2}[\ud \mu/\ud \epsilon]^2\ess  \ud \epsilon^2$ as singular as we want as $\epsilon \to 0$. This type of infinite distance singularity can be removed by working with the right variables, i.e. shifting $x \to x + \mu(\epsilon)$ so that expectation values remain finite as $\epsilon \to 0$.  It is important to reiterate that while the information metric does not depend on which variables we use to describe a given family of distributions $\condp{x}{\epsilon}$, it can change under $\epsilon$-dependent variable transformations since this changes the precise statistical family under consideration.

            \newpage
            However, we note that even these infinite distance limits are still factorization limits, even if the universality of the logarithmic singularity is now lost. For instance, let us consider a family that behaves as $\condp{x}{\epsilon} \sim \eta\big(x - \mu(\epsilon)\big)$, where $\eta(x)$ is a bump form and $\mu(\epsilon) \to \infty$ as $\epsilon \to 0$. The information metric behaves as $\ud s^2 = [\eta'(\epsilon)]^2 \ess \ud \epsilon^2$, while arbitrary expectation values behave as
            \begin{equation}
            	\langle x^k \rangle = \int\!\ud x\, x^k \, \eta\big(x - \mu(\epsilon)\big) = \mu(\epsilon)^{k} \int\!\ud y\, y^k \mu \ess \eta\big(\mu( y- 1)\big)\,.
            \end{equation}
            Obviously, we can identify $\mu \ess \eta\!\ess\big(\mu(y-1)\big)$ as a nascent delta function, centered around $y = 1$. As the mean diverges, $\mu(\epsilon) \to \infty$, probability localizes around $y =1$ and expectation values again factorize $\langle x^k \rangle \to \mu^k\big[1 + O(\mu^{\sminus 1})\big]$.

            Finally, we can also relax our assumption that $\condp{x}{\epsilon}$'s leading-order $\epsilon$-dependence is algebraic. Generalizing to an arbitrary $\epsilon$-dependence does not change our conclusions. For instance, if we allow the regular statistical family (\ref{eq:probExp}) to behave as
            \begin{equation}
                \condp{x}{\epsilon} = p_0(x)\Big(1 + \ell(\epsilon) f_0(x) + \cdots\Big)\,,
            \end{equation} 
            where the $\cdots$ denote higher order corrections and $\lim_{\epsilon \to 0} \ell(\epsilon) = 0$, 
            then the information metric behaves as
            \begin{equation}
                \ud s^2 \sim \left[\ell'(\epsilon) \ess \ud \epsilon\right]^2 \int_{\sminus \infty}^{\infty}\!\ud x\, f_0(x)^2 p_0(x)\,,\mathrlap{\qquad \epsilon \to 0\,.}
            \end{equation}
            Clearly, the statistical length from $\condp{x}{\epsilon}$ to $\condp{x}{0}$ is proportional to $\ell(\epsilon)$, and so the point at $\epsilon = 0$ is at finite distance. Similarly, if we assume that $\epsilon^\gamma$ is replaced with $\ell(\epsilon)$ in (\ref{eq:irregularFamily}),
            \begin{equation}
                \condp{x}{\epsilon} \sim \frac{1}{\ell(\epsilon)} \eta(x/\ell(\epsilon)) + \cdots\,,\mathrlap{\qquad \ell(\epsilon) \to 0\,,}
            \end{equation}
            then the information metric again behaves logarithmically, cf. (\ref{eq:classMetSing2}),
            \begin{equation}
                \ud s^2 \sim 2\left[\ud \!\ess\log \ell(\epsilon)\right]^2\left(1 - \frac{1}{2} \int_{\sminus \infty}^{\infty} \!\ud y \left[1 + y^2 \frac{\ud^2 \log \eta(y)}{\ud y^2}\right] \!\eta(y)\right)
            \end{equation}
            with $ y= \ell(\epsilon) x$, so the point $\epsilon = 0$ with $\lim_{\epsilon \to 0} \ell(\epsilon) = 0$ is still a logarithmic infinite distance singularity. In the next section, we will find that the same holds true for infinite distance singularities in the quantum information metric.

    \subsection{Quantum Infinite Distance Limits} \label{sec:quantumInfFact}

    	In the previous section, we saw that an infinite distance limit in the classical information metric implies that the family of distributions degenerates and expectation values factorize. How much of this story applies to quantum states and the quantum information metric? 

    	From (\ref{eq:qimCim}), we know that the quantum information metric is \emph{bounded from below} by the classical information metric, and that the two are equivalent if we can find a basis in which the wavefunction is purely real. We expect that it is generally possible to find such a basis for the ground state of a theory---unless, perhaps, there is a ground state degeneracy---and so all of the conclusions of the previous section apply \emph{mutatis mutandis}. However, we do not need to rely on this assumption and will argue that infinite distance limits in the quantum information metric, under similar reasonable assumptions, also imply factorization limits.

        Let us consider the quantum mechanical analog to the family of distributions (\ref{eq:probExp}), which is a one-parameter family of states
        \begin{equation}
            |\Psi(\epsilon) \rangle = |\Psi_0 \rangle + \epsilon^\alpha |\Psi_1 \rangle + \cdots,
        \end{equation}
        that asymptotes to a fixed fiducial state $|\Psi_0 \rangle$ as $\epsilon \to 0^+$, where the $\cdots$ denote corrections that are subleading in $\epsilon$. Again, we will drop the $+$ superscript as the direction of this limit will always be from above. As such, the leading $\epsilon$-dependence has a positive exponent, $\alpha > 0$. Here, we use $\cdots$ to denote terms that are subleading in $\epsilon$. Without loss of generality, we will assume that this family is normalized for all $\epsilon$, $\langle \Psi(\epsilon) |\Psi(\epsilon) \rangle = 1$. In particular, the fiducial state is also normalized $\langle \Psi_0 |\Psi_0\rangle = 1$,  and so the leading-order correction obeys $\lab{Re}\, \langle \Psi_0 | \Psi_1 \rangle = 0$. It is always possible to locally rephase the family $|\Psi(\epsilon)\rangle \to e^{i \alpha(\epsilon)} |\Psi(\epsilon) \rangle$ such that $\langle \Psi_0 | \Psi_1 \rangle$ is purely real and thus $\langle \Psi_0 | \Psi_1 \rangle = 0$, though such a rephasing may not exist globally in the presence of Berry curvature.

       	With the ``tangent'' state, $|\es \ud \Psi \rangle = \alpha \epsilon^{\alpha - 1} \ess \ud \epsilon \ess |\Psi_1 \rangle + \cdots\,$
        the quantum information metric is trivial to evaluate, 
        \begin{equation}
            \ud s^2 = \langle \ud \Psi |\big( \mathbbm{1} - |\Psi \rangle \langle \Psi | \big) |\es \ud \Psi \rangle = \frac{\alpha^2 \ess \ud \epsilon^2}{\epsilon^{2 - 2 \alpha}}\left[\langle \Psi_1 | \Psi_1 \rangle - \left|\langle \Psi_0 | \Psi_1 \rangle \right|^2 \right] + \cdots
        \end{equation}
        where the $\cdots$ denote terms subleading in $\epsilon$. In direct analogy to (\ref{eq:notInfDist}), we see that this type of ``regular'' family of states can never produce an infinite distance singularity. This provides a simple explanation for why quantum critical points are never  (\ref{eq:infDistInequality}) associated to infinite distance singularities. At most, the long-ranged fluctuations and closing energy gaps can generate a non-analytic dependence on $\epsilon$, such that $0 < \alpha < 1$, and so they may be associated with a finite-distance metric singularity. However, the family of vacua still approaches a fixed fiducial state---the one associated with the critical point---and so $\epsilon \to 0$ cannot be at infinite distance.

        As in the classical case, we could have that the ``fiducial'' state also depends on $\epsilon$, but since $|\Psi_0\rangle$ is normalized it must also have some non-trivial basis dependence. Projecting onto an arbitrary basis which we denote $|\mb{x} \rangle$, we must have that
        \begin{equation}
            \langle \mb{x} | \Psi_0(\epsilon)\rangle \sim \frac{1}{\epsilon^{\gamma N/2}} \, \eta(\mb{x}/\epsilon^\gamma)\,,\mathrlap{\qquad \epsilon \to 0\,,}
        \end{equation}
        where  $\eta(\mb{y})$ is a complex-valued function of $N$ variables. Again, $|\Psi_0(\epsilon)\rangle$ is meant to capture the leading behavior of the family as $\epsilon \to 0$, and so any corrections to this are taken to be subleading in $\epsilon$ and captured by an expansion like $|\Psi(\epsilon) \rangle = |\Psi_0(\epsilon) \rangle + \epsilon^\alpha |\Psi_1 \rangle + \cdots$. The leading behavior of the quantum information metric is then,
        \begin{equation}
            \ud s^2 \sim \frac{\gamma^2\, \ud \epsilon^2}{\epsilon^2} \left[\int\!\ud^N y \left|y^i \partial_i \eta(\mb{y}) + \tfrac{1}{2} N  \eta(\mb{y})\right|^2  - \left|\int\!{\ud^N y} \, \bar{\eta}(\mb{y}) \!\left[y^i \partial_i \eta(\mb{y}) + \tfrac{1}{2} N  \eta(\mb{y})\right]\right|^2\right] \label{eq:qimInf}
        \end{equation}
        as $\epsilon \to 0$, where we have changed integration variables to $\mb{y} = \mb{x}/\epsilon^\gamma$.

        In general, it is not possible to write (\ref{eq:qimInf}) in a form similar to (\ref{eq:cimInfRewrite}) which vanishes on an arbitrary Gaussian with complex covariance. However, as a reference for later sections it will be useful to compute the leading behavior of (\ref{eq:qimInf}) for a general Gaussian bump function $\eta(\mb{y})$, which we will write as
        \begin{equation}
            \eta(\mb{y}) = \left[\det(a/2\pi)\right]^{\frac{1}{4}} \e^{-\frac{1}{4} A_{ij} y_i y_j}\,.
        \end{equation}
        Here, $A_{ij} = a_{ij} + i b_{ij}$ is a complex symmetric $N \times N$ matrix whose real and imaginary components we call $a_{ij}$ and $b_{ij}$, with $a_{ij}$ positive definite. By inserting this into (\ref{eq:qimInf}), a tedious but straightforward calculation show that
        \begin{equation}
            \ud s^2 = \frac{N \es \gamma^2 \es \ud \epsilon^2}{2\es \epsilon^2}\left[1 + \frac{1}{N} b_{ij} b_{k \ell} a^{jk} a^{\ell i}\right] \label{eq:qimGaussian}
        \end{equation}
        where we use $a^{ij}$ to denote the inverse of $a_{ij}$, $a^{ij} a_{j k} = \delta^{i}_k$. As expected, when this state is purely real, $b_{ij} = 0$, this reduces to a quarter of the classical information metric evaluated on a Gaussian distribution, (\ref{eq:cimInfRewrite}). Again, the coefficient of the infinite distance singularity is proportional to the number of degrees of freedom factorizing, though the exact proportionality depends on how the factorization limit is approached. 
        
        As we described in \S\ref{sec:ftInfoGeo}, we can define a metric on arbitrary, continuous families of quantum field theories by studying the quantum information metric associated to their vacua $|\es \Omega(\epsilon)\rangle$. The expectation values that the metric characterizes are equal-time correlators, which  do not necessarily contain direct information about the dynamics of the theory. However, we will only be interested in Lorentz invariant theories, with Lorentz invariant vacua. Equal-time correlators in these theories can then be boosted or analytically continued to inequal-time correlators, and thus the quantum information metric also characterizes the dynamics of these theories. In the next section, we will study this metric and its infinite distance limits in a variety of examples.

\newpage
\section{Illustrative Examples} \label{sec:examples}
    
    In this section, we compute and study the information metric in a variety of examples of increasing complexity, with a particular focus on identifying both the infinite distance limits and their corresponding factorizing degrees of freedom. Furthermore, we will propose a ``physical'' definition of the intensive information metric, defined so that it correctly captures the number of factorizing Gaussian degrees of freedom in a scalar field.

    \subsection{Scalar Fields} \label{sec:scalarFields}

    		It will first be helpful to study the information metric associated with the mass $m$ of a complex scalar field $\Phi$ in $d$ spatial dimensions, defined by the Euclidean action
    		\begin{equation}
            	S_\slab{e} = \int\!\ud^{d+1} x \left[\partial_\mu \bar{\Phi} \ess \partial_\mu \Phi +  m^2 \ess \bar{\Phi} \Phi\right] . \label{eq:complexScalarAction}
        	\end{equation}
        	Intuitively, we expect that the limit $m^2 \to \infty$ is at infinite distance in the information metric because all $n$-point correlation functions of $\Phi$ become ultra-localized,
        	\begin{equation}
        		\langle \bar{\Phi}(x) \Phi(0) \rangle \equiv \int\!\!\frac{\ud^{d+1} k}{(2 \pi)^{d+1}} \frac{e^{i k x}}{k^2 + m^2} \sim \int\!\!\frac{\ud^{d+1}k}{(2 \pi)^{d+1}} \frac{e^{i k x}}{m^2} = \frac{1}{m^2} \ess \delta^{(d+1)}(x)\,, \quad m^2 \to \infty\,,
        	\end{equation} 
            and this is a type of factorization limit.

        	It is easiest to see that this is a factorization limit by studying the vacuum wavefunctional of the theory. We first expand the scalar field in the Fourier modes,
        	\begin{equation}
            	\Phi(t, \mb{x}) = \frac{1}{L^{\frac{d}{2}}} \sum_{\mb{k} \neq 0} \phi_{\mb{k}}(t) \,\e^{i \mb{k} \cdot \mb{x}}\,,
        	\end{equation}
        	where regulate the theory by placing it in a box of side-length $L = \Lambda_\slab{ir}^{\sminus 1}$ and restrict to momenta~$|\mb{k}| \leq \Lambda_\slab{uv}$, with $\mb{k} = 2 \pi \mb{n}/L$ and $n \in \mathbb{Z}^d$. Furthermore, we have explicitly excluded the zero-mode $\mb{k} = 0$, as we want this degree of freedom to be \emph{fixed} in the above family (\ref{eq:complexScalarAction}), cf.~\cite{Stout:2021ubb}. That is, we are only interested in mass deformations, and we want to keep the boundary condition $\Phi(x) \to 0$ as $|x| \to \infty$ fixed as we vary $m^2$. The theory then reduces to a set of decoupled harmonic oscillators, labeled by their momentum $\mb{k}$, with vacuum wavefunctional given by the product state
        	\begin{equation}
        		\Psi[\{\phi_{\mb{k}}\}] \propto \prod_{\mb{k} \neq 0}^{\mb{k} \leq \Lambda_\slab{uv}} \exp\left(\minus \sqrt{\mb{k}^2 + m^2} \es |\phi_{\mb{k}}|^2\right)\,. \label{eq:csVacWF}
        	\end{equation}
        	When $m^2 \gg \Lambda^2_\slab{uv}$ and the decay length of the field $\xi \propto m^{\sminus 1}$ becomes much smaller than our ability to spatially resolve points, this wavefunctional asymptotes to the product of $(L \Lambda_\slab{uv})^d$ copies of the simple Gaussian wavefunction $(2 m/\pi)^{\frac{1}{2}}\exp(\minus m |\phi|^2)$. Clearly, $m^2 \to \infty$ corresponds to a factorization limit, as each of the Fourier modes of the field completely freezes out. 

            \newpage
        	The metric associated with the parameter $m^2$ for each of these independent factors is just  $\ud s^2 = (\ud m^2)^2/(16 m^4)$, so the extensive metric for the complex scalar field  must behave as
        	\begin{equation}
        		\ud s^2 \sim \frac{N}{16} \frac{\big(\ud m^2\big)^2}{m^4}\,, \mathrlap{\qquad m^2 \to \infty\,,}
        	\end{equation}
        	while this result should be divided by two for a real scalar field. This limit gives a precise definition of what the intensive information metric \emph{should} be for a complex scalar field in the infinite mass limit,
        	\begin{equation}
        		\ud \tilde{s}^2 \sim \frac{1}{16} \frac{\big(\ud m^2\big)^2}{m^4}\,,\mathrlap{\qquad m^2 \to \infty\,,} \label{eq:csIntMet}
        	\end{equation}
			which can then be used to calibrate the intensive information metric in different regularization schemes. We caution, though, that this choice of scheme also determines precisely which family of theories we travel along as we vary the parameter $m^2$, since they also make different predictions. There is no reason for the intensive information metrics computed in different regularization schemes to agree away from the $m^2 \to \infty$ limit. Such an ambiguity is very familiar in quantum field theoretic calculations---we could say that we choose~(\ref{eq:csIntMet}) as our ``renormalization condition,'' and $m^2 \to \infty$ as our ``renormalization scale.''
            
        	For instance, the information metric associated with $m^2$ in the continuum is found to be~\cite{Stout:2021ubb}
        	\begin{equation}
        		\ud s^2 = \frac{L^d}{16} \frac{\big(\ud m^2\big)^2}{m^4} \! \int\!\!\es\frac{\ud^d k}{(2 \pi)^d} \frac{m^4}{(\mb{k}^2 + m^2)^2}\,. \label{eq:complexScalarMetMom}
        	\end{equation}
        	If we regulate this integral with a hard UV cutoff $|\mb{k}| \leq \Lambda_\slab{uv}$, we find that
        	\begin{equation}
        		\ud s^2 = \frac{1}{16}\frac{(L \Lambda_\slab{uv})^d}{(4 \pi)^{{d}/{2}} \Gamma\big(\frac{d+2}{2}\big)} \frac{\big(\ud m^2\big)^2}{m^4}  \,  {}_2 F_1\Big(2; \tfrac{d}{2}; \tfrac{d+2}{2}; \minus \tfrac{\Lambda_{\slab{uv}}^2}{m^2}\Big) \sim \frac{1}{16}\frac{(L \Lambda_\slab{uv})^d}{(4 \pi)^{{d}/{2}} \Gamma\big(\frac{d+2}{2}\big)} \frac{\big(\ud m^2\big)^2}{m^4}
        	\end{equation} 
        	as $m^2 \to \infty$. The number of factorizing degrees of freedom in this regularization scheme is $N = 2(L \Lambda_\slab{uv})^d (4 \pi)^{\sminus \frac{d}{2}}/\Gamma\big(\frac{d+2}{2}\big)$, so the intensive information metric at arbitrary $m^2$ is defined as
        	\begin{equation}
        		\ud \tilde{s}^2 = \frac{(4 \pi)^{\frac{d}{2}} \Gamma\big(\frac{d+2}{2}\big)}{(L \Lambda_\slab{uv})^d} \, \ud s^2 = \frac{1}{16} \frac{\big(\ud m^2\big)^2}{m^4}  \,  {}_2 F_1\Big(2; \tfrac{d}{2}; \tfrac{d+2}{2}; \minus \tfrac{\Lambda_{\slab{uv}}^2}{m^2}\Big)\,. \label{eq:intMetCSDef}
        	\end{equation}
        	Furthermore, as $m^2 \to 0$, this behaves as
        	\begin{equation}
        		\ud \tilde{s}^2 \sim \frac{1}{16}\frac{(\ud m^2)^2}{\Lambda_\slab{uv}^4}\begin{cases} \frac{\pi d}{4} (2 - d)  \csc \big(\frac{\pi d}{2}\big)\, \big(m/\Lambda_\slab{uv}\big)^{d-4} & d < 4 \\ \log (\Lambda_{\slab{uv}}^4/{m^4}) - 2 & d = 4 \\ d/(d - 4) & d > 4 \end{cases} \,,
        	\end{equation}
        	which is never at infinite distance, as expected.

        	Before we move on to other examples, we should also note it is possible to remove these particular infinite distance limits with our choice of regularization scheme. For instance, if we compute the extensive information metric from the correlator (\ref{eq:infMetCorr}),
        	\begin{equation}
           		\ud s^2 = \big(\ud m^2\big)^2\! \int_{\varepsilon}^{\infty} \!\ud \tau_1 \int_{\sminus \infty}^{\sminus \varepsilon}\!\ud \tau_2 \int_{L^d}\!\ud^d x_1\, \ud^2 x_2 \,\langle \bar{\Phi}\Phi(\tau_1, \mb{x}_1) \bar{\Phi}\Phi(\tau_2, \mb{x}_2) \rangle_\lab{c}\,, 
        	\end{equation}
        	with $\varepsilon = \Lambda_\slab{uv}^{\sminus 1}$, this is equivalent to regulating the momentum space integral with an exponential suppression factor and integrating over all spatial momenta,
        	\begin{equation}
            	\ud s^2 = \frac{L^d}{16} \frac{\big(\ud m^2\big)^2}{m^4}\! \int\!\!\frac{\ud^d k}{(2 \pi)^d} \frac{m^4}{(\mb{k}^2 + m^2)^2}\ess \e^{-4 \varepsilon \sqrt{\mb{k}^2 + m^2}} \,. \label{eq:complexScalarMetMomReg}
        	\end{equation}
        	This regularization scheme places $m^2 \to \infty$ at finite distance, even as $\varepsilon \to 0$. The infinite distance singularity appeared in (\ref{eq:complexScalarMetMom}) because we could take $m^2$ much larger than any $\mb{k}^2$ in the integral and effectively reduce the integrand to a constant $m^4/(\mb{k}^2 + m^2)^2 \to 1$ as $m^2 \to \infty$. This does not work in (\ref{eq:complexScalarMetMomReg}), as the integral always has support in the region where $\mb{k}^2 > m^2$ and this region dominates the $\varepsilon \to 0$ behavior of the integral. Intuitively, this type of cutoff ``blurs'' the vacuum wavefunctional on length scales set by $\varepsilon$ and so we never actually approach the state whose correlation functions are $\delta$-function localized, even as $m^2 \to \infty$.

     	\subsection{Towers of Fields} \label{sec:towers}

     		We now consider a complex scalar field $\phi(x^\mu, y)$ on $\mathbb{R}^{1,d} \times \lab{S}^1$, where $y \sim y + 2\pi R$ is the coordinate on the circle $\lab{S}^1$ with circumference $2 \pi R$ and $x^\mu \in \mathbb{R}^{1,d}$, defined by the Euclidean action
     		\begin{equation}
     			S_\slab{e} = \int\!\ud^{d+1} x \int_{0}^{2 \pi R}\!\!\ud y \left[\partial_\mu \bar{\phi}\ess  \partial_\mu \phi + \partial_y \bar{\phi}\ess \partial_y \phi +  m^2 \bar{\phi} \phi \right]. \label{eq:extraDimAction}
     		\end{equation}
     		This $(d+2)$-dimensional theory can be described by an infinite tower of $(d+1)$-dimensional complex scalar fields labeled by the integer $n \in \mathbb{Z}$ with masses $m_n^2  = m^2 + \mu^2 n^2$, where for convenience we have introduced the ``mass-spacing parameter'' $\mu^2 = R^{\sminus 2}$.  The goal of this section is to compute the information metric associated to the mass-spacing parameter $\mu^2$ in the decompactification limit $\mu^2 \to 0$, or equivalently $R \to \infty$. This is a curious limit because, on one hand, the theory seems to be changing very dramatically---there are an infinite number of fields that are becoming light---while on the other hand, an extremely large radius is (not so dramatically) becoming larger. If the information metric measures how ``different'' theories are from one another, why and how does it assign the same distance to the point $\mu^2 = 0$ in~each~of~these~pictures?

     		We will begin with the lower $(d+1)$-dimensional perspective. As usual, we can describe this theory in terms of an infinite tower of Kaluza-Klein (KK) modes,
     		\begin{equation}
     			\phi(x^\mu, y) = \frac{1}{\sqrt{2 \pi R}} \sum_{n \in \mathbb{Z}} \varphi_n(x)\, \e^{i n y/R}\,, \label{eq:kkDecomp}
     		\end{equation}
            where the $\varphi_n(x)$ are canonically normalized $(d+1)$-dimensional complex scalar fields.  In terms of these fields, (\ref{eq:extraDimAction}) becomes 
     		\begin{equation}
            	S_\slab{e} \big[\mu^2\big] = \sum_{\ell \in \mathbb{Z}} \int\!\ud^{d+1} x \left[ \partial_\mu \bar{\varphi}_n \partial_\mu \varphi_n + (m^2 + \mu^2 n^2) \bar{\varphi}_n \varphi_n\right].
        	\end{equation}
        	As we shift $\mu^2 \to \mu^2 + \ud \mu^2$, this action is perturbed by a local operator
            \begin{equation}
                S_\slab{e}\big[\mu^2 + \ud \mu^2\big] = S_\slab{e}\big[\mu^2\big] + \ud \mu^2 \sum_{n \in \mathbb{Z}} \int\!\ud^{d+1} x\, n^2 \bar{\varphi}_n \varphi_n \,. \label{eq:towerPert}
            \end{equation} 
   			Since the KK modes are all independent of one another, we may use (\ref{eq:complexScalarMetMom}) to immediately write the metric as
            \begin{equation}
                \ud s^2 = \frac{L^d}{16} \frac{\left(\ud \mu^2\right)^2}{\mu^4} \sum_{n \in \mathbb{Z}} \int\!\!\frac{\ud^d k}{(2 \pi)^d} \frac{\mu^4 n^4}{(\mb{k}^2 + m^2 + \mu^2 n^2)^2}\,. \label{eq:towerInfMetFirst}
            \end{equation}
            Explicitly, this is the metric associated with the mass-spacing parameter $\mu^2$, \emph{keeping the $(d+1)$-dimensional fields $\varphi_n$ fixed}. 

            We can evaluate this in the decompactification limit $\mu^2 \to 0$ as follows~\cite{Heidenreich:2019bjd,Stout:2021ubb}. We first introduce a Schwinger parameter $s$ to convert the summand into a Gaussian which makes it more straightforward to judge the dominance of different contributions,
            \begin{equation}
                \sum_{n \in \mathbb{Z}} \int\!\!\frac{\ud^d k}{(2 \pi)^d} \frac{\mu^4 n^4}{(\mb{k}^2 + m_0^2 + \mu^2 n^2)^2} = \int_{0}^{\infty}\!\ud s \int\!\!\frac{\ud^d k}{(2 \pi)^d}\, s \, \e^{-s(\mb{k}^2 + m_0^2)} \, \partial_s^2 \biggl[\ess\sum_{n \in \mathbb{Z}} \e^{-s \mu^2 n^2}\biggr].
            \end{equation}
            As $\mu \to 0$, this sum converges very slowly. This suggests that we use Poisson resummation
            \begin{equation}
                \sum_{n \in \mathbb{Z}} \e^{-s \mu^2 n^2} = \sum_{\ell \in \mathbb{Z}} \int_{\mathbb{R}}\!\ud n\, \e^{-s \mu^2 n^2 + 2 \pi i \ell n} = \sqrt{\frac{\pi}{s \mu^2}} \sum_{\ell \in \mathbb{Z}}  \e^{-\pi^2 \ell^2/({s \mu^2})}
            \end{equation}
            to rewrite the sum in a much more rapidly convergent form, in which case the metric becomes 
            \begin{equation}
               \ud s^2 = \frac{L^d}{16} \frac{(\ud \mu^2)^2}{\mu^4} \sum_{\ell \in \mathbb{Z}}\int_{0}^{\infty}\!\ud s \int\!\!\frac{\ud^d k}{(2 \pi)^d} \, \frac{3 \sqrt{\pi}}{4 s^{3/2} \mu} \e^{-s(\mb{k}^2 + m^2) - \pi^2 \ell^2/(s \mu^2)}\left[1 -\frac{4\pi^2 \ell^{2}}{s\mu^2} + \frac{4\pi^4 \ell^4}{3s^2 \mu^4} \right]\,.
            \end{equation}
            Clearly, only the $\ell = 0$ term survives as $\mu^2 \to 0$ since the $\ell \neq 0$ terms are exponentially suppressed. However, the remaining term diverges for small $s$ and must be regulated by a UV cutoff, $s \geq \Lambda_{\slab{uv}}^{\sminus 2}$. We thus have 
            \begin{equation}
               \ud s^2 \sim \frac{L^d}{16} \frac{(\ud \mu^2)^2}{\mu^4} \int_{\Lambda_\slab{uv}^{\ssminus 2}}^{\infty}\!\ud s \int\!\!\frac{\ud^d k}{(2 \pi)^d} \, \frac{3 \sqrt{\pi}}{4 s^{3/2} \mu} \e^{-s(\mb{k}^2 + m^2)}\sim \frac{3}{32} \frac{\sqrt{\pi}(4 \pi)^{\sminus \frac{d}{2}} }{d+2} \frac{L^d \Lambda_\slab{uv}^{d+1}}{\mu}\,  \frac{(\ud \mu^2)^2}{\mu^4}
            \end{equation}
            as $\mu^2 \to 0$. As we will argue shortly, cf.~\cite{Stout:2021ubb}, the extra factor of $\mu^{\sminus 1} \Lambda_{\slab{uv}}$ should be interpreted exactly like $L \Lambda_\slab{uv}$ was in the previous section. This divergence is a symptom of the fact that the theories we are studying have an infinite number of quantum fields, and so our definition of the intensive information metric should properly account for this. Stripping off this factor, the intensive information metric then displays the familiar logarithmic singularity $\ud \tilde{s}^2 \propto \big(\ud \mu^2\big)^2/\mu^4$, and so we expect that $\mu^2 \to 0$ is a factorization limit.

            This factorization limit is fairly similar to that of the single complex scalar field discussed in the previous section. As before, we can compute the vacuum wavefunctional of the theory in terms of the KK modes' Fourier components, 
            \begin{equation}
                \Psi[\{\varphi_{n, \mb{k}}\}] = \prod_{n \in \mathbb{Z}} \prod_{\substack{\mb{k} \neq 0}} \exp\!\left(-\sqrt{\mb{k}^2 + m^2 + \mu^2 n^2} \, |\varphi_{n, \mb{k}}|^2\right)\,, \label{eq:towerWF}
            \end{equation}
            where $\varphi_{n}(x) = L^{\sminus d/2} \sum_{\mb{k}} \varphi_{n,\mb{k}}(t) \, \e^{i \mb{k} \cdot \mb{x}}$.
            This wavefunctional has modes $|n| \gg 1$ with very large masses that are ``close'' to factorizing, much in the same way the complex scalar field had in the limit $m^2 \to \infty$. Furthermore, these are the modes whose frequencies $\omega_{n, \mb{k}} = \sqrt{\mb{k}^2 + m^2 + \mu^2 n^2}$ change most dramatically as we vary $\mu^2 \to \mu^2 + \ud \mu^2$ and, as such, these are the modes which provide the dominant contribution to the information metric. The factorization of these large $|n|$ modes are then collectively responsible for the infinite distance singularity.

            We can gain more insight by sticking with a description of this theory that remains valid in the decompactification limit. Instead of expanding in the Kaluza-Klein modes (\ref{eq:kkDecomp}), we may instead directly Poisson resum the field $\phi(x, y)$ and expand in ``images'' of a higher $(d+2)$-dimensional complex scalar field $\Phi(x, y)$ defined on the decompactified space $\mathbb{R}^{1, d+1}$,
            \begin{equation}
            	\phi(x, y) = \frac{1}{\sqrt{2 \pi R}} \sum_{n \in \mathbb{Z}} \varphi_n(x) e^{i n y/R} = \sum_{\ell \in \mathbb{Z}} \Phi(x, y - 2 \pi \ell R)\,,
            \end{equation}
            where the two descriptions are explicitly related by,
            \begin{equation}
            	\Phi(x, y) = \frac{1}{\sqrt{2 \pi R}} \int_{\sminus \infty}^{\infty}\!\ud n \, \varphi_n(x) \, \e^{i n y/R} 
              \,. \label{eq:fieldRedef}
            \end{equation} 
            Expressed in terms of this noncompact field, the Euclidean action (\ref{eq:extraDimAction}) appears nonlocal,
            \begin{equation}
                \begin{aligned}
                    S_\slab{e}\big[\mu^2\big] =& \sum_{\ell \in \mathbb{Z}} \, \int\!\ud^{d+1} x \int_{\sminus \infty}^{\infty}\!\ud y\, \bar{\Phi}(x, y- 2\pi \ell R) (-\partial^2 + m^2 ) \Phi(x, y) \,,
                \end{aligned} \label{eq:noncompactAction}
            \end{equation} 
            but this simply encodes the fact that the points $y$ and $y + 2 \pi \ell R$ are physically the same. 

            Importantly, the map (\ref{eq:fieldRedef}) involves factors of $R$ or, equivalently, $\mu^2$. Varying $\mu^2$ while simultaneously keeping the lower-dimensional fields $\varphi_n(x)$ fixed involves a non-obvious perturbation to the higher-dimensional action (\ref{eq:noncompactAction}), as the field $\Phi$ changes non-trivially. From (\ref{eq:fieldRedef}), we find that increasing $R$ while keeping $\varphi_n(x)$ fixed \emph{shrinks} the overall amplitude of the field $\Phi$, such that
            \begin{equation}
            	\frac{\partial}{\partial \mu^2} \Phi(x, y - 2 \pi \ell R) = \frac{1}{2 \mu^2} \left(y \ess \frac{\partial}{\partial y}+ \frac{1}{2}\right) \!\Phi(x, y - 2 \pi \ell R)\,,
            \end{equation} 
			and modes with higher momentum in the $y$-direction are affected more strongly. In this description, the perturbation to the action becomes
            \begin{equation}
                \delta S_\slab{e} = \ud \mu^2 \sum_{n \in \mathbb{Z}} \int\!\ud^{d+1}x \, n^2 \bar{\varphi}_n \varphi_n =- \frac{\ud \mu^2}{\mu^2}\sum_{\ell \in \mathbb{Z}} \int\!\ud^{d+2} x \,\bar{\Phi}(x, y-2 \pi \ell R) \ess \partial_y^2 \Phi(x, y)\,.
            \end{equation} 
           	As $\mu^2 \to 0$, the amplitude of the noncompact field shrinks very quickly, though this is obscured by the canonical normalization. We expect that the corresponding conjugate momentum field $\Pi(x, y)$ \emph{grows} in amplitude, and so the vacuum wavefunctional in the conjugate momentum basis becomes extremely strongly peaked around $\Pi(x, y) = 0$. The complicated, collective factorization limit seen in (\ref{eq:towerWF}) takes a much simpler form in this higher-dimensional description.

            We can also use (\ref{eq:infMetCorr}) to compute the information metric associated with this perturbation. As $R \to \infty$, only the $\ell = 0$ contribution matters as all others are exponentially suppressed due to propagation around the large spatial circle. We thus have
            \begin{equation}
                \ud s^2 \sim \frac{L^{d+1}}{16} \frac{\big(\ud \mu^2\big)^2}{\mu^4} \int\!\!\frac{\ud^{d+1} k}{(2 \pi)^{d+1}} \frac{k_{y}^4}{(\mb{k}^2 + k_y^2 + m^2)^2}\,, \mathrlap{\qquad \mu^2 \to 0\,,}
            \end{equation}
            where $\mb{k}$ and $k_y$ are the spatial momenta in the noncompact $x^\mu$ and $y$ directions, respectively.
            This is exactly what we find if we replace the sum in (\ref{eq:towerInfMetFirst}) with an integral
            \begin{equation}
            	\sum_{n \in \mathbb{Z}} \frac{(\mu n)^4}{(\mb{k}^2 + m^2 + (\mu n)^2)^2}  \to \frac{2 \pi}{\mu} \int\!\frac{\ud k_y}{2 \pi} \frac{k_y^4}{(\mb{k}^2 + k_y^2 + m^2)^2}
            \end{equation}
            and identify the extra factor of $(2 \pi/\mu) \equiv L$ with the additional IR cutoff needed to regulate this higher-dimensional theory. This justifies our earlier identification that $ \mu^{\sminus 1} \Lambda_\slab{uv} \sim L \Lambda_\slab{uv}$. 

            Regulating this integral in the UV by restricting to $(d+1)$-dimensional spatial momenta $k^2 = \mb{k}^2 + k_y^2 \leq \Lambda^2_\slab{uv}$, we find that in the limit $\Lambda_\slab{uv} \to \infty$ , the information metric behaves as
              \begin{equation}
                \ud s^2 \sim \frac{3 \ess (L \Lambda_\slab{uv})^{d+1}}{2 \ess (d+1)}  \frac{(4 \pi)^{\sminus \frac{d+1}{2}}}{\Gamma\big(\frac{d+5}{2}\big)} \frac{1}{16}\frac{\big(\ud \mu^2\big)^2}{\mu^4} \,,\mathrlap{\qquad \mu^2 \to 0\,.}
            \end{equation}
            This higher-dimensional picture also allows us to use the infinite distance point of a $(d+2)$-dimensional complex scalar field to explicitly define the intensive information metric. Using (\ref{eq:intMetCSDef}), we can explicitly compute the coefficient of the intensive information metric to be
            \begin{equation}
            	\ud \tilde{s}^2 = \frac{(4 \pi)^{\frac{d+1}{2}} \Gamma\big(\frac{d+3}{2}\big)}{(L \Lambda_\slab{uv})^{d+1}}\,  \ud s^2 \sim \frac{3}{(d+3)(d+1)} \frac{1}{16} \frac{\big(\ud \mu^2\big)^2}{\mu^4}\,, \mathrlap{\qquad \mu^2 \to 0\,.} 
            \end{equation}
            Thus, in terms of the intensive distance, the masses of the tower behave as
            \begin{equation}
            	m_n^2(\tilde{s}) \sim m^2 + \mu_0^2\ess   n^2\ess \e^{\sminus \sqrt{\frac{16 (d+3)(d+1)}{3}} (\tilde{s}-\tilde{s}_0)}\,,\mathrlap{\qquad \tilde{s} \to \infty\,,} 
            \end{equation}
            where $\mu_0^2$ is the mass spacing at $\tilde{s}_0$.
            Interestingly, the fact that not all KK modes factorize at the same ``rate'' suppresses the overall coefficient of infinite distance singularity in the intensive information metric, when compared to the complex scalar field.
            It would be useful to understand how this normalization relates to the standard normalization of the metric on moduli space to compare to other results about these towers of fields, though this is beyond the scope of this paper and so we leave it to future work.

    \subsection{Tensionless Strings} \label{sec:tensionless}

    	Infinite distance limits are often associated with the appearance of a tensionless string~\cite{Lee:2019wij}. In this section, we will argue from the worldsheet that a tensionless string also realizes a type of factorization limit. Our arguments will be more indirect than in previous sections, as we must rely on a first-quantized description of the theory and we do not yet know how to compute the information metric of the full theory directly from this perspective.

    	Let us consider closed bosonic string theory on flat space $\mathbb{R}^{1, d}$, defined by the Polyakov action,
        \begin{equation}
        	S_\slab{p} = -\frac{T_\lab{s}}{2} \int\!\ud^2 \sigma \, \sqrt{\minus \gamma} \ess \gamma^{ab}\ess \eta_{\mu \nu} \ess \partial_a X^\mu \partial_b X^\nu \,. \label{eq:polyakovAction}
        \end{equation}
        Here, we use $(\tau, \sigma)$ to denote coordinates on the worldsheet, $\gamma_{ab}$ the worldsheet metric,  $a, b, \ldots = 0, 1$ to denote worldsheet indices, $X^\mu(\tau, \sigma) = X^\mu(\tau, \sigma + 2\pi)$ to describe the string's embedding in spacetime, $\mu, \nu, \ldots = 0, \ldots, d$ to denote spacetime indices, and $\eta_{\mu \nu} = \lab{diag}(\minus 1, +1, \ldots)$ to denote the standard Lorentzian flat metric. Finally, $T_\lab{s} = (2 \pi \alpha')^{\sminus 1} = 2 \pi/\ell_\lab{s}^2$ is the tension of the string, which may be interchanged with the Regge slope $\alpha'$ or the (reduced) string length $\ell_\lab{s}$. Our arguments should survive if (\ref{eq:polyakovAction}) is only a sector of a larger worldsheet theory that includes, for instance, the supersymmetric partners of the $X^\mu$ or an internal CFT describing the string's propagation along a compact manifold. Similarly, the extension to the open string is straightforward.

        We will not review the full quantization of this theory, cf.~\cite{Polchinski:1998rq,Green:2012oqa,Zwiebach:2004tj}, but will instead focus on the features relevant to the tensionless limit and its interpretation as a factorization limit. Working in the gauge in which the worldsheet metric is flat, $\gamma_{ab} = \eta_{ab} = \lab{diag}(\sminus 1, +1)$, and in worldsheet lightcone coordinates $\sigma^\pm = \tau \pm \sigma$, the equations of motion simplify to
        \begin{equation}
        	\partial_\subp \partial_{\!\ess \subm} X^\mu = 0\,,
        \end{equation}
        where $\partial_{\pm} \equiv \partial/\partial {\sigma^\pm}$. Solutions can be divided into left- and right-moving modes, $X^\mu_\slab{l}(\sigma^\subp)$ and $X^\mu_\slab{r}(\sigma^\subm)$ respectively,
        \begin{equation}
            	X^\mu(\tau, \sigma) = X^\mu_\slab{l}(\sigma^\subp) + X^\mu_\slab{r}(\sigma^\subm) = \frac{1}{\sqrt{2 \pi}}\sum_{n \in \mathbb{Z}} x_n^\mu(\tau)\ess  \e^{\sminus i n \sigma}\,.
        \end{equation}
       	Since the string is closed, these solutions must also be periodic as $\sigma \to \sigma + 2 \pi$ and so admit a Fourier expansion in terms of the $\tau$-dependent coefficients $x^\mu_n(\tau)$. The left- and right-moving modes also admit such a decomposition,
       	\begin{equation}
                \begin{aligned}
                    X_\slab{l}^{\mu}(\sigma^\subp) &= \frac{1}{2} x^\mu + \sqrt{\frac{\alpha'}{2}} \tilde{\alpha}_0^\mu \sigma^\subp + i \sqrt{\frac{\alpha'}{2}} \sum_{n \neq 0} \frac{1}{n} \tilde{\alpha}_n^\mu \ess \e^{\sminus i n \sigma^+} \\
                    X_\slab{r}^{\mu}(\sigma^\subm) &= \frac{1}{2} x^\mu + \sqrt{\frac{\alpha'}{2}} \alpha_0^\mu \sigma^\subm + i \sqrt{\frac{\alpha'}{2}} \sum_{n \neq 0} \frac{1}{n} \alpha_n^\mu \ess \e^{\sminus i n \sigma^-}
                \end{aligned}
        \end{equation}
        where the $\tilde{\alpha}_n^\mu$ and $\alpha_n^\mu$ are the Fourier coefficients of the left- and right-moving modes, respectively, $\alpha_0^\mu = \tilde{\alpha}_0^\mu =  \sqrt{\alpha'/2}\, p^\mu$ is the center-of-mass momentum of the string and $x^\mu$ the initial position of the string at $\tau = 0$.

        The momentum density canonically conjugate to $X^\mu$ is
        \begin{equation}
			P^\mu(\tau, \sigma) = \frac{1}{2 \pi \alpha'} \dot{X}^\mu(\tau, \sigma) = \frac{1}{\sqrt{2 \pi}}\sum_{n \in \mathbb{Z}} p_n^\mu(\tau)\ess  \e^{\sminus i n \sigma}\,, \label{eq:stringMom}
        \end{equation}
        where we again define the $\tau$-dependent Fourier coefficients $p_n^\mu(\tau)$. Quantization of this theory then proceeds by promoting $X^\mu$ and $P^\mu$ to operators and imposing the canonical commutation relations,
        \begin{equation}
            [X^\mu(\tau, \sigma), P^\nu(\tau, \sigma')] = i \es \delta(\sigma - \sigma')\es \eta^{\mu \nu} 
        \end{equation}
        with $[X^\mu(\tau, \sigma), X^\nu(\tau, \sigma')] = [P^\mu(\tau, \sigma), P^\nu(\tau, \sigma')] = 0$. This then implies that Fourier coefficients $\tilde{\alpha}_n^\mu$ and $\alpha_n^\mu$ obey a ladder-like structure when promoted to operators,
        \begin{equation}
        	[\tilde{\alpha}_m^\mu, \tilde{\alpha}_n^\nu] = [\alpha_m^\mu, \alpha_n^\nu] =  m \ess \delta_{m+n, 0} \ess \eta^{\mu \nu} \quad \text{and} \quad [\tilde{\alpha}^\mu_m, \alpha_n^\nu] = 0\,.
        \end{equation}
        Finally, the vacuum state $|0 \rangle$ of this theory is defined as the state which annihilated by the positive frequency operators
        \begin{equation}
        	\tilde{\alpha}_n^\mu | 0 \rangle = \alpha_n^\nu | 0 \rangle = 0\,, \quad \forall n > 0\,, \label{eq:stringVac}
        \end{equation}
       	while the rest of the states in the Fock space are generated by the creation operators $\alpha_{n}^{\mu}$ and $\tilde{\alpha}_{n}^{\mu}$ for $n < 0$. Of course, many of these states are unphysical and must be removed by imposing the Virasoro constraints. However, this complication does not play a role in our story and so we will not detail it here.

        The worldsheet description of the theory is incomplete, as it does not provide an off-shell understanding of the vacuum state of this quantum gravitational theory, at the same level the vacuum wavefunctional did in the previous two sections. However, it should help us understand the behavior of this particular one-string sector of the theory (which becomes a particularly well-defined concept as the string becomes more and more weakly coupled) and it should give us a clue about how the theory behaves in this limit. Said differently, we can use this worldsheet description to gain insight into how distinguishable the theory is becoming if one only measures quantities that have to do with a single string or a non-interacting ensemble.

       	The information metric we have described throughout this paper characterizes our ability to distinguish between nearby theories based on the predictions their vacuum states make. An infinite distance point arises when the vacuum state ``degenerates'' and vacuum expectation values factorize. We can see how this plays out for the tensionless string by expressing the creation and annihilation operators in terms of the Fourier modes of both $X^\mu$ and $P^\mu$,
       	\begin{equation}
       		\begin{aligned}
       			(2 T_\lab{s})^{\frac{1}{2}} \ess \alpha_{\sminus n}^\mu &=   p_{n}^{\mu} - i  T_\lab{s} \ess n \ess x_{n}^{\mu} \\
            	(2T_\lab{s})^{\frac{1}{2}} \ess \tilde{\alpha}_n^\mu &= p_{n}^{\mu} - i T_\lab{s} \ess n \ess x_{n}^{\mu}
       		\end{aligned}
       	\end{equation}
       	which allows us to see explicitly the role the string tension $T_\lab{s}$ plays. These Fourier coefficients are normalized such that $[x_n^\mu, p_m^\nu] = i \delta_{n+m,0} \ess \eta^{\mu \nu}$ and so, ignoring gauge issues, $p_m^\mu = -i \partial/\partial x_{\sminus m}^\mu$. In terms of these operators, the vacuum state (\ref{eq:stringVac})  satisfies~\cite{Isberg:1993av,Gustafsson:1994kr}
       	\begin{equation}
       		\Big(p_n^\mu - i T_\lab{s} |n| x_n^\mu\Big) | 0 \rangle = 0, \qquad n \neq 0\,,
       	\end{equation}
       	so that the vacuum wavefunctional \emph{on the string} is proportional to
       	\begin{equation}
       		\Psi[\{x_n^\mu\}] \propto \exp\big(\minus |n| T_\lab{s}  |x_n^\mu|^2 \big) 
       	\end{equation}
       	This can be compared to the analogous expression for the simple harmonic oscillator with coordinate $q$ and momentum $p$,
       	\begin{equation}
       		\big(p - i \omega q\big) | 0 \rangle = 0 \to \Psi(q) \propto \exp\big(\minus \tfrac{1}{2} \omega q^2\big)\,.
       	\end{equation}
       	We see that the string is delocalized on length scales of order the string length $\ell_\lab{s} \propto T_\lab{s}^{\sminus 1/2}$. As the string tension goes to zero, the string becomes infinitely long and floppy, occupying all of space. However, as is familiar from the simple harmonic oscillator, the canonical commutation relations ensure that the string's momentum must simultaneously become extremely well-defined, and expectation values of this momentum should then factorize. As we mentioned previously, we expect that also applies to strings with more complicated worldsheet descriptions. As long as there is a noncompact direction over which the string can delocalize, we expect that the tensionless limit is at infinite distance and there are some observables that factorize.

        This argument is unfortunately very indirect, and it would be helpful to explicitly identify which observables factorize (or even exist) in this limit. This is much easier to do if the string lives in anti-de Sitter (AdS) space with curvature radius $\ell_\lab{AdS}$. For instance, it is believed~\cite{Sundborg:1999ue,Haggi-Mani:2000dxu,Sundborg:2000wp} that $\mathcal{N}=4$  super Yang-Mills with gauge group $\lab{SU}(N)$ at weak 't Hooft coupling $\lambda = g_\slab{ym}^2 N \sim (\ell_\lab{AdS}/\ell_\slab{s})^4 \ll 1$ is dual to the tensionless limit $\ell_s \to \infty$ of Type IIB string theory in $\lab{AdS}_5 \times \lab{S}^5$. The bulk theory contains an infinite tower of massless higher spin gauge fields. The boundary theory then has a completely free sector~\cite{Sundborg:2000wp,Maldacena:2011jn,Maldacena:2012sf,Boulanger:2013zza,Alba:2015upa,Hartman:2015lfa}, and correlators of gauge-invariant operators like $\Phi^{IJ} \equiv \lab{tr}[\es \phi^I \ess \phi^J]$ factorize, where $\phi^I$ are the scalars of  $\mathcal{N} = 4$ super Yang-Mills  and $I, J=1,\ldots, 6$ are $\lab{SO}(6)$ R-symmetry indices.

       It is strongly suspected~\cite{Gross:1987kza,Gross:1987ar,Gross:1988ue,Witten:1988xi,Isberg:1992ia,Isberg:1993av,Gustafsson:1994kr,Sundborg:1994py} that the tensionless string (or equivalently the high energy or high temperature phase of string theory) in flat space is a topological theory,\footnote{This is similar to the proposal that our very early universe existed in a topological phase~\cite{Brandenberger:1988aj,Agrawal:2020xek}, though the states considered there are not necessarily the vacuum state of the theory.} in which the string is invariant under spacetime diffeomorphisms. Such strings are only characterized by their topology, so the tensionless string seemingly realizes a factorization limit by removing the concept of space altogether. Interestingly, the scattering of strings is known~\cite{Gross:1987kza,Gross:1989ge} to be dominated at high energies by a non-trivial saddle in the string path integral, which is related to the appearance of a higher spin symmetry~\cite{Sagnotti:2003qa}. Following the discussion surrounding (\ref{eq:generalSaddle}), it is natural to suppose that this saddle is indicative of a factorization limit, though how these are related in practice is far beyond the scope of this paper. We instead leave the study of this deep question to future work and move on to other examples.

    \subsection{Classical Limits} \label{sec:classicalLimits}

        Classical limits are those in which quantum fluctuations vanish and expectation values collapse to a classical trajectory. We thus expect the classical limit of any quantum system to be at infinite distance in the information metric. However, a classical limit is more complicated than simply taking $\hbar \to 0$. For instance, the same quantum field theory can behave as either a collection of localized classical particles or a classical wave, depending on which state the system is in and precisely how the $\hbar \to 0$ limit is taken. Neither is the $\hbar \to 0$ limit sufficient to ensure that quantum fluctuations vanish since the system can be in a highly squeezed state or something analogous. The classical limit of a quantum system thus requires a precise specification of the state of the system as we take $\hbar \to 0$.

    	Whatever classical limit we take, it must be true that the uncertainty in a set of coordinates $x_i$ and their corresponding conjugate momenta $p_i$, with $i=1, \ldots, N$, and $[\hat{x}_i, \hat{p}_j] = i \hbar \delta_{ij}$, simultaneously vanish as $\hbar \to 0$. This is the definition of a classical limit. To this end, we may introduce a set of Gaussian coherent states, whose wavefunctions in the $|\mb{x}\rangle$ basis are
		\begin{equation}
			\langle\es  \mb{x}\es | \ess \mb{p}\ess, \mb{q} \ess \rangle_\hbar  = \frac{1}{(\pi \hbar)^{N/4}} \exp\!\left(\frac{1}{\hbar}\left[i \mb{p}\cdot \mb{x} - \frac{1}{2}(\mb{x} - \mb{q})^2 \right]\right). \label{eq:coherentStates}
		\end{equation}
        It will be useful to review the properties of these coherent states, following~\cite{Yaffe:1981vf}, as this discussion will also be relevant for the next section in which we analyze large-$N$ limits.

        Coherent states (\ref{eq:coherentStates}) have a number of extremely useful properties. They are non-orthogonal,
        \begin{equation}
            \langle \ess \mb{p}, \mb{q}\ess  | \ess \mb{p}', \mb{q}'\rangle = \exp\left(-\frac{1}{4 \hbar}\left[(\mb{p}- \mb{p}')^2 + (\mb{q}- \mb{q}')^2 + 2 i (\mb{p} - \mb{p}') \cdot (\mb{q} - \mb{q}')\right]\right)
        \end{equation}
        and thus provide an overcomplete basis on the Hilbert space spanned by $|\mb{x}\rangle$,
        \begin{equation}
            \mathbbm{1} = \int\!\frac{\ud^N p\, \ud^N q}{(2 \pi \hbar)^{N}}\,\,  |\ess \mb{p}\ess ,\es \mb{q}\ess  \rangle \langle \ess \mb{p}\ess , \es\mb{q}\ess|\,.
        \end{equation}
        This basis can be used to represent any quantum state $\psi(\mb{q}, \mb{p}) \equiv \langle \mb{q}, \mb{p} | \psi \rangle$, regardless of whether $\hbar$ is small or how these states evolve in time. Furthermore, the fact that it is overcomplete allows one to recover \emph{all} matrix elements of an operator $\hat{A}$ just from its diagonal elements in the coherent state basis,
        \begin{equation}
            A(\mb{p}, \mb{q}) = \langle \ess \mb{p}, \mb{q}\ess |\ess \hat{A}\ess |\ess \mb{p}, \mb{q}\ess\rangle\,,
        \end{equation}
        where $A(\mb{p}, \mb{q})$ is called the \emph{symbol} of the operator $\hat{A}$. Importantly, since these coherent states become orthogonal in the classical limit,  this structure makes it very simple to argue that expectation values factorize as $\hbar \to 0$. 

        For instance, if we consider the symbol of the operator product $\hat{A} \hat{B}$, which may be written as
        \begin{equation}
            (AB)(\mb{p}, \mb{q}) = \int\!\frac{\ud^N p'\ess \ud^N q'}{(2 \pi \hbar)^N}\,  \big|\langle\ess  \mb{p} \ess , \mb{q} \ess | \ess \mb{p}', \mb{q}' \rangle \big|^2 \,  \frac{\langle \mb{p} \ess , \mb{q} \ess | \es \hat{A}\ess | \ess \mb{p}', \mb{q}' \rangle}{\langle \mb{p} \ess , \mb{q} \ess | \ess \mb{p}', \mb{q}' \rangle}\, \frac{\langle \mb{p}', \mb{q}' \es | \es  \hat{B} \ess |\ess\mb{p} \es, \mb{q} \rangle}{\langle \mb{p}', \mb{q}' \es | \ess \mb{p}\es , \mb{q} \rangle}\,,
        \end{equation}
        then as long as both $\hat{A}$ and $\hat{B}$ are \emph{classical operators},\footnote{The set of classical operators includes arbitrary polynomials in $\hat{\mb{x}}$ and $\hat{\mb{p}}$.} meaning that the ratios like 
        \begin{equation}
			\frac{\langle \mb{p} \ess , \mb{q} \ess | \es \hat{A}\ess | \ess \mb{p}', \mb{q}' \rangle}{\langle \mb{p} \ess , \mb{q} \ess | \ess \mb{p}', \mb{q}' \rangle} \label{eq:classical}
		\end{equation}
		remain smooth as $\hbar \to 0$, then the small-$\hbar$ asymptotics of this integral are dominated by the highly-peaked factor
        \begin{equation}
             \big|\langle \mb{p} \ess , \mb{q} \ess | \ess \mb{p}', \mb{q}' \rangle \big|^2 = \exp\left(-\frac{1}{2 \hbar} \left[ (\mb{p} - \mb{p}')^2 + (\mb{q} - \mb{q}')^2\right]\right)\,.
        \end{equation}
        This implies that these operators factorize in the $\hbar \to 0$ limit,
        \begin{equation}
            \lim_{\hbar \to 0} (AB)(\mb{p}, \mb{q}) = (A)(\mb{p}, \mb{q}) (B)(\mb{p}, \mb{q})\,, 
        \end{equation}
        as do expectation values in these states, $\langle \hat{A} \hat{B} \es \rangle \to \langle \hat{A} \rangle \langle \hat{B} \rangle$, up to terms that are subleading in $\hbar$.

        If the quantum Hamiltonian $\hat{\mathcal{H}}$ is also a classical operator, then the quantum dynamics \emph{also} classicalizes, evolving an initial coherent state into a different coherent state. The dynamics of the system can then be described by the time-dependent labels $\big(\mb{p}(t), \mb{q}(t)\big)$ for the coherent states, which evolve according to Hamilton's equations with Hamiltonian given by the symbol $\mathcal{H}(\mb{p}, \mb{q})$. Crucially, this means that the ground state of the system can be found by minimizing this classical Hamiltonian, and so the quantum mechanical vacuum state asymptotes to a particular coherent state, $|\Omega \rangle \equiv |\mb{p}_0, \mb{q}_0\rangle$.

        Given this vacuum state or any other coherent state, we can compute the information metric associated to $\hbar$ and thus to this classical limit. A short calculation shows this limit is indeed at infinite distance,
        \begin{equation}
            \ud s^2 = \langle \ud \Omega | \ud \Omega \rangle - \langle \Omega | \ud \Omega \rangle \langle \ud \Omega | \Omega\rangle = \frac{N}{8} \frac{\ud \hbar^2}{\hbar^2} \left(1 + \frac{4}{N} \frac{\mb{p}_0^2}{\hbar}\right)\,. \label{eq:classicalLimitMet}
        \end{equation}
        Interestingly, in the presence of non-zero momentum $\mb{p}_0 \neq 0$, the $\hbar \to 0$ infinite distance singularity is more severe than logarithmic. This is not in conflict with the arguments of the previous section, however, as expectation values and operators purely involving $\mb{x}$ and not $\hbar$, i.e. $\mb{\hat{p}}/\hbar \equiv -i \nabla$, can diverge as $\hbar \to 0$. This can be easily seen from the fact that (\ref{eq:coherentStates})'s dependence on $\hbar$ cannot be removed by a simple rescaling of $\mb{x}$.  

    \subsection{Abelian Gauge Theories} \label{sec:gaugeTheory}

        The previous section argued that any classical limit in which $\hbar \to 0$ is at infinite distance in the information metric. It stands to reason that the information metric in any parameter which enters a theory like the Planck constant $\hbar$ should also display an infinite distance point as that parameter is taken to zero. A natural example is provided by abelian gauge theory in the ``charge-quantized basis,'' whose Lorentzian action is  
        \begin{equation}
            S = -\frac{1}{4 e^2} \int\!\ud^{d+1} x\, F_{\mu \nu} F^{\mu \nu}\,, 
        \end{equation}
        where $F_{\mu \nu} = \partial_\mu A_\nu - \partial_\nu A_\mu$, and $A_\mu$ is the vector potential.
        In this section, we will compute the information metric associated with the gauge coupling $e$ and show that it indeed displays a logarithmic infinite distance singularity as $e \to 0$. Since this is a weak-coupling limit, adding additional charged fields should not change this conclusion.

        Following \cite{Jackiw:1988sf}, we will work in temporal gauge $A_0 = 0$, in which case the Lagrangian becomes
        \begin{equation}
            \mathcal{L} = \frac{1}{2 e^2} \delta^{ij} \dot{A}_i \dot{A}_j - \frac{1}{4 e^2} \delta^{ij} \delta^{k \ell} F_{ik} F_{j \ell}\,. \label{eq:photonLag}
        \end{equation}
        The momentum conjugate to the vector potential $A_i$ is then
        \begin{equation}
            \Pi^{i}(t, \mb{x}) = \frac{\partial \mathcal{L}}{\partial A_i(t, \mb{x})} = \frac{1}{e^2} \delta^{ij} \dot{A}_j(t, \mb{x})\,,
        \end{equation}
        and once we impose the canonical commutation relations $[\hat{A}_i(t, \mb{x}), \hat{\Pi}^j(t, \mb{y})] = \delta^{j}_i \delta^{(d)}(\mb{x} - \mb{y})$,  the momentum operator becomes a functional derivative $\hat{\Pi}^j(\mb{x}) \equiv -{\delta}/{\delta A_j(\mb{x})}$ 
        when acting on wavefunctionals in the $A_i(\mb{x})$ basis, $\Psi[A_i(\mb{x})] = \langle A_i(\mb{x}) | \Psi \rangle$.

        Temporal gauge is not a complete gauge-fixing, as we may still transform $A_i(t, \mb{x}) \to A_i(t, \mb{x}) + \partial_i \Lambda(t, \mb{x})$ without affecting the temporal gauge condition $A_0 = 0$. To this end, we must explicitly impose the Gauss' Law constraint $\nabla \cdot \mb{E}\ess |\Psi\rangle = 0$ on the physical states of the system. If we decompose the vector potential into longitudinal and transverse components $A_i(\mb{x}) = A_i^\slab{l}(\mb{x}) +  A_i^\slab{t}(\mb{x})$ with $\nabla_j A_j^\slab{t}(\mb{x}) = 0$, this constraint translates into the requirement that
        \begin{equation}
            \nabla_j \frac{\delta}{\delta A_i^\slab{l}(\mb{x})} \Psi[A] = 0\,,
        \end{equation}
        where $\nabla_j = \partial/\partial x^j$. We can implement this constraint by demanding that $\Psi[A]$ is only a functional of transverse field configurations.

        With this in mind, the Hamiltonian
        in the $A_i^\slab{t}(\mb{x})$ basis may be written as
          \begin{equation}
            \mathcal{H} = \frac{e^2}{2} \int\!\ud^d k \left[ -\frac{\delta}{\delta A^\slab{t}_i(\mb{k})} + \frac{|\mb{k}|}{(2\pi)^{d} e^2} A_i^\slab{t}(\minus \mb{k})\right]\left[\frac{\delta}{\delta A^\slab{t}_i(\minus \mb{k})} + \frac{|\mb{k}|}{(2\pi)^{d} e^2} A_i^\slab{t}(\mb{k})\right], \label{eq:photonHam}
        \end{equation}
        where we have dropped the zero-point contribution and introduced the Fourier transform of the transverse potential,
        \begin{equation}
            A_i^\slab{t}(\mb{x}) = \int\!\!\frac{\ud^d k}{(2 \pi)^d} A^\slab{t}_i(\mb{k}) \ess \e^{i \mb{k} \cdot \mb{x}}\,,
        \end{equation}
        such that $k_i A_{i}^\slab{t}(\mb{k}) = 0$.
		The vacuum state $|\Omega(e) \rangle$ of the theory is annihilated by the right half of~(\ref{eq:photonHam}) for all $\mb{k}$, and this first-order equation can be solved to find the vacuum wavefunctional
        \begin{equation}
            \langle A^\slab{t}_i(\mb{k}) | \Omega(e)\rangle  = \mathcal{C} \exp\left[-\frac{1}{2 e^2} \int\!\!\frac{\ud^d k}{(2 \pi)^d} \,|\mb{k}| A_j^\slab{t}(\mb{k}) A_j^\slab{t}(\minus \mb{k})\right]\,,
        \end{equation}
        where $\mathcal{C}$ is an overall constant. Clearly, the vacuum fluctuations of the vector potential die off as $e \to 0$, and so we again expect that $e = 0$ is an infinite distance point.

        We can verify this by extracting the metric from the K\"{a}hler potential
        \begin{equation}
            e^{\mathcal{K}(e, e')} =  \langle \Omega(e) | \Omega(e')\rangle = \int\!\mathcal{D} A^\slab{t}_j(\mb{k})\,  \exp\!\left[-\!\ess\left(\frac{1}{2 e^2} + \frac{1}{2 e'^2}\right) \!\int\!\!\frac{\ud^d k}{(2 \pi)^d} \,|\mb{k}| A_j^\slab{t}(\mb{k}) A_j^\slab{t}(\minus \mb{k})\right].
        \end{equation}
        For each $\mb{k}$, the integral is over two transverse complex amplitudes, and so consists of four Gaussian integrals in total. The K\"{a}hler potential for this family is then given by
        \begin{equation}
            \mathcal{K}(e,e') = -4 \times \frac{1}{2} \log \!\left[\frac{1}{e^2} + \frac{1}{e'^2}\right] \times L^d \int\!\!\frac{\ud^d k}{(2 \pi)^d}\, + \cdots\,,
        \end{equation}
        where we use $\cdots$ to denote terms that do not contribute to the metric.
        Regularizing the integral over momenta with a UV cutoff $|\mb{k}| \leq \Lambda_\slab{uv}$ and defining the intensive information metric with respect to the conventions of the complex scalar (\ref{eq:intMetCSDef}), we find that
        \begin{equation}
            \ud \tilde{s}^2 = \frac{(4 \pi)^{\tfrac{d}{2}} \Gamma\big(\frac{d+2}{2}\big)}{(L \Lambda_\slab{uv})^d} \,  \ud e^2 \left.\partial_e \partial_{e'} \mathcal{K}(e, e')\right|_{e' = e} = \frac{\ud e^2}{e^2}\,.
        \end{equation}
        Pleasantly, this definition of the intensive information metric, which was found by demanding that the complex scalar field's factorization limit was at the correct intensive distance, is equivalent to demanding that the overall coefficient of the metric for an abelian gauge coupling is unity.

        We expect similar results to apply to non-abelian gauge theories as well, and we can provide evidence for this in the context of four-dimensional $\mathcal{N} = 2$ superconformal field theories. There the Zamolodchikov metric can be computed exactly and, since this is proportional to the intensive quantum information metrics, we can interpret results about one in terms of the other. An $\mathcal{N} = 2$ vector multiplet in the adjoint representation of the group $\mathcal{G}$ is characterized by the holomorphic gauge coupling
        \begin{equation}
        	\tau = \frac{4 \pi i}{g^2} + \frac{\theta}{2 \pi}\,.
        \end{equation}
        As $g \to 0$, the Zamolodchikov metric associated to this coupling behaves as~\cite{Baggio:2014ioa,Baume:2020dqd,Perlmutter:2020buo}
        \begin{equation}
        	\ud s_\slab{z}^2 \sim \left[24 \, \lab{dim} \, \mathcal{G}\right] \frac{\ud \tau \, \ud \bar{\tau}}{(\lab{Im}\, \tau)^2}  \sim \left[96 \, \lab{dim} \, \mathcal{G}\right] \frac{\ud g^2}{g^2} \, ,\mathrlap{\quad\, g \to 0\,.}
        \end{equation}
        The intensive quantum information metric is thus proportional to $\left[\lab{dim}\, \mathcal{G}\right] \ud g^2/g^2$. As our general analysis suggests, cf. Section~\ref{sec:infFact}, the logarithmic singularity in the information metric is proportional to the number of degrees of freedom factorizing, and here it represented by the dimension of the gauge group $\lab{dim}\, \mathcal{G}$.

        Finally, this abelian example illustrates an important point. If we instead worked in the canonically-normalized basis in which the gauge coupling enters the action, for instance, in covariant derivatives of matter fields like $S_\lab{m} = \int\!\ud^{d+1} x\, \big(\big|(\partial_\mu - i e A_\mu ) \Phi\big|^2 + m^2 |\Phi|^2\big)$ where $\Phi$ is a charged scalar, this infinite distance limit disappears because now the associated statistical family is regular as $e \to 0$. Such a parameter-dependent field transformation changes the information metric associated with that parameter because that transformation changes the statistical family---we are no longer comparing members of the family based on the same observables.

        The previous two examples illustrated that ``classical'' limits are at infinite distance in the information metric. Large-$N$ limits are another type of ``classical'' limit, in which a complicated statistical or quantum theory reduces to that of a reduced and emergent degree of freedom, which enjoys no uncertainty in its predictions. However, these are limits in which the total number of degrees of freedom of the family changes as we vary the parameter of interest $N$, and so defining the information metric---which assigns distances by comparing predictions of nearby theories for the \emph{same} degrees of freedom---is conceptually subtle and is the subject of the next section.

\section{Large-\texorpdfstring{$N$}{N} Limits are at Infinite Distance} \label{sec:largeN}
	
	Theories with large-$N$ limits are defined by a complete factorization of expectation values for a preferred class of operators as the number of degrees of freedom, which is proportional to the parameter $N$, is taken to infinity. That is, given two operators $\hat{A}$ and $\hat{B}$ in this preferred class, theories that have a large-$N$ limit satisfy
	\begin{equation}
		\langle \hat{A} \hat{B} \rangle \sim \langle \hat{A} \rangle \langle \hat{B} \rangle\,, \mathrlap{\qquad N \to \infty\,.} \label{eq:largeNFact}
	\end{equation} 
	Importantly, this also implies that fluctuations in these operators $\langle \hat{A}^2 \rangle - \langle \hat{A} \rangle^2 \to 0$  as $N \to \infty$.
	In Section~\ref{sec:infFact}, we argued that infinite distance limits and factorization limits are intrinsically tied to one another. We thus expect that these large-$N$ limits must be at infinite distance in the information metric associated with the parameter $N$.

	However, this is not so straightforward. The classical and quantum information metrics assign a distance between two members of a family based on how different the predictions they make for the \emph{same degrees of freedom} are. How do we compare the predictions of two distributions or states if they make predictions for different objects? 

	We must remember that the existence of a large-$N$ limit is also predicated on the existence of a preferred set of observables that are ``stable'' as we vary the number of degrees of freedom with $N$. The theory at $N = \infty$ makes well-defined predictions for such observables, and they correspond to operators like $\hat{A}$ and $\hat{B}$ above. Our strategy will be to compare theories with different numbers of degrees of freedom based on the predictions they make for these shared operators. That is, we will trace out all other degrees of freedom to derive a reduced probability distribution or state, in which $N$ appears as a continuous parameter, so that we may then define the classical or quantum information metric.

	This prescription is maximally ignorant, in the sense that it makes no assumptions about the other degrees of freedom in the system and is thus the most natural way (or least presumptuous) to define an information metric on such a family of theories. We will argue that the information metric defined this way has an infinite distance singularity as $N \to \infty$. We will illustrate this procedure and compute the information metric for $\lab{O}(N)$-symmetric classical probability distributions, their quantum mechanical counterparts, and then finally in quantum field theory for arbitrary $\lab{O}(N)$ vector models.

    \subsection{Classical Large-\texorpdfstring{$N$}{N} Limits} \label{sec:largeNClassical}
    
    	Let us begin with classical $\lab{O}(N)$ vector models, the family of probability distributions in $N$~variables $x_i$, $i = 1, \ldots, N$, defined by
        \begin{equation}
        	P_{N}(\{x_i\}) = \frac{1}{\mathcal{Z}_N} \exp\!\left( -N V\big( \delta^{ij} x_i x_j\big)\right)\,.
        \end{equation}
        with $V(\delta^{ij} x_i x_j)$ an arbitrary $N$-independent potential in terms of the ``radius'' $r^2 \equiv \delta_{ij} x_i x_j$, and
        \begin{equation}
        	\mathcal{Z}_N = \int\!\ud^N x \, \exp\!\left(-N V\big(\delta^{ij} x_i x_j\big)\right)\,.
        \end{equation}

        To compare distributions with different numbers of degrees of freedom, we need to reduce them to distributions that make predictions for the same variables. That is, we should compare the family of distributions based on the predictions they make for the $\lab{O}(N)$ invariant $r^2 = \delta^{ij} x_{i} x_j$. This is usually an implicit step in the analysis of these large-$N$ limits, but it helps to make it explicit here. To this end, we construct the family of probability distributions for $r$,
        \begin{equation}
        	\condp{r}{N} = \int\!\ud^N x\, r \, \delta\big(\tfrac{1}{2}[r^2 - \delta^{ij} x_i x_j]\big) P_N(\{x_i\})\,,
        \end{equation}
        where we have changed notation to make it clear that $\condp{r}{N}$ can now be interpreted as a continuous family of distributions in the number of degrees of freedom $N$.
        This distribution is trivial to compute if we rely on $P_N(\{x_i\})$'s $\lab{O}(N)$ symmetry and go to hyperspherical coordinates. However, it should be clear that this type of reduced distribution is always well-defined, regardless of $P_N(\{x_i\})$'s symmetry structure, since we can always derive a probability distribution for a composite observable like $r^2 = \delta^{ij} x_i x_j$ by imposing the appropriate constraint and tracing over all ``fundamental'' degrees of freedom. This then allows us to compute the information metric for that reduced family of distributions. Simply, if a family of distributions makes predictions for a particular observable, we can always compare members of that family based on those predictions.

        The reduced distribution is then
        \begin{equation}
        	\condp{r}{N} = \frac{1}{\tilde{\mathcal{Z}}_N} r^{N-1} \e^{\sminus N V(r)} \label{eq:largeNCondP}
        \end{equation}
        where we have defined the reduced partition function
        \begin{equation}
        	\tilde{\mathcal{Z}}(N) = \int_0^\infty\!\ud r\, r^{N-1} \e^{\sminus N V(r)}\,. \label{eq:vecPF}
        \end{equation}
        This family of distributions reproduces the same expectation values of $r$ that the full family does,
        \begin{equation}
        	\langle f(r) \rangle = \int_{\mathbb{R}^N}\!\!\ud^N x\, f\big(\delta^{ij} x_i x_j\big)\ess P_N(\{x_i\}) = \int_0^{\infty}\!\ud r\, f(r) \ess \condp{r}{N}\,, 
        \end{equation}
        so no information about the ``radial'' behavior of the family $P_N(\{x_i\})$ is lost or gained in this process. Furthermore, $N$ is now a parameter that can be varied continuously in this reduced family of distributions, a fact we signify by switching notation to $\condp{r}{N}$.

        This family falls into the same class as discussed in Section~\ref{sec:infFact}, cf. (\ref{eq:generalSaddle}), whose partition function can be approximated in the large-$N$ limit as, 
        \begin{equation}
            \tilde{\mathcal{Z}}(N) \sim e^{-N V_*} \sum_{i = 1}^{m} 
             \frac{2 \Gamma(1/p_i)}{p_i} \left[\frac{p_i!}{N V^\floq{p_i}(r_i)}\right]^{1/p_i} 
        \end{equation}
        where $r_i$ with $ i = 1, \ldots, m$ denote the dominant saddles of the integral, $V'(r_i) - r_i^{\sminus 1} = 0$, and $p_i$ represent their multiplicity. As before, we can compute the information metric via
        \begin{equation}
            \ud s^2 = \left[\partial_N^2 \log \tilde{\mathcal{Z}}(N)\right] \! \ess \ud N^2\,,
        \end{equation}
        and since the maximal $p_i = p$ contribution dominates, we have
        \begin{equation}
            \ud s^2 = \frac{\ud N^2}{p \es N^2}\,. \label{eq:vecInfMet}
        \end{equation}
        As we would expect from the general arguments in Section~\ref{sec:infFact}, the $N \to \infty$ limit in this family is at infinite distance because the distribution $\condp{r}{N}$ localizes about the dominant saddle(s).

        In the case of a Gaussian distribution, $P_{N}(\{x_i\}) \propto \exp\left(-\frac{1}{2} N \delta^{ij} x_i x_j\right)$, we can be more precise. The partition function (\ref{eq:vecPF}) in this case is simply
        \begin{equation}
            \tilde{\mathcal{Z}}(N) = \frac{\Gamma\big(\frac{N}{2}\big)}{2 (N/2)^{N/2}}
        \end{equation}
        with information metric
        \begin{equation}
            \ud s^2 = \frac{\ud N^2}{2\es  N^2}\! \left[ \frac{1}{2} N^2 \psi^\floqq{1}\big(N/2\big) - N\right] \sim \frac{\ud N^2}{2\es N^2} \,,{\qquad N \to \infty\,,}
        \end{equation}
        where $\psi^\floqq{1}(z) = {\ud^2}\log \Gamma(z)/{\ud z^2}$ is the polygamma function of order one. Clearly, the asymptotic scaling as $N \to \infty$ agrees with the more general result (\ref{eq:vecInfMet}).

    \subsection{Quantum Large-\texorpdfstring{$N$}{N} Limits} \label{sec:quantLargeN}

       Let us now consider perhaps the simplest example of a large-$N$ limit in a quantum mechanical theory: the humble hydrogen atom in $\mathbb{R}^{N+1}$, with Hamiltonian
        \begin{equation}
        	\mathcal{H} = \left[-\frac{1}{2 r^N} \frac{\partial}{\partial r} \left(r^N \frac{\partial}{\partial r}\right) + \frac{1}{2 r^2} \Delta_{\lab{S}^N} - \frac{1}{r}\right]. \label{eq:hydrogen}
        \end{equation}
        Here, $\Delta_{\lab{S}^N}$ is the Laplace-Beltrami operator on the $N$-dimensional sphere $\lab{S}^N$, whose eigenfunctions are the hyperspherical harmonics with eigenvalues $\ell(\ell+N - 1)$. We will focus on the ground state of this system, the $s$-wave state
        \begin{equation}
        	\langle r, \varphi_1, \cdots, \varphi_N| \psi_0(N) \rangle =  \frac{2}{N}\!\!\left[\frac{(2/\sqrt{\pi})^N}{ N^{N} \Gamma\big(\frac{N}{2}\big)}\right]^{\frac{1}{2}} \!\e^{-2 r/N}
        \end{equation}
        with energy $\mathcal{H} | \psi_0(N)\rangle = -(2/ N^{ 2}) |\psi_0(N)\rangle$.

        As in the classical case, our goal is to integrate the angular degrees of freedom out and compare the vacua at different $N$. Tracing out degrees of freedom is a very familiar operation in quantum mechanics---we can simply construct the reduced density matrix
        \begin{equation}
        	\langle r | \tilde{\rho} \ess| r'\rangle = \int\!\ud \Omega_N \, \langle r, \varphi_1, \cdots, \varphi_N | \tilde{\rho} \ess| r', \varphi_1, \cdots, \varphi_N \rangle =  \frac{4}{N^2} \frac{(4/N)^N}{\Gamma(N)} \e^{-2 r/N - 2 r'/N}\,.
        \end{equation}
        This is a pure state, and so we can identify the ``reduced'' ground state as
        \begin{equation}
        	\langle r | \tilde{\psi}_0(N) \rangle = \frac{2}{N} \left[\frac{(4/N)^N}{\Gamma(N)}\right]^{\frac{1}{2}}\! \e^{-2 r/N}\,,
        \end{equation}
        which is the quantum mechanical analog of (\ref{eq:largeNCondP}). In general, this reduction will yield a family of mixed states. We may still define an information metric associated with that family (the so-called Bures metric) but, fortunately, the $\lab{O}(N)$ symmetry of the Hamiltonian implies that energy eigenstates have no entanglement between the radial and angular variables degrees of freedom, and so we will not need to pursue this complication here. 

        However, there is a slight hiccup. This reduced state is defined on the Hilbert space of square-integrable functions on the semi-infinite line, $L^2(\mathbb{R}_+)$, but with weight function $r^{N}$. To carry out our previous program of treating $N$ as a continuous parameter, we need to project this state onto a new basis with a trivial weight function. To do this, we will define a new basis $|\rho\rangle$ such that 
        \begin{equation}
        	\int_0^{\infty}\!\ud \rho\, |\langle \rho | \psi \rangle|^2 = \int_{0}^{\infty}\!\ud r \, r^{N} |\langle r | \psi \rangle|^2 = 1\,.
        \end{equation}
        Assuming that $\langle \rho | r \rangle \propto \delta(r - \rho)$, we must have that $\langle \rho | r \rangle = r^{-\frac{N}{2}} \delta(r - \rho)\,$ and so the reduced state projected onto this basis is
        \begin{equation}
        	\langle \rho | \tilde{\psi}_0(N) \rangle = \int_0^{\infty}\!\ud r\, r^{N} \langle \rho | r \rangle \langle r | \psi_0 \rangle = \frac{2}{N} \left[\frac{(4/N)^{N}}{\Gamma(N)}\right]^{\frac{1}{2}} \! \rho^{\frac{N}{2}} \e^{\sminus 2 \rho/N}\,.
        \end{equation}
        Note that
        \begin{align}
        	\langle \rho | \ess \hat{r}^a \ess | \rho' \rangle &= \int_{0}^{\infty}\!\ud r \, r^N \langle \rho | r \rangle \langle r | \hat{r}^a | \rho \rangle = \rho^a \delta(\rho - \rho')\,.
        \end{align}
        so expectation values in $\hat{r}$ are mapped to expectation values of $\hat{\rho}$. Importantly, the reduced state $\langle \rho | \psi_0(N)\rangle$ can be used to compute arbitrary expectation values of $\hat{r}$ and its momentum.

        We are now in a position to calculate the quantum information metric associated with the dimension of the space $N$. We construct the K\"{a}hler potential
        \begin{equation}
        	\mathcal{K}(N, N') = \log \, \langle \psi_0(N) | \psi_0(N') \rangle = \frac{1}{2}\log\!\left[\frac{2^{N+N'} N^{N'} N'^{N} \Gamma\big(\frac{1}{2}(N+N')\big)}{(N + N')^{N+N'} \Gamma(N) \Gamma(N')}\right]\,,
        \end{equation}
        to find the information metric (\ref{eq:qimKahler}),
        \begin{equation}
        	\ud s^2 =  \frac{\ud N^2}{4 N} \left[N \psi^\floqq{1}(N) + 3\right] \sim \frac{\ud N^2}{N}\,,\mathrlap{\qquad N \to \infty\,.}
        \end{equation}
        Again, we see that $N \to \infty$ is at infinite distance. However, it is a more severe infinite distance singularity than we expect to be associated with a factorization limit, i.e. $\ud s^2 \sim \ud N^2/N^2$. How does this fit with our general story? If we look at the associated classical distribution in $\rho$,
        \begin{equation}
        	\condp{z}{N} = |\langle \rho | \tilde{\psi}_0(N)\rangle|^2 = \frac{4}{N^2} \frac{(4/N)^N}{\Gamma(N)} \rho^{N} \e^{-4 \rho /N}\,,
        \end{equation}
        we find that expectation values in $\rho$ (or equivalently $r$) scale as
        \begin{equation}
        	\langle \rho^k \rangle = \langle r^k \rangle = \frac{\Gamma(N+k + 1)}{\Gamma(N+1)} \frac{N^k}{4^k} \sim \left(\frac{N^2}{4}\right)^k\,,\quad N \to \infty\,.
        \end{equation}
        We see that $N \to \infty$ is again a factorization limit. However the predictions the theory makes also scale with $N$ as $N \to \infty$. As we discussed in Section~\ref{sec:infFact}, allowing the predictions of a family of probability distributions to diverge allows for a wider, albeit uninteresting, range of infinite distance limits. If we instead remove this scaling by defining the new coordinate $\rho = N^2 \zeta$ with
        \begin{equation}
        	\condp{\zeta}{N} = \frac{4 (4 N)^N}{\Gamma(N)} \zeta^{N} \e^{-4 N \zeta}\,,
        \end{equation}
        then expectation values are finite, $\langle \zeta^k \rangle \sim 4^{\sminus k} + O\big(N^{\sminus 1}\big)$,  and the associated information metric~is
        \begin{equation}
        	\ud s^2 = \frac{\ud N^2}{4 N^2} \left(N^2 \psi^\floqq{1}(N) - N +1 \right) \sim \frac{\ud N^2}{8 N^2} \,,\quad N \to \infty\,. 
        \end{equation}
        Accounting for the factor of $4$ between the classical and quantum information metric, this asymptotic behavior is exactly what we would expect for a Gaussian-like factorization limit (\ref{eq:vecInfMet}).

    \subsection{Field Theoretic Large-\texorpdfstring{$N$}{N} Limits} \label{sec:ftLargeN}

		In \S\ref{sec:classicalLimits}, we described how the classical limit $\hbar \to 0$ of a quantum system described by positions $\hat{x}_i$ and momenta $\hat{p}_i$, $[\hat{x}_i, \hat{p}_j] = i \hbar \delta_{ij}$, was at infinite distance in the information metric. Here, expectation values for ``classical'' operators---which included arbitrary polynomials of the positions and momenta---factorized,
		\begin{equation}
			\langle (\hat{x}_i)^n (\hat{p}_j)^m \rangle \sim \langle \hat{x}_i \rangle^n \langle \hat{p}_j\rangle^m\,, \mathrlap{\qquad \hbar \to 0\,,} \label{eq:classicalFact}
		\end{equation}
        and calculations in the theory reduced to those of classical mechanics. In particular,  we could identify the vacuum state of the theory by minimizing a classical Hamiltonian.

        The similarities between the classical (\ref{eq:classicalFact}) and large-$N$ (\ref{eq:largeNFact}) limits suggest that the latter can be viewed as a type of classical limit and that the $N = \infty$ theory can be solved by identifying an appropriate set of coherent states and a corresponding ``classical'' Hamiltonian, which can subsequently be minimized to solve the theory. This was the subject of the very nice paper~\cite{Yaffe:1981vf}, which describes the construction and properties of these coherent states in a variety of large-$N$ models. We will now quickly review the relevant properties of this construction for $\lab{O}(N)$~vector models and heavily rely on analogy with the discussion of \S\ref{sec:classicalLimits}; the interested reader should consult~\cite{Yaffe:1981vf} for proofs and details.

        An arbitrary $\lab{O}(N)$ vector model can be described by a set of coordinates $\hat{x}^a_i$ and their conjugate momenta $\hat{p}^a_i$, $[\hat{x}_i^a, \hat{p}_j^b] = i N^{\sminus 1} \delta_{ij} \delta^{ab}$. We use $a, b, \ldots = 1, \ldots, N$ to denote $\lab{O}(N)$ vector indices and $i, j, \ldots = 1, \ldots, N_\lab{s}$ to denote ``spatial'' indices, where $N_\lab{s}$ can be thought of as the total number of spatial sites when the $\hat{x}_i^a = \hat{\phi}^a(y_i)$ represents an $O(N)$ vector field at a lattice site $y_i$. In terms of the regulating cutoffs, $N_\lab{s} \propto (L \Lambda_\slab{uv})^d$. The canonical commutation relations have all been rescaled by a factor of $N^{\sminus 1}$, to avoid needing to rescale the Hamiltonian's coupling constants in order to exhibit a nontrivial large-$N$ limit.  

        As before, the Hamiltonian is assumed to be $\lab{O}(N)$ invariant and we will only consider the $\lab{O}(N)$-invariant sector of the theory. There are three basic $\lab{O}(N)$ invariants from which all others may be constructed,
        \begin{equation}
            \begin{aligned}
                \hat{A}_{ij} &= \tfrac{1}{2} \delta_{ab} \hat{x}^a_i \hat{x}^b_j \\
                \hat{B}_{ij} &= \tfrac{1}{2} \delta_{ab} \big(\hat{x}^a_i \hat{p}^b_j + \hat{p}^a_j \hat{x}^b_i \big)\\
                \hat{C}_{ij} &= \tfrac{1}{2} \delta_{ab} \hat{p}_i^a \hat{p}_j^b
            \end{aligned}
        \end{equation}
        where summation over repeated indices is implied. We will then assume that the Hamiltonian is a function of these invariants,
         \begin{equation}
            \hat{H}_N \equiv N h[\hat{A}_{ij}, \hat{B}_{ij}, \hat{C}_{ij}]
        \end{equation}
        where the function $h[\hat{A}, \hat{B}, \hat{C}]$ has no explicit $N$ dependence. In the example of the previous section, we can convert (\ref{eq:hydrogen}) to this form by taking $r \to N^2 r$ and $\mathcal{H} \to N^3 \mathcal{H}$, in which case $\hat{A} = \frac{1}{2} \hat{\mb{x}}^2$, $\hat{B} = \hat{\mb{x}} \cdot \hat{\mb{p}}$ and $\hat{C} = \frac{1}{2} \hat{\mb{p}}^2$, and $h[\hat{A}, \hat{B}, \hat{C}] = \hat{C} - (2 \hat{A})^{\sminus \frac{1}{2}}$.

        The coherent state basis in \S\ref{sec:classicalLimits} provided a particularly convenient way of representing the Hilbert space of the theory, in which the factorization of ``classical'' operators as $\hbar \to 0$ could be easily proved and whose labels provided a classical phase space for the $\hbar = 0$ dynamics. The strategy of~\cite{Yaffe:1981vf} was to argue that the group generated by the operators $\hat{A}_{ij}$ and $\hat{B}_{ij}$ could be used to construct a similar family of coherent states. This family provides a resolution of the identity on the $\lab{O}(N)$-invariant subspace of the theory and a classical phase space on which the $N = \infty$ dynamics localizes.

        The result of this construction is a set of coherent states, labeled by a complex symmetric $N_\lab{s}\times N_\lab{s}$ matrix $z_{ij}$, which in the $x_i^a$ basis takes the form
        \begin{equation}
            \langle \es \{x_i^a\} \ess | \ess z(N) \rangle =\det \!\left[\tfrac{N}{2\pi}\big(z + \bar{z}\big)\right]^{\frac{N}{4}} \exp\!\left(\!\ess-\tfrac{1}{2} N z_{ij} \delta_{ab} x_{i}^a x_{j}^b\right)\,. \label{eq:largeNCoherent}
        \end{equation}
        Operators which are $\lab{O}(N)$ invariant, like $\hat{A}_{ij}$, $\hat{B}_{ij}$, and $\hat{C}_{ij}$ and thus $h[\hat{A}, \hat{B}, \hat{C}]$, can be shown to classicalize, in the sense that expectation values factorize and are completely determined by the one-point functions,
        \begin{equation}
            \begin{aligned}
                [A(z)]_{ij} &= \tfrac{1}{2} w_{ij} \\ 
                [B(z)]_{ij} &= (v w)_{ij} \\
                [C(z)]_{ij} &= \left(\tfrac{1}{2} v w v + \tfrac{1}{8} w^{\sminus 1}\right)_{ij}
            \end{aligned} 
        \end{equation}
        where $[A(z)]_{ij} \equiv \langle z(N) \ess | \hat{A}_{ij} | \ess z(N) \rangle$ denotes the symbol of the operators and we have introduced the real symmetric $N_\lab{s} \times N_\lab{s}$ matrices $w = (z + \bar{z})^{\sminus 1}$ and $v = \frac{1}{2} i ( z - \bar{z})$, so that $z = \tfrac{1}{2} w^{\sminus 1} - i v$. Furthermore, $w$ is positive definite. The $N = \infty$ theory then reduces to a classical Hamiltonian,
        \begin{equation}
            h_\lab{cl}[v, w] = h\big[\tfrac{1}{2} w, v w, \tfrac{1}{2} v w v + \tfrac{1}{8} w^{\sminus 1}\big]\,, \label{eq:classicalHamLargeN}
        \end{equation} 
        and the vacuum state of the theory as $N \to \infty$ is given by a coherent state (\ref{eq:largeNCoherent}) with $z$ determined by the $v$ and $w$ that minimize this Hamiltonian.
        
        For instance, in previous section's example (\ref{eq:hydrogen}), the $N = \infty$ theory is that of a classical particle with angular momentum $\frac{1}{4}$ in a Coulombic potential,
        \begin{equation}
            h_\lab{cl}(v, w) = \frac{1}{2} v^2 w + \frac{1}{8 w} - \frac{1}{\sqrt{w}} = \frac{1}{2} p^2 + \frac{1}{8 r^2} - \frac{1}{r}\,,
        \end{equation}
        where we have identified $w = r^2$ and $v = p/r$. The vacuum state in the large-$N$ limit is then given by the coherent state~(\ref{eq:largeNCoherent}) with $p = 0$ and $r = \frac{1}{4}$, or $z = 8$. In the language of the previous section, we can identify the rescaled coordinate $\zeta$ with $r$ and $\langle \zeta^k \rangle = w^{k/2}$.

        However, this formalism can also be used to describe more complicated $\lab{O}(N)$ theories in the large-$N$ limit. For instance, the $\lab{O}(N)$ vector model in $d$ spatial dimensions is defined by the Hamiltonian 
        \begin{equation}
            \mathcal{H} = N \int\!\ud^d  x \left[\tfrac{1}{2} \pi^a \pi^a + \tfrac{1}{2} \nabla \phi^a \nabla \phi^a + \tfrac{1}{2} \mu^2 \phi^a \phi^a + \tfrac{1}{4} \lambda \big(\phi^a \phi^a\big)^2\right]
        \end{equation}
        and equal-time canonical commutation relations $[\phi^a(\mb{x}), \pi^b(\mb{y})] = i N^{\sminus 1} \delta^{ab} \delta^\floq{d}(\mb{x} - \mb{y})$. The corresponding family of coherent states still take the form (\ref{eq:largeNCoherent}) but now $z_{ij} \to z(\mb{x}, \mb{x}')$ is a bilocal scalar field, as are $v(\mb{x}, \mb{x}')$ and $w(\mb{x}, \mb{x}')$. The  associated $N = \infty$ Hamiltonian (\ref{eq:classicalHamLargeN}) is 
        \begin{equation}
        	\begin{aligned}
            	h_\lab{cl}[v, w] &= \int\!\ud^d x\, \ud^d x'\, \ud^d x'' \left[\tfrac{1}{2} v(\mb{x}, \mb{x}') w(\mb{x}', \mb{x}'') v(\mb{x}'', \mb{x})\right] \nonumber \\
            	&+ \int\!\ud^d x \left[\tfrac{1}{8} w^{\sminus 1}(\mb{x}, \mb{x}) + \tfrac{1}{2}\!\left(-\nabla_{\mb{x}}^2 + \mu^2\right) \! w(\mb{x}, \mb{x}')|_{\mb{x}' = \mb{x}} + \tfrac{1}{4} \lambda w(\mb{x}, \mb{x})^2 \right]\,,
            \end{aligned}
        \end{equation}
        and the vacuum coherent state can be found by minimizing this Hamiltonian~\cite{Yaffe:1981vf}, yielding $v_0(\mb{x}, \mb{x}') = 0$ and
        \begin{equation}
       		w_0(\mb{x}, \mb{x}') = \frac{1}{2} \int\!\!\frac{\ud^d k}{(2 \pi)^d} \frac{\e^{i \mb{k} \cdot (\mb{x} - \mb{x}')}}{\sqrt{\mb{k}^2 + \mu^2 + \lambda \sigma}} = \langle \phi^a(\mb{x}) \phi^a(\mb{x}') \rangle\,,
        \end{equation}
        where $\sigma$ is defined by the gap equation $\sigma = w_0(\mb{x}, \mb{x})$. At large distances, the $\lab{O}(N)$-invariant two-point function $w_{0}(\mb{x}, \mb{x}') = \langle \phi^a(\mb{x}) \phi^a(\mb{x}') \rangle \sim \exp \left(-m |\mb{x} - \mb{x}'|\right)$ with $m^2 = \mu^2 + \lambda \sigma$, and so this limiting theory contains an $\lab{O}(N)$ vector multiplet of particles with mass $m$. Furthermore, note that it is the bi-local \emph{two-point function} $w(\mb{x}, \mb{x}')$ that factorizes as $N \to \infty$, and not a local degree of freedom like a scalar field~$\Phi(\mb{x})$ as in \S\ref{sec:scalarFields}.

        This minimization identifies a particular value for the complex symmetric $N_\lab{s} \times N_\lab{s}$ matrix $z_{ij}$, which we will call $\Omega = \frac{1}{2} w_0^{\sminus 1} - i v_0$, that yields the vacuum state of the $\lab{O}(N)$ vector model as $N \to \infty$,
        \begin{equation}
            \langle \es \{x_i^a\} \ess | \ess \Omega(N) \rangle =\det \!\left[N \Omega/\pi\right]^{\frac{N}{4}} \exp\left(\!\ess-\tfrac{1}{2} N \Omega_{ij} \delta_{ab} x_{i}^a x_{j}^b\right)\,. \label{eq:largeNCoherentVac}
        \end{equation}
        Typically, these vacua have no ``classical momenta,'' $[v_0]_{ij} = 0$, and so $\Omega_{ij}$ is both real and symmetric and can therefore be diagonalized. This diagonal basis is just the Fourier basis in the above because (\ref{eq:largeNCoherentVac}) describes the vacuum state of a translationally invariant quantum field theory. In what follows, we will first compute the information metric associated with $N$ assuming that $\Omega_{ij}$ is real. However, we will later relax this, and only assume that $\Omega_{ij}$ can be diagonalized, as we might expect of any translationally invariant state.

        The state (\ref{eq:largeNCoherentVac}) is defined on the full $L^2(\mathbb{R}^{N \!\ess N_\lab{s}})$ Hilbert space and, as in the previous section, we must project it onto the $\lab{O}(N)$-invariant subspace $L^2(\mathbb{R}_+^{N_\lab{s}})$ with $N$-independent measure. To do so, we first introduce the diagonal spatial basis $\Omega_{ij} \delta_{ab} x_i^a x_j^b = \Omega_\alpha \delta_{ab} y_\alpha^a y_\alpha^b = \Omega_\alpha r_\alpha^2$, and then project onto the $N$-independent radial basis denoted by $\rho_\alpha$ to find the ``reduced'' pure state,
        \begin{equation}
            \langle \es \{\rho_\alpha\} \es | \ess \tilde{\Omega}(N)\rangle= \prod_{\alpha = 1}^{N_\lab{s}} \, \frac{\left[ N \Omega_\alpha \right]^{N/4}}{\left[\frac{1}{2}\Gamma\big(\frac{N}{2}\big)\right]^{1/2}} \, \rho_{\alpha}^{\frac{1}{2}(N-1)} \e^{\sminus \frac{1}{2}N \es \Omega_\alpha \rho_\alpha^2}\,. \label{eq:reducedStateVec}
        \end{equation}
        The K\"{a}hler potential is then $\Omega$-independent, 
        \begin{equation}
            \mathcal{K}(N, N') = \log \,\langle \ess\es \tilde{\Omega}(N) \ess | \ess \es \tilde{\Omega}(N') \rangle = N_\lab{s} \ess \log\!\!\ess\es\left[\frac{\Gamma\big(\frac{1}{4}[N+N']\big)}{\big(N + N'\big)^{\frac{1}{4}(N + N')}} \frac{(2N)^{N/4} (2N')^{N'/4}}{[\Gamma\big(\frac{1}{2}N\big) \Gamma\big(\frac{1}{2} N'\big)]^{\frac{1}{2}}}\right]\,,
        \end{equation}
        and the information metric associated with $N$ is (\ref{eq:qimKahler}) then
        \begin{equation}
        	\ud s^2 = \frac{N_\lab{s}}{8} \frac{\ud N^2}{N^2} \left[\frac{1}{2} N^2 \psi^\floq{1}(N/2) - N\right] \sim \frac{N_\lab{s}}{8} \frac{\ud N^2}{N^2}\,, \mathrlap{\qquad N \to \infty\,.}
        \end{equation}
        As we would expect, there are exactly $N_\lab{s}$ degrees of freedom factorizing in a Gaussian way, and so the quantum information metric displays the familiar logarithmic singularity as $N \to \infty$ with the standard coefficient $N_\lab{s}/8$.

        We can also perform the reduction (\ref{eq:reducedStateVec}) even if $\Omega_{ij}$ is complex albeit still diagonalizable, as we should expect it to be if $|\Omega(N)\rangle$ describes a translationally invariant state. The quantum information metric then becomes
        \begin{equation}
            \ud s^2 = \frac{N_\lab{s}}{8} \frac{\ud N^2}{N} \left(\frac{1}{2} N \psi^\floq{1}(N/2) - 1 + \frac{1}{N_\lab{s}} \sum_{\alpha = 1}^{N_\lab{s}} \left[\frac{\lab{Im}\, \Omega_\alpha}{\lab{Re}\,\Omega_\alpha}\right]^2\right) \sim \frac{ \mathcal{C} N_\lab{s}}{8} \frac{\ud N^2}{N}\,,
        \end{equation}
        where we have introduced the coefficient $\mathcal{C} = N_\lab{s}^{\sminus 1} \sum_{\alpha=1}^{N_s} \left[\lab{Im}\, \Omega_\alpha/\lab{Re}\, \Omega_\alpha\right]^2 \propto \int\!\ud^d x\, \ud^d x' \langle \hat{\phi}^a(\mb{x}) \hat{\pi}^a(\mb{x}')\rangle^2$. Like (\ref{eq:classicalLimitMet}), we see that a non-trivial phase causes the infinite distance singularity to become more severe. This is due to the fact that, when $v(\mb{x}, \mb{x}') \neq 0$, expectation values of the derivatives $\partial/\partial \rho_\alpha$ (which are like expectation values of the ``radial'' canonical momenta) in the state (\ref{eq:reducedStateVec}) diverge linearly with $N$, much in the same way expectation values of $\hat{\mb{p}}/\hbar = - i \nabla$ diverged as $\hbar \to 0$ in a coherent state~(\ref{eq:coherentStates}) with nontrivial momentum $\mb{p}_0 \neq 0$. Regardless, the information metric always has an infinite distance singularity as $N \to \infty$ in $\lab{O}(N)$-symmetric~models. 

\section{Quantum Gravitational Implications} \label{sec:swamp}

	One of the primary motivations for this work was to better understand the physical interpretation of, and bottom-up motivations for, the Swampland Distance Conjecture (SDC)~\cite{Ooguri:2006in,Brennan:2017rbf,Palti:2019pca,vanBeest:2021lhn,Grana:2021zvf}. To establish notation, let us consider a $(d+1)$-dimensional effective field theory of $n$ massless scalar fields $\phi^a$ coupled to gravity, with action~\cite{vanBeest:2021lhn}
    \begin{equation}
        S = \frac{1}{2} \ess M_\lab{pl}^{d-1} \!\int\!\ud^{d+1} x\, \sqrt{\minus h} \left(R - G^\slab{ms}_{\smash{ab}}(\phi) \nabla_\mu \phi^a \nabla^\mu \phi^b\right),
    \end{equation}
    where $h_{\mu \nu}$ is the spacetime metric, $R$ its associated scalar curvature, and $M_\lab{pl}$ is the Planck mass. The field space metric $G^\slab{ms}_{\smash{ab}}(\phi)$ defines a notion of distance between theories with different vacuum expectation values $\langle \phi^a \rangle = \varphi^a$, and two theories connected by a geodesic path $\varphi^a(t)$, with $t \in [0, 1]$ and endpoints $\varphi^a(0) = \varphi_0^a$ and $\varphi^a(1) = \varphi_1^a$, are said to be separated by a distance
    \begin{equation}
        d(\varphi_1, \varphi_0) = \int_0^1\!\ud t\, \sqrt{G^\slab{ms}_{{ab}}(\varphi) \dot{\varphi}^a(t) \dot{\varphi}^b(t)}\,.
    \end{equation}
    The most basic form of the SDC states that, in a consistent quantum gravitational theory, a tower of weakly-coupled fields must become exponentially light as we approach an infinite distance point,
    \begin{equation}
        M(\varphi_1) \sim M(\varphi_0)\ess \e^{\sminus \lambda \ess d(\varphi_1, \varphi_0)}\,,\mathrlap{\qquad d(\varphi_1, \varphi_0) \to \infty\,,} \label{eq:sdc}
    \end{equation}
    where $\lambda$ is an $O(1)$ constant. A similar conjecture holds for the Zamolodchikov metric, whose infinite distance points are tied to towers of higher spin fields~\cite{Seiberg:1999xz,Baume:2020dqd,Perlmutter:2020buo}.

	This conjecture presents a number of puzzles. The SDC (\ref{eq:sdc}) demands that any effective description of the theory in terms of local fields fails as we approach an infinite distance point---there is a spectacular breakdown of effective field theory since it is impossible to consistently couple Einstein gravity to infinitely many weakly-coupled fields. In these limits, it seems as if consistent quantum gravitational theories must become extremely complicated. But, why then are these infinite distance points always associated to an emergent and simple weakly-coupled description? Decompactification limits are particularly innocuous---why should a low-energy observer care if they have to use a well-controlled higher-dimensional description of the physics that is both simple and local? Why would they bother with the lower-dimensional description in the first place? More importantly, \emph{why} is the SDC satisfied in so many examples? It is simply a lamppost effect, only a property of controlled supersymmetric string compactifications, or is it a more fundamental property of all quantum gravitational theories?

    A similar situation occurs when a system is brought near a phase transition. Here, an ``effective field theory'' description also breaks down in the sense that mean-field theory fails~\cite{Goldenfeld:2018lop,Sachdev:2011qpt,Kardar:2007spf}. We understand why: an energy gap closes and whatever order parameter used to describe the system necessarily exhibits large fluctuations that invalidate the mean-field approach. Even though our effective description breaks down here, we now also know the correct or most convenient ``organizational principle'' to use to understand the physics of a system near a critical point. That is, these large fluctuations are scale-invariant, and so we should leverage this symmetry to describe such systems. We should use conformal field theory, not mean field theory. As we reviewed in \S\ref{sec:ftInfoGeo}, such quantum critical points are always at finite distance in the information metric; what is the correct ``organizational principle'' to use at an infinite distance point?

	In this paper, we argued that unitarity ties infinite distance points in the information metric to factorization limits. One cannot have an infinite distance limit in the information metric unless there are some ``fundamental'' degrees of freedom whose expectation values factorize. Since the field space and Zamolodchikov metrics are both proportional to the information metric, they inherit this interpretation. We gave a general argument for this fact in Section~\ref{sec:infFact} and then explicitly demonstrated it in a variety of examples in both Sections~\ref{sec:examples} and \ref{sec:largeN}. We will assume that this holds broadly, and now describe how this relationship between infinite distance and factorization resolves a number of conceptual puzzles surrounding the SDC and provides a simple bottom-up explanation for why it is satisfied in consistent quantum gravitational theories.

	For instance, this relationship between infinite distance and factorization yields an information-theoretic explanation for why infinite distance limits are always associated with an emergent weakly-coupled description. A factorization limit \emph{is} a weak-coupling limit, as by definition there necessarily exists a description of the theory whose connected correlation functions or expectation values vanish. This can be made more explicit by considering a particular limiting distribution for a single random variable $x$, wherein expectation values take the form
	\begin{equation}
		\langle x^k \rangle = \int\!\ud x\, x^k \, \epsilon^{\sminus 1} \eta\big(\tfrac{x - \mu}{\epsilon}\big) \big(1 + \cdots \big) = \int\!\ud y\, (\mu + \epsilon y)^k \eta(y) \big(1 + \cdots\big) \sim \mu^k + \cdots\,,
	\end{equation}
	with $\cdots$ denoting terms that are subleading as $\epsilon \to 0$. Connected moments are thus proportional to $\epsilon$, $\langle x^k \rangle_c = O(\epsilon)$, so this is rightfully a weak-coupling limit. Factorization, and thus weak coupling, is necessary to generate an infinite distance limit. However, as the example of~\S\ref{sec:gaugeTheory} illustrates, not all weak-coupling limits are factorization limits, though they can be phrased as such with an appropriate $\epsilon$-dependent field redefinition.

    Unfortunately, the information metric is not sensitive enough to determine what the correct weakly-coupled description is, and this description is not always obvious. It may exist in a different number of spacetime dimensions, as it does in a decompactification limit, and involve a complicated rearrangement of the ``fundamental'' degrees of freedom, as it does for strongly-coupled, large-$N$ gauge theories. Furthermore, the weakly-coupled description may even be nonlocal, as it is for the tensionless string.  In cases where we have a full theory that is consistent in all limits of parameter (or moduli) space, like string theory, it is often the case that we can use string dualities to identify this weakly-coupled description, but it does not seem like this is possible just by studying the information metric. Interestingly, the existence of string dualities themselves is implied by these purely information-theoretic concerns. They are the consequence of a unitary theory consistently realizing multiple non-trivial infinite distance limits which, by their relation to factorization limits, must have a weakly-coupled description.\footnote{It is remarkable, however, that so many of these dualities are relatively simple.}

	Why, then, does consistency with quantum gravity seemingly demand that infinite distance limits are tied to towers of exponentially light fields? It is notoriously difficult to completely sequester or decouple two sectors of a gravitational theory from one another---there are generically Planck-suppressed interactions mediated by gravity that couple the two sectors to one another. More importantly, though, a field will couple to \emph{itself} through its \emph{own} self-gravity.  How, then, can \emph{any} degree of freedom become free and factorize in the presence of dynamical gravity? Said differently, we expect that there is always a type of gravitational noise that prevents factorization. This is why we expect it to be impossible to realize a factorization limit in quantum gravity unless something dramatic happens in the theory. Thus, a simple bottom-up motivation for the Swampland Distance Conjecture is that \emph{gravity abhors factorization.}

    One way a factorization limit can be realized is to simply decouple gravity by sending the Planck mass $M_\lab{pl} \to \infty$. This is ultimately the job of the tower of fields that compresses in mass (in units of $M_\lab{pl}$) as we approach an infinite distance point in (\ref{eq:sdc}). We generally expect the appearance of $N$ light fields with masses $m_n$ to renormalize and drive up the Planck mass, $M_\lab{pl}\sim N^{1/({d-1})} \Lambda_\slab{qg}$, where $\Lambda_\slab{qg}$ is the scale, which we hold fixed, at which quantum effects begin to significantly modify gravity. Once gravity decouples, observables can consistently factorize. 

    We can thus  ``derive'' the SDC (\ref{eq:sdc}) as follows. We demand that a family of quantum gravitational theories in a single parameter $\epsilon$ has observables that factorize as $\epsilon \to 0$. Furthermore, we require that these theories have a large number of initially ``heavy'' fields with masses $m_n$.  Unitarity dictates that the intensive information metric has a logarithmic singularity $\ud \tilde{s}^2 \sim  \tilde{\lambda}^{\sminus 2} \, \ud \epsilon^2/\epsilon^2$ in this limit, with $\tilde{\lambda}$ an undetermined constant that depends on both the family and the precise definition of $\epsilon$.  We can then express $\epsilon$ in terms of the intensive statistical distance $\tilde{s} = \int\ud \tilde{s}$, $\epsilon(\tilde{s}) = \epsilon_0\exp(\minus \tilde{\lambda} \tilde{s})$ as $\tilde{s} \to \infty$.  If we now demand that the gravity decouples, so that this factorization limit is consistent, then the Planck mass $M_\lab{pl}(\epsilon) \sim M_\lab{pl}(\epsilon_0) (\epsilon_0/\epsilon)$ must diverge as $\epsilon \to 0$.\footnote{Our choice of the Planck mass' dependence on $\epsilon$ is degenerate with our choice the constant $\lambda$. As such, we can restrict to the case in which $M_\lab{pl} \propto \epsilon^{\sminus 1}$  without loss of generality. Furthermore, our assumption that the masses $m_n$ remain fixed can be relaxed---as long as they do not diverge as quickly as the Planck mass, we will have $m_n(\tilde{s}) \sim m_n(0) \e^{\sminus \hat{\lambda} \tilde{s}}$ with $\hat{\lambda}$ yet another constant, assuming that the entire tower has the same asymptotics.} As long as the masses $m_n$ approach a fixed value as $\epsilon \to 0\,$, we have that
    \begin{equation}
        \frac{m_n(\tilde{s})}{M_\lab{pl}(\tilde{s})} = \frac{m_n(0)}{M_\lab{pl}(0)} \e^{\sminus \tilde{\lambda} \tilde{s}}\,. 
    \end{equation}
    That is, if we measure everything in units in which $M_\lab{pl}$ is fixed, then we find that the masses in the tower behave as $m_n(\tilde{s}) \sim m_n(0) \ess \e^{\sminus \tilde{\lambda} \tilde{s}}$. This is the distance conjecture (\ref{eq:sdc}), with $\lambda$ replaced with $\tilde{\lambda}$  and the distance on field space $d(\varphi_1, \varphi_0)$ replaced with the intensive information distance $\tilde{s}$, cf. (\ref{eq:intMetMS}). Furthermore, it is consistent to view the tower as driving the Planck mass to~infinity.

    \subsection{Holographic Perspective} \label{sec:holographic}

	This picture can be made more precise using holography. Let us consider a $d$-dimensional conformal field theory equipped with a stress-energy tensor
	\begin{equation}
		\langle T_{\mu \nu}(x_1) T_{\rho \sigma}(x_2) \rangle =  \frac{C_\slab{t}}{x_{12}^{2 d}} I_{\mu \nu, \rho \sigma}(x_{12})\,, \label{eq:setCorr}
	\end{equation}
	where $I_{\mu \nu, \rho \sigma}(x)$ is a well-known tensor~\cite{Osborn:1993cr} whose structure is fixed by conformal symmetry and $x_{12}^2 \equiv |x_1 - x_2|^2$. We will assume that this CFT is holographically dual to a quantum gravitational theory in anti-de Sitter space $\lab{AdS}_{d+1}$ with AdS radius $\ell_\lab{AdS}$ and Planck length $\ell_\lab{pl} = M_\lab{pl}^{\sminus 1}$. 

	The constant $C_\slab{t}$ in (\ref{eq:setCorr}) characterizes the number of degrees of freedom in the CFT. For instance, in two-dimensional CFTs it is directly proportional to the central charge $C_\slab{t} = 2 c$, though in higher dimensions it loses this property. We can also find a more physical measure of the number of degrees of freedom by heating the CFT to a temperature $T$ and asking how its entropy changes. By conformal symmetry, the entropy density $s$ takes the form~\cite{Kovtun:2008kw}
	\begin{equation}
		s = C_\slab{s} \left[\frac{1}{4}\frac{d-1}{d+1}\frac{\Gamma\big(\frac{d}{2}\big)^3}{\Gamma(d)} \frac{4^d \pi^{d/2}}{d^d} \right]  T^{d-1}\,. \label{eq:entropyDensity}
	\end{equation}
	The coefficient $C_\slab{s}$ is similarly proportional to the central charge $C_\slab{s} = c$ when $d =2$, and provides another way of measuring how many degrees of freedom are ``in'' the CFT. In a holographic CFT with many degrees of freedom and a large spectral gap, these two measures are equal and related to the Planck mass in the bulk,
    \begin{equation}
        C_\slab{t} = C_\slab{s} = \frac{2 \pi^{\frac{d}{2}}}{d-1} \frac{\Gamma(d+2)}{\Gamma\big(\frac{d}{2}\big)^3} \left(\frac{\ell_\lab{AdS}}{\ell_\lab{pl}}\right)^{d-1}\,,  \label{eq:cTStress}
    \end{equation}
    up to corrections of $O\big(C_\slab{t}^{\sminus 1}\big)$. In contrast, for weakly-coupled CFTs $C_\slab{t}$ and $C_\slab{s}$ are equal to one another up to an $O(1)$ coefficient. Taking $\mathcal{N} = 4$ super Yang-Mills as an example, this ratio changes as we move along the moduli space, and $C_\slab{t}/C_\slab{s} = 1$ at strong coupling while $C_\slab{t}/C_\slab{s} = 3/4$ at weak coupling~\cite{Gubser:1996de,Gubser:1998nz}. The two are never very different and so we will call them both the ``central charge'' of the CFT.  Crucially, $C_\slab{s} \to \infty$ as $C_\slab{t} \to \infty$, and vice versa, so suppressing bulk gravitational effects that are $O(C_\slab{t}^{\sminus 1})$ requires the CFT to have many degrees of~freedom. 

    For simplicity, let us assume that this CFT has a single scalar primary operator $\mathcal{O}(x)$ with scaling dimension $\Delta$ and normalized two-point function $\langle \mathcal{O}(x_1) \mathcal{O}(x_2) \rangle = 1/{x_{12}^{2 \Delta}}\,$, which is ``single-trace'' so that it corresponds to a bulk field of mass $m^2 \ell_\lab{AdS}^2 = \Delta(\Delta - d)$. We will also assume that its higher-order correlation functions factorize,
    \begin{equation}
        \langle \mathcal{O}(x_1) \mathcal{O}(x_2) \cdots \mathcal{O}(x_{n-1}) \mathcal{O}(x_n) \rangle = \langle \mathcal{O}(x_1) \mathcal{O}(x_2) \rangle \cdots \langle \mathcal{O}(x_{n-1}) \mathcal{O}(x_n)\rangle + \lab{permutations}\,,
    \end{equation}
    as we take some parameter $\epsilon \to 0$. Note that this parameter $\epsilon$ might interpolate between a sequence of different CFTs whose spectra, central charges, and OPE coefficients depend on $\epsilon$. It does not necessarily parameterize motion along the moduli space of the CFT or parent string compactification---the information metric can be assigned to any continuous (or even discrete, cf. Section~\ref{sec:largeN}) family of theories, and so it is conceptually sound to consider the most general type of factorization or infinite distance limit.  Since $\epsilon \to 0$ corresponds to a factorization limit, we generically expect that it is also at infinite distance in the intensive information metric and, assuming that observables are finite in this limit, $\ud \tilde{s}^2 \propto \ud \epsilon^2/\epsilon^2$.

    When the scaling dimension of this operator exceeds the unitarity bound, $\Delta > (d-2)/2$, $\mathcal{O}$~is called a generalized free field when $\epsilon = 0$. Following~\cite{El-Showk:2011yvt}, let us consider the operator product expansion (OPE) of the factorized four-point function,
    \begin{equation}
        \langle \mathcal{O}(x_1) \mathcal{O}(x_2) \mathcal{O}(x_3) \mathcal{O}(x_4) \rangle 
        =  \frac{1}{x_{12}^{2 \Delta} x_{34}^{2 \Delta}}\left(1 + \frac{1}{u^\Delta} + \frac{1}{v^\Delta}\right), \label{eq:factoredFour}
    \end{equation}
    in the $(12) \to (34)$ channel,
    \begin{equation}
        \langle \mathcal{O}(x_1) \mathcal{O}(x_2) \mathcal{O}(x_3) \mathcal{O}(x_4) \rangle = \frac{1}{x_{12}^{2 \Delta} x_{34}^{2 \Delta}} \sum_{\phi} f^2_{\mathcal{O} \mathcal{O} \phi} \, g_{\Delta_\phi, \ell_\phi}(u, v)\,, \label{eq:ope}
    \end{equation}
    where $u = (x_{12}^2 x_{34}^2)/(x_{13}^2 x_{24}^2)$ and $v = (x_{23}^2 x_{14}^2)/(x_{13}^2 x_{24}^2)$ are the conformal cross-ratios, $\phi$ runs over all primary operators in the theory, and $g_{\Delta_{\phi}, \ell_{\phi}}(u, v)$ is the conformal block of the primary operator $\phi$ with dimension $\Delta_\phi$ and spin $\ell_\phi$, and $f_{\mathcal{O}\mathcal{O}\phi}$ are the OPE coefficients of $\mathcal{O}$ with these operators. By demanding crossing symmetry~\cite{Fitzpatrick:2011dm,Fitzpatrick:2012yx,El-Showk:2011yvt}, one can show that factorization implies the existence of a tower of spin-$\ell$ symmetric traceless primaries of the form $\mathcal{O}_{n, \ell} = \nord{\mathcal{O} \hat{\partial}_{\ess \smash{(}\mu_1} \! \cdots \hat{\partial}_{\mu_\ell\smash{)_{\slab{t}}}} (\hat{\partial}^2)^n \mathcal{O}}$ with scaling dimension $\Delta_{n, \ell} = 2 \Delta + 2 n  + \ell$, where $\hat{\partial} = \vec{\partial} - \cev{\partial}$ is used to project out descendants, $(\mu_1 \cdots \mu_\ell)_\slab{t}$ denotes the symmetric traceless combination of indices, and the $\nord{\,\,\,}$ denotes normal ordering. These are the ``double-trace'' two-particle states that must exist in the spectrum because $\mathcal{O}$ is free, and they alone conspire to generate the full factorized four-point function (\ref{eq:factoredFour}).

	However, there is a problem. We assumed that the CFT has a stress-energy tensor $T_{\mu \nu}$, with scaling dimension $\Delta = d$ and spin $\ell = 2$, so that it was dual to a bulk theory with dynamical gravity. The stress-energy tensor also contributes to the OPE (\ref{eq:ope}), but its contribution is entirely fixed by conformal symmetry, since the OPE coefficient $f^2_{\mathcal{O} \mathcal{O} T} \propto \langle \mathcal{O} \mathcal{O} T\rangle^2 \propto \Delta^2/C_\slab{t}$ is fixed by the conformal Ward identities, and so the ``gravitational'' contribution to the four-point function (\ref{eq:factoredFour}) is similarly fixed
    \begin{equation}
        \langle \mathcal{O}(x_1) \mathcal{O}(x_2) \mathcal{O}(x_3) \mathcal{O}(x_4) \rangle \supset \frac{\Delta^2}{C_\slab{t}}\frac{1}{x_{12}^{2\Delta} x_{34}^{2 \Delta}}\frac{d^2}{(d-1)^2} u^{\frac{1}{2}(d-2)} G^\floq{2}\big(\tfrac{1}{2}(d-2), \tfrac{1}{2}(d-2), d; u, v\big)\,, \label{eq:gravInt}
    \end{equation} 
    where $G^{\floq{2}}(a, b, c; u, v)$ is another function determined by conformal invariance and defined in~\cite{Dolan:2000ut}. This contribution is not present in (\ref{eq:factoredFour}), and so this generalized free field (and thus this factorization limit) cannot be consistently coupled to a stress-energy tensor except in the limit $C_\slab{t} \to \infty$.\footnote{The conformal partial waves are orthogonal to one another, as evidenced by the existence of an inversion formula~\cite{Caron-Huot:2017vep}. In particular, this implies that other primaries with $(\Delta, \ell) \neq (d, 2)$ cannot conspire and sum up to cancel the contribution (\ref{eq:gravInt}). } Since this limit apparently requires an infinite number of degrees of freedom,  this reasoning led \cite{El-Showk:2011yvt} to conclude that CFTs with a small number of light operators, $\Delta \sim \mathcal{O}(1)$,  that nearly factorize and are thus holographically dual to Einstein gravity with a finite number of weakly-interacting light fields can only consistently arise as a subsector of a much larger CFT. The other degrees of freedom serve to consistently drive up the Planck mass and realize the factorization limit. These states must sit above a certain conformal dimension, which we denote~$\Delta_*$\footnote{Note that $\Delta_*$ is different than what is often referred to in the literature as $\Delta_\lab{gap}$, the dimension of the lightest higher spin single trace operator~\cite{Camanho:2014apa,Alday:2019qrf}, which governs the size of higher-curvature corrections in the bulk.} \cite{El-Showk:2011yvt}, in order to have a local bulk dual with a finite number of light fields, and both their number and~$\Delta_*$ must diverge as $C_\slab{t} \to \infty$.\footnote{If we think of these states with $\Delta \gtrsim \Delta_*$ as black hole microstates~\cite{El-Showk:2011yvt}, then their degeneracy is predicted by the Bekenstein-Hawking formula, $S_\slab{bh} = A/(4 \ell_\lab{pl}^2)$. As $C_\slab{t} \to \infty$, $\ell_\lab{pl} \to 0$ and their number diverges.}

\begin{figure}
		\centering
		\includegraphics[scale=1.1]{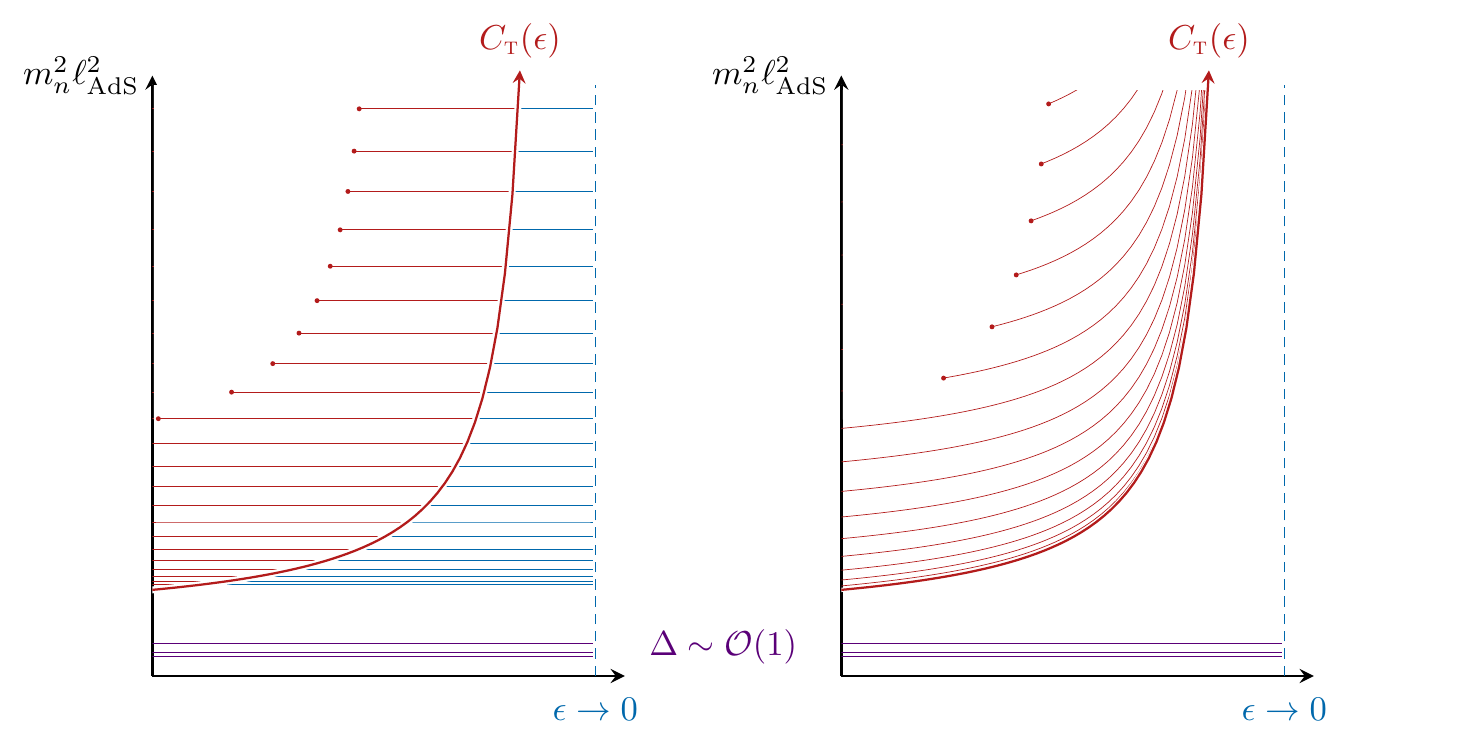}

		\caption{Schematic spectrum of scaling dimensions $\Delta_n(\epsilon) = m_n^2 \ell_\lab{AdS}^2$ in the limit of  infinite distance $\epsilon \to 0$ for a family of theories in which the SDC is satisfied [\emph{left}] and one in which it is not~[\emph{right}]. Consistency of the factorization limit requires that the bulk Planck Mass $M_\lab{pl}$, and thus the ``central charge'' of the theory $C_\slab{t}(\epsilon)$, diverges as $\epsilon \to 0$. Since $C_\slab{t}$ measures the number of degrees of freedom in the theory, new degrees of freedom must be added to the theory as $\epsilon \to 0$ for it to diverge. In principle, these new degrees of freedom can either become light, $\Delta_n(\epsilon) \gtrsim \Delta_* \ll C_\slab{t}(\epsilon)$, in which case a bulk observer would find a tower of light fields, or they could remain both strongly-coupled and heavy, $\Delta_n(\epsilon) \gtrsim \Delta_* \sim C_\slab{t}(\epsilon)$, in which case the bulk observer would only see a few fields [{\color{regal}purple}] that remain light as $\epsilon \to 0$.    \label{fig:spectrum}}
	\end{figure}

	We can then consider a very general type of infinite distance limit. If we begin with a theory that has a finite Planck mass and try to approach a factorization limit by taking some parameter $\epsilon \to 0$, then it seems necessary that $C_\slab{t} = C_\slab{t}(\epsilon)$ must also diverge in this limit, as do the number of degrees of freedom in the theory. In fact, following Section~\ref{sec:largeN}, we could simply replace $\epsilon$ for $C_\slab{t}$ and be confident that $C_\slab{t}\to \infty$ is at infinite distance. However, note that this does not necessitate the appearance of a tower of exponentially light fields, but only demands that these degrees of freedom are introduced to pump up the Planck mass. As illustrated in Figure~\ref{fig:spectrum}, this can happen as long as the gap $\Delta_*$ also diverges as $\epsilon \to 0$, so that this sequence of theories is holographically dual~\cite{Heemskerk:2009pn,Heemskerk:2010ty} to a sequence of bulk theories that are more and more weakly-coupled to Einstein gravity. This seems to allow for a factorization limit, and thus an infinite distance point, \emph{without} an associated tower of light fields.  Here, the extra degrees of freedom would be strongly-coupled with one another, ``heavy'' with $\Delta \gtrsim \Delta_*$ scaling dimensions, and would thus correspond to bulk degrees of freedom with Planck-scale masses $m_n(\epsilon) \gtrsim M_\lab{pl}(\epsilon)$. 

	Note that since $C_\slab{t} \propto (\ell_\lab{AdS}/\ell_\lab{pl})^{d-1}$, sending $C_\slab{t} \to \infty$ can be equivalent to taking the limit $\ell_\lab{AdS} \to \infty$ in Planck units, as long as we are specific about what other parameters are being held fixed. This limit is the domain of the AdS Distance Conjecture~\cite{Lust:2019zwm}, which posits that $\ell_\lab{AdS} \to \infty$ is both at infinite distance and is always associated with a tower of light fields with masses that behave as $m_n \propto \ell^{\raisebox{2pt}{$\scriptstyle \sminus \alpha$}}_{\smash{\raisebox{0pt}{$\scriptstyle \lab{AdS}$}}}$ as $\ell_\lab{AdS} \to \infty$, with $\alpha$ a positive exponent. The fact that the limit $\ell_\lab{AdS}\to \infty$ is at infinite distance is very reasonable, considering that it is holographically equivalent to the large-$N$ limits discussed in Section~\ref{sec:largeN}. 

    However, this conjecture also implies that the aforementioned mechanism of realizing an infinite distance limit sans a tower of light fields, by adding the ``extra'' degrees of freedom above a scaling dimension $\Delta_*$ which diverges with the Planck scale, is forbidden by some as-of-yet unknown quantum gravitational consistency requirement. If that is true, it would require that this well-studied class of holographic large-$N$ CFTs are somehow internally inconsistent, a fact perhaps apparent from the conformal bootstrap or hidden in the difference between $N \to \infty$ and finite-$N$ CFTs. If we think of $g = C_\slab{t}^{\ess \sminus 1/2}$ as an expansion parameter, as it appears in conformal perturbation theory~\cite{Heemskerk:2009pn}, these states have masses that are nonperturbative in $g$, $\Delta \sim g^{\sminus 2}$, and so perhaps such an inconsistency arises at the nonperturbative level.\footnote{Similar non-perturbative inconsistencies have been found by studying the thermodynamics of bulk black hole thermodynamics, ruling out extremely teeny gauge couplings~\cite{Montero:2017mdq}.} The validity of a very general form of the SDC is then linked to whether or not arbitrary scale separation (or an arbitrarily large gap in the dimensions of operators $\Delta_*$) in AdS vacua is fundamentally disallowed. From the boundary side, however, we already know of robust mechanisms to ensure a large gap in the spectrum of operators---large-$N$ factorization and strong coupling. Why can $\Delta_*$ not be arbitrarily large~\cite{Polchinski:2009ch,Heemskerk:2009pn,Alday:2019qrf,Meltzer:2017rtf,Belin:2019mnx,Kologlu:2019bco,Conlon:2020wmc}, especially if gravity in the bulk is simultaneously arbitrarily weak?\footnote{As \cite{Conlon:2020wmc} points out, this can instead be phrased as a statement about the behavior of the density of states $\rho(\Delta)$ for $1 \ll \Delta \ll C_\slab{t}$ in holographic CFTs.} Examples remain elusive~\cite{Polchinski:2009ch}, so perhaps there is a reason why this is inconsistent~\cite{Collins:2022nux}.

    \subsection{Conformal Manifolds} \label{sec:conformalManifolds}

	There is a prominent loophole in the conclusion that a factorization limit can only be realized by taking $C_\slab{t} \to \infty$~\cite{Baume:2020dqd,Perlmutter:2020buo}. It could be that, as $\epsilon \to 0$, the factorizing primary $\mathcal{O}$ begins to saturate the unitarity bound $\Delta = \frac{1}{2} (d-2)$ and becomes a legitimately free field. In this case, the OPE (\ref{eq:ope}) contains a contribution from $\mathcal{O}_{0, 2} = \nord{\mathcal{O} \hat{\partial}_{\smash{(\mu}} \hat{\partial}_{\smash{\nu)_{\slab{t}}}} \mathcal{O}}$, which is the stress-energy tensor of the completely decoupled free sector associated to $\mathcal{O}$. There is thus no need to suppress the contribution (\ref{eq:gravInt}), and we can consistently factorize these operators without sending $C_\slab{t}$, and thus the bulk Planck mass $M_\lab{pl}$ if we keep $\ell_\lab{AdS}$ fixed, to infinity. Such an infinite distance limit would thus not be associated to a divergent Planck mass.\footnote{Of course, whether or not the Planck mass diverges depends on what exactly is being held fixed. The examples discussed in this section are more general versions of the tensionless string limit discussed in \S\ref{sec:tensionless}. We can either think of the weak-coupling limit $\lambda \sim (\ell_\lab{AdS}/\ell_\lab{s})^4 \to 0$ of Type IIB string theory on $\lab{AdS}_5 \times \lab{S}^5$ as the tensionless limit $\ell_\lab{s} \to \infty$ with $\ell_\lab{AdS}$ held fixed, or the $\ell_\lab{AdS} \to 0$ limit with $\ell_\lab{s}$ held fixed. To keep $N \sim (\ell_\lab{AdS}/\ell_\lab{pl})^{3/2}$ held fixed and to match onto the discussion of this section, we must then simultaneously send $M_\lab{pl} = \ell_\lab{pl}^{\sminus 1} \to \infty$. }

	This is particularly relevant if the parameter $\epsilon$ interpolates along the conformal manifold of a family of CFTs via deformation by an exactly marginal operator. In this case, the information metric associated with $\epsilon$ is proportional to the Zamolodchikov metric, and so a factorization limit should imply an infinite distance singularity in both. However, marginal deformations of a CFT will not be able to send $C_\slab{t} \to \infty$.  It may evolve as we move along the conformal manifold, as it does as we move along the moduli space of $\mathcal{N} = 4$ super Yang-Mills from weak- to strong-coupling, but it does not diverge.\footnote{In two dimensions, this is a consequence of the fact that exactly marginal deformations do not change the central charge $c= C_\slab{t}$. In four-dimensional CFTs, the Wess-Zumino consistency conditions~\cite{Osborn:1991gm} imply that the Weyl anomaly coefficient $a \propto C_\slab{t}$ also does not change under marginal deformations, and this fact extends to higher (even) dimensions. Intuitively, we should only be able to change the number of degrees of freedom by flowing to a new fixed point, not by deforming the theory by an exactly marginal operator.} This implies that any infinite distance point in the Zamolodchikov metric is necessarily associated to an operator becoming properly free, $\Delta = \tfrac{1}{2}(d - 2)$. However, such limits are necessarily~\cite{Maldacena:2011jn,Maldacena:2012sf,Boulanger:2013zza,Alba:2015upa,Hartman:2015lfa} tied to the appearance of a higher spin symmetry and an infinite tower of higher-spin ``single-trace'' conserved currents, and thus an infinite tower of massless higher-spin bulk fields~\cite{Baume:2020dqd,Perlmutter:2020buo}, wherein the bulk theory becomes Vasiliev-like. In this case, the factorization limit is protected from gravity by an enormous gauge symmetry in the bulk, in which the spacetime dependence of the fields can be gauged away~\cite{Giombi:2010vg,Giombi:2016ejx}. We thus expect that any infinite distance point in a conformal moduli space is tied to an emergent higher-spin symmetry and an infinite tower of higher-spin fields. 

    There seem to be two qualitatively different types of infinite distance limits allowed by quantum gravity---those in which spacetime fluctuations are removed by sending $M_\lab{pl} \to \infty$ and those in which they are removed by gauge symmetry. If we think of gravity's obstruction to factorization as the fluctuations of spacetime providing an irreducible statistical ambiguity in correlators, then it makes sense that this ambiguity can be removed by either turning the fluctuations off or rendering them unphysical gauge artifacts. This is essentially the content of the Emergent String Conjecture though, in light of the loophole presented in \S\ref{sec:holographic}, we can say nothing about any towers of light fields.

	Our information-theoretic arguments do not make any claims about what type of operator must factorize to realize an infinite distance limit. It could be that a class of non-primary operators (or non-trivial combinations of primaries and their descendants) factorize to create an infinite distance point, in which case the conformal bootstrap arguments that we relied on do not apply and there may be more general factorization limits in conformal theories. Furthermore, we could consider quantum gravitational theories that do not live in asymptotically anti-de Sitter space and are not dual to a boundary conformal field theory. Regardless of the specifics, we expect consistent quantum gravitational theories to change qualitatively as we approach a factorization limit, as there must be something that disrupts gravity's universal attraction.

\newpage

\section{Conclusions} \label{sec:conclusions}
    
    Information theory's unique power lies in its ability to characterize a family of physical theories, or anything that \emph{predicts}, without having to restrict to a particular set of predictions or observables. It is thus a natural tool for theoretical physics, whose goal is to characterize the space of consistent physical theories and understand the range of possible physical behaviors among them. It is astonishing that so much information-theoretic structure can be captured geometrically via the information metric, and doubly so that this matches onto more familiar notions of distance between theories when restricted to supersymmetric or conformal field theories.  

    Singularities in the information metric denote exceptional theories which are qualitatively different from other members of the family. For example, quantum critical points are associated with finite distance metric singularities. The goal of this paper was to understand what behavior a family of theories must display to generate an infinite distance metric singularity. We argued that infinite distance points in the metric necessarily correspond to limits in which the family degenerates, assigning finite probability to a measure zero set of events. In these limits, expectation values factorize and statistical ambiguity vanishes. This matches with the intuition that since the information metric places theories further apart if they make very different predictions, infinite distance singularities represent those theories that are immediately distinguishable from other members in the family. Under reasonable assumptions, we found that unitarity implies that infinite distance singularities are universally logarithmic.
    
    We illustrated this relationship in a variety of examples in Sections~\ref{sec:examples} and~\ref{sec:largeN}, and used it in Section~\ref{sec:swamp} to provide a bottom-up motivation for the Swampland Distance Conjecture. Gravity universally interacts with stress-energy and thus disrupts factorization limits, and so families of consistent gravitational theories must effectively decouple gravity to realize an infinite distance limit. This perspective suggests a potential way around the Swampland Distance Conjecture consistent with quantum gravity.

    	We now close with some observations and potential future directions: 
        \begin{itemize}
            \item From a purely information-theoretic perspective, it would be interesting to understand how ``far'' away from a factorization limit one can be. A classical probability distribution cannot have an arbitrary shape---there is a ``duality'' between how broad the distribution is and how tall it is. We also know that the infinite breadth and infinite height limits are both factorization limits in ``conjugate'' variables---if we have a parameter that interpolates between the two, is there a notion of being perfectly in the ``middle'' of these infinite distance points? Said differently, how strongly-coupled can a unitary theory be? Does the fact that our universe permits a weakly-coupled description imply that we are ``near'' an infinite distance point? 

            \item In Section~\ref{sec:largeN}, we studied the information metric associated with the large-$N$ limits of a number of relatively simple theories and used the apparent equivalence between factorization and infinite distance limits to argue that all limits which enjoy large-$N$ factorization are at infinite distance. It would be interesting to explicitly show this for large-$N$ gauge theories.

            \item In the previous section, we argued that a factorization limit, in which the information metric takes the form $\ud \tilde{s}^2 \sim \tilde{\lambda}^{\sminus 2} \ud \epsilon^2/\epsilon^2$ as $\epsilon \to 0$, combined with both the ``consistency condition'' that the Planck mass diverged and the assumption that the masses of some fields did not diverge with it, implied that those fields became exponentially light $m_n \propto \e^{\sminus \tilde{\lambda} \tilde{s}}$ as $\tilde{s} \to \infty$ in units of the Planck mass. Is there a fundamental information-theoretic reason why the coefficient $\tilde{\lambda}$ must be bounded from below? The answer is no: the coefficient may be as small as we like and is only required to be positive. This is evidenced by the saddle-point calculation (\ref{eq:universalCIM}), where we may take the order of the saddle $p \propto \tilde{\lambda}^{2}$ smaller and smaller, which is equivalent to approaching the factorization limit with a sequence of distributions that have heavier and heavier tails. 

            However, it could be that consistency under dimensional reduction~\cite{Etheredge:2022opl} restricts all factorization limits to be at least Gaussian, i.e. $p \geq 2$ in (\ref{eq:universalCIM}). It could also be that this is simply a lamppost effect and (as we might expect) the factorization limits that we have observed are Gaussian, in which case the infinite distance singularity appears with a universal coefficient. It would be interesting to understand how additional constraints placed on a family of theories might dictate how that family may approach a factorization limit. For instance, this has been done for the higher-spin points discussed in \S\ref{sec:conformalManifolds} in~\cite{Maldacena:2012sf}.

            Relatedly, it would be extremely useful to determine the precise numerical relationship between the intensive information metric defined in Section~\ref{sec:examples} and the moduli space and Zamolodchikov metrics used elsewhere. This would require a careful accounting of the regularization and renormalization schemes used for each.

            \item It is widely expected that a global symmetry always emerges at an infinite distance point, and there are a plethora of examples in which this emergent symmetry can be explicitly identified. For instance, a massless scalar field necessarily enjoys an infinite-dimensional higher-spin symmetry~\cite{Maldacena:2011jn}, while a free massive scalar field also has an infinite number of conserved charges corresponding to the number of particles in each momentum mode~\cite{Frishman:2010zz}. Perhaps the most obvious example is when a gauge coupling vanishes $e \to 0$, cf. \S\ref{sec:gaugeTheory}, in which case the gauge field decouples and the gauge symmetry ``reduces'' to a global symmetry. A more advanced version of this occurs at any infinite distance point in the moduli spaces of Calabi-Yau threefolds, where the two-form dual of an axion decouples and a global two-form symmetry emerges~\cite{Lanza:2021udy,Grimm:2018ohb,Corvilain:2018lgw,Gendler:2020dfp}. Is there a systematic way of identifying symmetries that emerges as we move towards an infinite distance point? 

            Intuitively, the prohibition of factorization and global symmetries in quantum gravity are intrinsically related to one another: gravity forces everything to fluctuate. However, we should be careful to distinguish between two possible cases: those in which the emergent symmetry acts on the factorizing degrees of freedom and those in which it does not. The last two examples mentioned above are instances of the latter, where a gauge field freezes out and those degrees of freedom that were consistently coupled to it then enjoy an emergent global symmetry. However, from the bottom-up, there seems to be no reason why this particular type of symmetry must always appear at infinite distance---an infinite distance limit does not care \emph{which} degrees of freedom are factorizing. Is this type of emergent symmetry a lamppost effect---the byproduct of focusing on a particular factorization limit, like in \S\ref{sec:conformalManifolds}? Or is there another quantum gravitational principle that demands its existence?

            \item Related to the above, it is strongly suspected~\cite{Grimm:2018ohb,Corvilain:2018lgw,Gendler:2020dfp,Heidenreich:2020ptx} that the Swampland Distance Conjecture is related to the Weak Gravity Conjecture~\cite{Arkani-Hamed:2006emk,Harlow:2022gzl} for any infinite distance point in which a gauge coupling vanishes. Is there a way of using gravity's obstruction to factorization to give a bottom-up argument for the Weak Gravity Conjecture and its sibling the Repulsive Force Conjecture~\cite{Palti:2017elp,Lust:2017wrl,Lee:2018spm,Heidenreich:2019zkl}? If taking a gauge coupling $e \to 0$ is a factorization limit, we expect that the Planck mass must also diverge $M_\lab{pl} \to \infty$, and so that there will be states that become light with respect to the Planck mass in the limit as $e \to 0$. But why is the specific bound $\sqrt{2} e |Q| \geq M/M_\lab{pl}$ in four dimensions, which dictates that the Planck mass diverges as $M_\lab{pl} \propto e^{\sminus 1}$ if the charge-to-mass ratio is held fixed~\cite{Gendler:2020dfp}, satisfied by states (or a tower) with quantized charge~$Q$ and mass $M$ in quantum gravitational theories?  

            \item Similarly, can factorization provide a better bottom-up motivation for the Axionic Weak Gravity Conjecture? It is interesting to note that the axion decay constant in either the $0$-form frame, $\mathcal{L} \supset f^2 |\ud \varphi|^2$, or the $2$-form frame, $\mathcal{L} \supset f^{\sminus 2} |\ess \ud B_2|^2$, plays the role of $\hbar^{\sminus 1}$ or $\hbar$, respectively, and so we would expect that both $f \to 0$ and $f \to \infty$ (with the appropriate fields held fixed) are at infinite distance. But why should ``instanton actions'' behave as $S \sim M_\lab{pl}/f$~\cite{Arkani-Hamed:2006emk} in this limit? Or why should an energy gap close~\cite{Stout:2020uaf}? 
         
            \item The emergence proposal \cite{Heidenreich:2017sim,Heidenreich:2018kpg,Harlow:2015lma,Grimm:2018ohb,Palti:2019pca,vanBeest:2021lhn} supposes that the kinetic terms for all fields in the infrared emerge from integrating out towers of states in the ultraviolet, and are thus responsible for infinite distance singularities. If we take factorization as the fundamental driver of these singularities, then whether the emergence proposal is true depends on whether or not counterexamples like the one presented in \S\ref{sec:holographic} are consistent. 

            However, even if such counterexamples are inconsistent it seems like this ``emergence'' is a matter of perspective. We can see this in the extra-dimensional example of \S\ref{sec:towers}, in which we could either think of the infinite distance singularity as arising from a tower of lower-dimensional fields or the factorization of a \emph{single} higher-dimensional field. Likewise, in \S\ref{sec:holographic} and \S\ref{sec:conformalManifolds}, the towers of higher-dimensional bulk fields ``generating'' the infinite distance singularity are tied to the factorization of a single lower-dimensional boundary field. Who is to say which is more fundamental? Relatedly, it would be very interesting to understand why the quantum information metric is related to the kinetic terms of moduli in the first place. Is there a computational or complexity~\cite{Chapman:2017rqy,Crooks:2007mtl} interpretation to this?

            \item Finally, it would be extremely useful to better understand how the appearance of a tensionless string consistently permits a factorization limit, cf. \S\ref{sec:tensionless} and \cite{Isberg:1993av,Sundborg:1994py,Giombi:2010vg,Giombi:2016ejx}. Besides sending the Planck mass to infinity via a tower of weakly-coupled fields, are there other routes to consistently realizing factorization limits? Said differently, is the Emergent String Conjecture~\cite{Lee:2019wij} true? Are holographic CFTs with arbitrarily large $\Delta_*$ inconsistent?

        \end{itemize}

    We hope to return to some of these questions in the future.

\subsection*{Acknowledgements}
It is a pleasure to thank Lars Aalsma, Alex Alemi, Tarek Anous, Alex Cole, Naomi Gendler, Austin Joyce, Severin L\"{u}st,  Cody Long, Gr\'{e}goire Mathys, Rashmish Mishra, Miguel Montero, Aditya Parikh, Matt Reece, and Max Wiesner for very helpful conversations. We are especially indebted to Naomi Gendler for collaboration during the early stages of this project, and both Gr\'{e}goire Mathys and Matt Reece for comments on a draft. This work is supported by NASA grant \texttt{80NSSC20K0506}. 

\appendix

\phantomsection
\addcontentsline{toc}{section}{References}
\bibliographystyle{utphys}
\bibliography{factor}

\end{document}